\documentclass[12pt,titlepage]{book}

\usepackage[totalwidth=448pt,totalheight=655pt,centering]{geometry}

\usepackage{amssymb}
\usepackage[german,english]{babel}
\usepackage{slashed}
\usepackage[dvips]{graphicx}
\usepackage{chngpage}

\newcommand{\lagr}{\mathcal{L}}
\newcommand{\lint}{\mathcal{L}^{\mathrm{int}}}
\newcommand{\nn}{\nonumber}
\newcommand{\nutau}{\nu_{\tau}}
\newcommand{\beq}{\begin{equation}}
\newcommand{\eeq}{\end{equation}}
\newcommand{\beqa}{\begin{eqnarray}}
\newcommand{\eeqa}{\end{eqnarray}}
\newcommand{\sigmabar}{\bar{\sigma}}
\newcommand{\half}{{1 \over 2}}
\def\stilde{\widetilde}
\def\sbar{\overline}

\begin{document}

\frontmatter

% usage of \newpage and \input{...} instead of \include{...} due to LaTeX processing issues with arXiv

\newpage
\selectlanguage{german}

% \changetext{textheight}{textwidth}{evensidemargin}{oddsidemargin}{columnsep}
\changetext{2.0cm}{}{}{}{}

\begin{titlepage}

\begin{center}
\vspace*{-3.0cm}
\hspace*{-1.95cm}
\includegraphics[width=0.92\paperwidth]{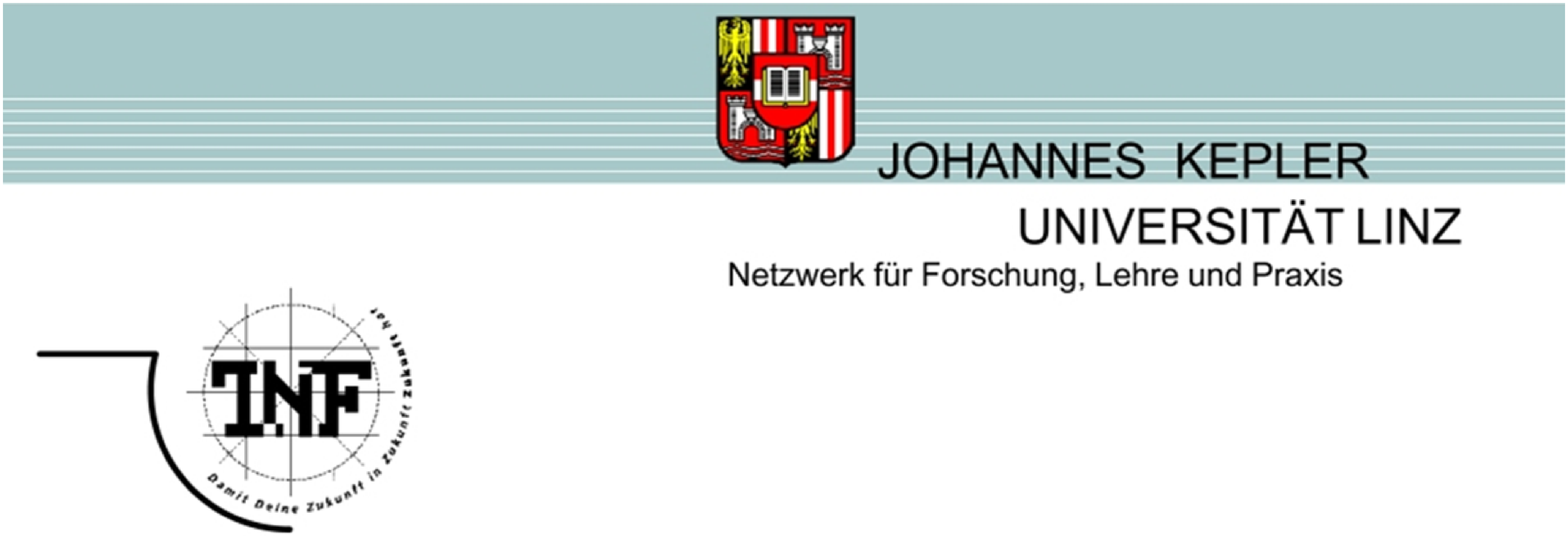}
\vspace{0.1cm}\\
\LARGE \textbf{CP Violating Asymmetries \vspace{0.2cm}\\ Induced by Supersymmetry}
\vspace{1.0cm}\\
\Large \scshape Diplomarbeit
\vspace{0.4cm}\\
\large \upshape zur Erlangung des akademischen Grades
\vspace{0.4cm}\\
\Large \scshape Diplomingenieur
\vspace{0.4cm}\\
\large \upshape in der Studienrichtung
\vspace{0.4cm}\\
\Large \scshape Technische Physik
\vspace{0.4cm}\\
\large \upshape Angefertigt am
\vspace{0.4cm}\\
\large \scshape Institut f"ur Hochenergiephysik\\
"Osterreichische Akademie der Wissenschaften
\end{center}
\vspace{1.0cm}
\large \upshape Betreuung:\\
\large \scshape Univ.~Prof.~Dr. Walter Majerotto
\vspace{0.5cm}\\
\large \upshape Eingereicht von:\\
\large \scshape Sebastian Frank
\vspace{1.0cm}\\
\large \upshape Linz, Mai 2008
\vfill
\hrule
\vspace{-0.3cm}
\begin{center}
\footnotesize \textsf{\textbf{Johannes Kepler Universit"at}}\\
\textsf{A-4040 Linz $\cdot$ Altenbergerstra"se 69 $\cdot$ Internet: http://www.jku.at/ $\cdot$ DVR 0093696}
\end{center}
\vspace{-0.3cm}
\hrule
\normalsize

\end{titlepage}

\changetext{-2.0cm}{}{}{}{}

% 2. Seite leer!
\thispagestyle{empty}
\mbox{}
\newpage

\selectlanguage{english}

\newpage
\thispagestyle{empty}
\vspace*{4.0cm}
\hspace{11.0cm}
\Large{For P\i nar}
\normalsize

\newpage
% 3. Seite leer!
\thispagestyle{empty}
\mbox{}
\newpage

\thispagestyle{plain}
\selectlanguage{german}

\section{Eidesstattliche Erkl"arung}

Ich erkl"are an Eides statt, da"s ich die vorliegende Diplomarbeit
selbstst"andig und ohne fremde Hilfe verfasst, andere als die
angegebenen Quellen und Hilfsmittel nicht benutzt bzw. die w"ortlich
oder sinngem"a"s entnommenen Stellen als solche kenntlich gemacht
habe.\\

\vspace{2.5cm}
\noindent Linz, Mai 2008
\hspace{5.5cm}
Sebastian Frank

\selectlanguage{english}

\newpage
\thispagestyle{plain}
\selectlanguage{german}

\section{Zusammenfassung}

Im Minimal Supersymmetrischen Standardmodell (MSSM) mit komplexen
Parametern ergeben Einschleifen-Strahlungskorrekturen des Zerfalls
eines Stop in ein Bottom-Quark und ein Chargino eine CP verletzende
Asymmetrie der Zerfallsbreite
\begin{displaymath}
\delta^{CP} = \frac{\Gamma^+(\tilde t_i \to b \, \tilde \chi^+_k) -
\Gamma^-(\tilde t^*_i \to \bar b \, \tilde \chi^{+
c}_k)}{\Gamma^+(\tilde t_i \to b \, \tilde \chi^+_k) +
\Gamma^-(\tilde t^*_i \to \bar b \, \tilde \chi^{+ c}_k)} \, .
\end{displaymath}
Wir f"uhren eine detaillierte, numerische Analyse von $\delta^{CP}$
als auch von $\delta^{CP} \times BR$ (wobei $BR$ das
Verzweigungsverh"altnis des Zerfalls ist) durch, wobei wir die
Abh"angigkeit von den beteiligten Parametern und komplexen Phasen
untersuchen. Dabei nehmen wir die Yukawa-Kopplungskonstanten des
Top- und Bottom-Quarks als laufend an. Wir ber"ucksichtigen die
Einschr"ankungen der Parameter, welche von der experimentellen
Obergrenze des elektrischen Dipolmoments des Elektrons ausgehen,
indem wir das Dipolmoment w"ahrend der Analyse automatisch
mitberechnen und damit kontrollieren.\\
Als Resultate erhalten wir f"ur die Asymmetrie $\delta^{CP}$ einen
Wert bis zu $\sim 24 \, \%$, abh"angig vom gew"ahlten Punkt im
Parameterraum. Die kombinierte Gr"o"se $\delta^{CP} \times
\mathrm{BR}$ wird bis zu $\sim 3.5 \, \%$ gro"s.\\
Wir kommentieren auch die M"oglichkeit einer Messung dieser
Asymmetrie am Large Hadron Collider (LHC) des CERN. Es wird m"oglich
sein, unsere Asymmetrie der Zerfallsbreite $\delta^{CP}$ am LHC zu
messen.

\selectlanguage{english}

\newpage
\thispagestyle{plain}
\section{Abstract}

In the Minimal Supersymmetric Standard Model (MSSM) with complex
parameters, one-loop corrections to the decay of a stop into a
bottom-quark and a chargino lead to the CP violating decay rate
asymmetry
\begin{displaymath}
\delta^{CP} = \frac{\Gamma^+(\tilde t_i \to b \, \tilde \chi^+_k) -
\Gamma^-(\tilde t^*_i \to \bar b \, \tilde \chi^{+
c}_k)}{\Gamma^+(\tilde t_i \to b \, \tilde \chi^+_k) +
\Gamma^-(\tilde t^*_i \to \bar b \, \tilde \chi^{+ c}_k)} \, .
\end{displaymath}
We perform a detailed numerical analysis of $\delta^{CP}$ and also
$\delta^{CP} \times BR$ (where $BR$ is the branching ratio of the
decay) for $\tilde t_1 \to b \, \tilde \chi^+_1$ and $\tilde t_2 \to
b \, \tilde \chi^+_1$, analyzing the dependence on the parameters
and complex phases involved. In addition, we take the Yukawa
couplings of the top- and bottom-quark running. We account for the
constraints on the parameters coming from the experimental limit of
the electric dipole moment of the electron by calculating and thus
checking it automatically along the way.\\
We obtain as results that the asymmetry $\delta^{CP}$ rises up to
$\sim 24 \, \%$, depending on the point in parameter space. The
combined quantity $\delta^{CP} \times \mathrm{BR}$ reaches up to
$\sim 3.5 \, \%$.\\
We also comment on the feasibility of measuring this asymmetry at
the Large Hadron Collider (LHC) at CERN. It will be possible to
measure our decay rate asymmetry $\delta^{CP}$ at LHC.

\newpage
\thispagestyle{plain}
\section{Acknowledgements}

First and foremost I especially want to thank my wife P{\i}nar and
my parents for supporting me, always believing in me and being there
for me whenever I needed them in the past years. Without them, I
could have never finished my study.\\
I want to thank Prof.~Dr.~Walter Majerotto for his very friendly
mentoring and guidance, his advice, his continuous encouragement and
for giving me the possibility to write a diploma thesis as an
untrained external student.\\
Further I want to thank Dr.~Helmut Eberl for his exceptional
supervision, his great help, his patience, the many discussions and
that he took so much time for me. He answered all my questions
promptly and friendly supported me in every way.\\
I also want to thank Dipl.Ing.~Bernhard Schrau{\ss}er for his kind
helpfulness, his discussions and his rapid responses to pressing
difficulties.\\
In addition I want to thank Dipl.Ing.~Robert Sch{\"o}fbeck,
Dipl.Ing.~Georg Sulyok, Dr.~Karol Kovarik, Dr.~Christian Weber and
Mag.~Hana Hlucha for their help and useful hints.\\
Furthermore I want to thank Eveline Ess for her cordial and helpful
administrative support. Moreover I want to thank Dr.~Gerhard Walzel
for his assistance and his efforts in establishing a working
connection between the institute and my home computer.\\
Finally I want to thank all members of the Institute of High Energy
Physics for their numerous support and for welcoming me so kindly
from the very beginning.

\tableofcontents

\mainmatter

\newpage
\chapter{Introduction}

Although the Standard Model (SM) of elementary particle physics is a
remarkably successful description of presently known phenomena, it
has to be extended to describe physics at high energies properly.
Among the theoretical and phenomenological issues that the SM fails
to address properly is the baryon asymmetry of the
universe~\cite{Chung2005}.\\
Phenomenologically, there are many reasons to believe that we live
in a baryon asymmetric universe, so that there exists much more
matter than anti-matter. One strong piece of evidence comes from the
acoustic peaks --- early universe baryon-photon plasma oscillations
--- inferred from Cosmic Microwave Background measurements (see e.g.~\cite{Bennett2003}),
which give the baryon-to-photon ratio
\begin{equation}
\eta \equiv \frac{n_B}{s} \equiv \frac{n_b - n_{\bar b}}{s} =
(6.1^{+0.3}_{-0.2}) \times 10^{-10},
\end{equation}
in which \( s \) is the entropy density (roughly the photon
density), and \( n_{b} \) and \( n_{\overline{b}} \) are the number
densities of baryons and antibaryons, respectively. This data agrees
well with big bang nucleosynthesis, which requires the baryon-to
entropy density ratio to be (see e.g.~\cite{Olive2000,Fields2002})
\begin{equation}
2.6 \times 10^{-10} \leq \eta \leq 6.2 \times 10^{-10}.
\end{equation}
The problem of baryogenesis~\cite{Sakharov1967} is to explain the
origin of this small number starting from the natural initial
condition of \( \eta =0 \), which in most cases is attained at high
enough temperatures.\\
Assuming CPT is preserved, there are three necessary conditions for
baryogenesis, usually referred to as the Sakharov
requirements~\cite{Sakharov1967}:
\begin{enumerate}
\item Baryon number violation
\item Departure from thermal equilibrium
\item Charge (C) and {\it Charge-Parity (CP) violation}.
\end{enumerate}
Although the Sakharov criteria can be met in the SM, the baryon
asymmetry generated at the electroweak phase transition is too
small. However, supersymmetric extensions of the SM can contain new
sources of CP violation which can increase and
thus might explain the baryon asymmetry of the universe.\\
Supersymmetric models introduce superpartners to every particle,
that just differ in the spin quantum number. Since we do not observe
such superpartners to the known particles with the same mass,
supersymmetry must be broken. This supersymmetry breaking leads to a
couple of new parameters. If these parameters are chosen to be
complex, supersymmetric models contain new sources of CP violation.
In the Minimal Supersymmetric Standard Model (MSSM) --- the most
promising extension of the SM --- the $U(1)$ and $SU(2)$ gaugino
(superpartner of gauge bosons) mass parameters $M_1$ and $M_2$,
respectively, the higgsino (superpartner of Higgs-boson) mass
parameter $\mu$, as well as the trilinear couplings $A_f$
(corresponding to a fermion f) may be complex. Radiative corrections
at one-loop level can then lead to new CP violating asymmetries in
addition to the small CP asymmetries in the SM coming from the
Cabibbo-Kobayashi-Maskawa-(CKM)-matrix.\\
Although large complex phases are desirable to explain
baryogenesis~\cite{Carena2003,Cohen1993,Riotto1999,Trodden1999},
there are constraints on these phases coming from the experimental
limits on the electric dipole moments (EDMs) of the electron,
neutron and Hg. Especially the complex phase of $\mu$ is highly
constrained for a typical SUSY mass scale of the order of a few
hundred GeV.\\
In this thesis we study the CP violating decay rate asymmetry
\begin{equation}
\delta^{CP} = \frac{\Gamma^+(\tilde t_i \to b \, \tilde \chi^+_k) -
\Gamma^-(\tilde t^*_i \to \bar b \, \tilde \chi^{+
c}_k)}{\Gamma^+(\tilde t_i \to b \, \tilde \chi^+_k) +
\Gamma^-(\tilde t^*_i \to \bar b \, \tilde \chi^{+ c}_k)}
\end{equation}
of the decay of a stop (bosonic superpartner of the top-quark) into
a bottom-quark and a chargino (mass eigenstate of the superpartners
of charged Higgs bosons and charged gauge bosons) in the MSSM with
complex parameters at full one-loop level.\\
The asymmetry is of course zero if CP is conserved and also vanishes
at tree-level in the case of CP violation. The complete list of all
graphs at one-loop level which contribute to this asymmetry can be
found in the Appendix~\ref{chapter:all-contributions}. They give a
contribution only if they have an absorptive part, i.e. at least a
second decay channel of $\tilde t_i$ must be kinematically possible
in addition to that into $b \, \tilde \chi^+_k$. In spite of the
overall 47 graphs one, however, expects that two graphs will
dominate. They are the two possible processes involving a gluino
$\tilde g$ (see Chapter~\ref{chapter:contributions}). Because the
gluino couples like its superpartner the gluon with the strong
interaction force, these contributions are expected to dominate over
all others, if the decay channel $\tilde t_i \to t \, \tilde g$ is
open ($m_{\tilde t_i} \geq m_t + m_{\tilde g}$).\\
As a loop-level quantity the decay rate asymmetry $\delta^{CP}$
depends on the phases of all complex parameters involved. One,
however, expects that the dependence on the phase of $A_t$ and $A_b$
is strongest (taking $\mu$ real because of the stringent EDM
constraints). We neglect the mixing in the neutral Higgs sector due
to the complex trilinear couplings of the third generation $(A_t,
A_b, A_\tau)$ (see
e.g.~\cite{Pilaftsis1998,Pilaftsis1999,Christova2002}) because they
are very small~\cite{Eberl2005}.

We perform a detailed numerical analysis for $\tilde t_1 \to b \,
\tilde \chi^+_1$ and $\tilde t_2 \to b \, \tilde \chi^+_1$ analyzing
the dependence on the parameters and phases involved. In addition,
we take the Yukawa couplings of the top- and bottom-quark running.
We account for the constraints coming from the EDM of the electron
by calculating and thus checking it automatically along the way. We
also comment on the feasibility of measuring this asymmetry at the
Large Hadron Collider (LHC) at CERN.\\
The work is organized as follows:
\begin{itemize}
\item In Chapter 2 we give an introduction into the basic concept of supersymmetry (SUSY).
\item In Chapter 3 we derive supersymmetric Lagrangians as well
as the soft supersymmetry breaking interactions.
\item In Chapter 4 we present the particle content of the Minimal Supersymmetric Standard
Model (MSSM).
\item In Chapter 5 the mass matrices of the sfermions, charginos and
neutralinos are derived.
\item In Chapter 6 we list the relevant couplings of the
most important interactions involved.
\item Chapter 7 contains some definitions which are important for the
study of CP violating asymmetries.
\item Chapter 8 provides a deeper insight into CP violation and the
decay rate asymmetry $\delta^{CP}$.
\item In Chapter 9 we list the most important processes who are
expected to yield the highest $\delta^{CP}$.
\item In Chapter 10 we discuss the numerical results and present the
conclusions.
\item Appendix A holds the complete list of all processes at
full one-loop level who can contribute to $\delta^{CP}$.
\item In Appendix B we define the Passarino--Veltman integrals and provide
them with a special argument set.
\item In Appendix C and D we derive a generic structure and a
tree-level coupling, respectively.
\item Appendix E shows how to transform from Weyl to Dirac spinors
and vice versa.
\item Finally, in Appendix F we derive the electric dipole moment
(EDM) of a fermion and calculate the EDM of the electron.
\end{itemize}

\newpage
\chapter{Supersymmetry (SUSY)}

\section{Beyond the Standard Model}

The Standard Model of elementary particle physics (SM)
\cite{Weinberg1967,Glashow1961,Salam1969} is a spectacularly
successful theory of the known particles and their electroweak and
strong forces. The SM is a gauge theory, in which the gauge group
$SU(3)_C \times SU(2)_L \times U(1)_Y$ is spontaneously broken to
$SU(3)_C \times U(1)_{EM}$ by the non-vanishing vacuum expectation
value (VEV) of a fundamental scalar field, the Higgs field, at
energies of order 100~GeV. Although the SM provides a correct
description of  all known microphysical non-gravitational phenomena
(except neutrino masses and oscillations), there are a number of
theoretical and phenomenological issues that the SM fails to address
adequately~\cite{Chung2005}:
\begin{itemize}

\item {\it Hierarchy problem.} Phenomenologically the mass of
the Higgs boson associated with electroweak symmetry breaking must
be in the range of $\cal O$(100~GeV). However, radiative corrections
to the Higgs mass are quadratically dependent on the ultra-violet
cutoff $\Lambda$, since the masses of fundamental scalar fields are
not protected by chiral or gauge symmetries. The ``natural"  value
of the Higgs mass is therefore of ${\cal O}(\Lambda)$ rather than
$\cal O$(100~GeV), leading to a destabilization of the Higgs mass
and the hierarchy of the mass scales in the SM. To achieve $m \sim
{\cal O}$(100~GeV) it is necessary to fine-tune the scalar
mass-squared parameter $m^2_0 \sim \Lambda^2$ of the fundamental
ultraviolet theory to a precision of $m^2 / \Lambda^2$.  If, for
example, $\Lambda = 10^{16}$~GeV and $m = 100$~GeV, the precision of
fine-tuning must be $10^{-28}$, which is very unnatural.

\item {\it Electroweak symmetry breaking.} In the SM, electroweak
symmetry breaking is parameterized by the Higgs boson $h$ and its
general potential $V = \mu^2 |h|^2 + \lambda |h|^4$. In order to
make this symmetry breaking happen, one has to set $\mu^2 < 0$ by
hand. This postulate is rather artificial.

\item {\it Gauge coupling unification.} The idea that the gauge couplings
undergo renormalization group evolution in such a way that they meet
at a point at a high scale lends credence to the picture of grand
unified theories (GUTs) and certain string theories. However,
precise measurements of the low energy values of the gauge couplings
demonstrated that the SM cannot describe gauge coupling unification
(see e.g.~\cite{Marciano1988}) accurately enough to imply it is more
than an accident.

\item {\it Family structure and fermion masses.}
The SM does not explain the existence of three families and can only
parameterize the strongly hierarchical values of the fermion masses.
Massive neutrinos imply that the theory has to be extended, as in
the SM the neutrinos are strictly left-handed and massless.
Right-handed neutrinos can be added, but achieving ultra-light
neutrino masses from the seesaw
mechanism~\cite{Gell-Mann1979,Yanagida1979} requires the
introduction of a new scale much larger than $\cal O$(100~GeV).

\item {\it Cosmological challenges.}
Several difficulties are encountered when trying to build
cosmological models based solely on the SM particle content. As
already mentioned in the introduction, the SM cannot explain the
baryon asymmetry of the universe. The SM also does not have a viable
candidate for the cold dark matter of the universe, nor a viable
inflaton. The most difficult problem the SM has when trying to
connect with the gravitational sector is the absence of the expected
scale of the cosmological constant.

\end{itemize}
Therefore, the Standard Model must be extended to be valid at higher
energies. Theories with {\it low energy supersymmetry} have emerged
as the strongest candidates for physics beyond the SM. The main idea
behind supersymmetry (SUSY) is an underlying symmetry between bosons
and fermions. In the simplest supersymmetric world, each SM particle
has a {\it superpartner} which differs only in the spin by $1/2$ and
is related to the original particle by a supersymmetry
transformation.\\
Since no superpartners were discovered yet, SUSY is not an exact
symmetry. They seem to have a significant higher mass, which can be
explained by spontaneous breaking of the supersymmetry. In other
words, the underlying model should have a Lagrangian density that is
invariant under supersymmetry, but a vacuum state that is not.

\section{Predictions and Successes of SUSY}

The main reasons that low energy supersymmetry is taken very
seriously are not its elegance or its likely theoretical
motivations, but its successful explanations and predictions. Of
course, these successes may just be remarkable coincidences because
there is as yet no direct experimental evidence for SUSY.
Superpartners and a light Higgs boson must be discovered or
demonstrated not to exist at the Large Hadron Collider (LHC) at
CERN. The main successes are as
follows~\cite{Chung2005,bernhard-dipl}:
\begin{itemize}

\item {\it Hierarchy problem.}
SUSY provides a solution to the hierarchy problem by protecting the
Higgs mass from large radiative corrections coming from heavy
particles. Due to the symmetry between bosons and fermions, each
contribution to the radiative correction can be completely canceled
in the case of exact SUSY. In the case of broken SUSY the
cancelation is not complete but still strong enough to render the
corrections to the Higgs mass harmless, if the masses of the new
superpartners are $\lesssim {\cal O}$(1~TeV).

\item {\it Radiative electroweak symmetry breaking.}
With plausible boundary conditions at a high scale, low energy
supersymmetry can provide the explanation of the origin of
electroweak symmetry
breaking~\cite{Ibanez1982,Alvarez-Gaume1982,Inoue1982,Inoue1984}. To
oversimplify a little, the SM effective Higgs potential has the form
$V = m^2 h^2 + \lambda h^4$. First, supersymmetry requires that the
quartic coupling $\lambda$ is a function of the $U(1)_Y$ and
$SU(2)_C$ gauge couplings $\lambda = ({g'}^2 + g^2)/2$. Second, the
$m^2$ parameter runs to negative values at the electroweak scale,
driven by the large top quark Yukawa coupling. Thus the ``Mexican
hat'' potential with a minimum away from $h=0$ is derived rather
than assumed.

\item {\it Gauge coupling unification.}
In contrast to the SM, the MSSM allows for the unification of the
gauge couplings, as first pointed out in the context of GUT models
by~\cite{Dimopoulos1981,Dimopoulos1981a,Sakai1981}. The
extrapolation of the low energy values of the gauge couplings using
renormalization group equations and the MSSM particle content shows
that the gauge couplings unify at the scale $M_G \simeq 3 \times
10^{16}$~GeV~\cite{Giunti1991,Amaldi1991,Langacker1991,Ellis1991}.

\item {\it Cold dark matter.} In supersymmetric theories, the lightest
superpartner (LSP) can be stable. This stable superpartner provides
a nice cold dark matter candidate~\cite{Pagels1982,Goldberg1983}.
Simple estimates of its relic density are of the right order of
magnitude to provide the observed amount. LSPs were noticed as good
candidates before the need for nonbaryonic cold dark matter was
established.

\item {\it Supergravity.} Gauged SUSY includes a coupling between
gravity and matter. The invariance of the Lagrangian density under a
local supersymmetry transformation leads to a quantized form of
Einstein's general relativity. However, like all known theories that
include general relativity, supergravity is still non-renormalizable
as a quantum field theory.

\item {\it Baryon asymmetry.} As already mentioned in the introduction,
SUSY may help explain the baryon asymmetry of the universe. At least
three different approaches can provide the observed baryon
asymmetry: (i) generating the asymmetry at the electroweak phase
transition via the electroweak baryogenesis mechanism (see
e.g.~\cite{Trodden1999,Riotto1999}), (ii) generating it via
leptogenesis, and (iii) the Affleck-Dine
mechanism~\cite{Affleck1985} using the decay of a scalar field in
the early universe into matter. All three mechanism need the
existence of CP violation, as the Sakharov criteria demand. The CP
violation in the SM is too small to explain baryogenesis, but SUSY
can provide sufficient additional sources of CP violation.

\end{itemize}
Supersymmetry has also made several correct
predictions~\cite{Chung2005}:
\begin{itemize}

\item  Supersymmetry predicted in the early 1980s that the
top quark would be heavy~\cite{Ibanez1983,Pendleton1981}, because
this was a necessary condition for the validity of the electroweak
symmetry breaking explanation.

\item Supersymmetric grand unified theories (GUTs) with a high fundamental
scale accurately predicted the present experimental value of
$\sin^{2} \theta_{W}$ before it was
measured~\cite{Dimopoulos1981a,Dimopoulos1981,Ibanez1981,Einhorn1982}.

\item Supersymmetry requires a light Higgs boson to
exist~\cite{Kane1993,Espinosa1993}, consistent with current
precision measurements, which suggest $M_{h} < 200$~GeV.

\item When the Large Electron-Positron Collider (LEP) began to run
in 1989 it was recognized that either LEP would discover
superpartners if they were very light or there would be no
significant deviations from the SM (all supersymmetry effects at LEP
are loop effects and supersymmetry effects decouple as superpartners
get heavier). In nonsupersymmetric approaches with strong
interactions near the electroweak scale it was natural to expect
significant deviations from the SM at LEP.

\end{itemize}
Together these successes provide a powerful indirect sign that low
energy supersymmetry is indeed part of the correct description of
nature.\\
Remarkably, supersymmetry was not invented to explain any of the
above physics. Supersymmetry was discovered as a beautiful theory
and was studied for its own sake in the early 1970s. Only after
several years of studying the theory did it become clear that
supersymmetry solved the above problems, one by one. Furthermore,
all of the above successes can be achieved simultaneously, with one
consistent form of the theory and its parameters. Low energy
supersymmetry also has no known incorrect predictions; it is not
easy to construct a theory that explains and predicts certain
phenomena and has no conflict with other experimental observations.

\section{The Superalgebra}

A supersymmetry transformation turns a bosonic state into a
fermionic state, and vice versa~\cite{primer}. The operator $Q$ that
generates such transformations must be an anticommuting spinor with
\begin{equation}
Q |{\rm Boson}\rangle = |{\rm Fermion }\rangle \qquad \qquad Q |{\rm
Fermion}\rangle = |{\rm Boson }\rangle \, .
\end{equation}
Using the Minimal Supersymmetric Standard Model (MSSM), we only deal
with a $N = 1$ supersymmetry, that means only one generator-pair is
needed. Spinors are intrinsically complex objects, so $Q^\dagger$
(the hermitian conjugate of $Q$) is also a symmetry generator.
Because $Q$ and $Q^\dagger$ are fermionic operators, they carry spin
angular momentum $1/2$, so it is clear that supersymmetry must be a
spacetime symmetry. The possible forms for such symmetries in an
interacting quantum field theory are highly restricted by the
Haag-Lopuszanski-Sohnius extension of the Coleman-Mandula
theorem~\cite{Coleman1967,Haag1975}. For realistic theories like the
SM, this theorem implies that the generators $Q$ and $Q^\dagger$
must satisfy an algebra of anticommutation and commutation relations
with the form
\begin{eqnarray}
& \{ Q_\alpha , Q^\dagger_{\dot{\alpha}} \} = 2
\sigma^\mu_{\alpha\dot{\alpha}} P_\mu \, , &  \label{eq:susy-algebra} \\
& \{ Q_\alpha, Q_\beta \} = \{ Q^\dagger_{\dot{\alpha}},
Q^\dagger_{\dot{\beta}} \} = 0 \, , & \\
& \lbrack Q_\alpha, P^\mu \rbrack = \lbrack
Q^\dagger_{\dot{\alpha}}, P^\mu \rbrack = 0 \, &
\end{eqnarray}
where $P^\mu$ is the four-momentum generator of spacetime
translations, the indices $\alpha,\dot{\alpha} = 1,2$ are the left-
and right-handed spinor indices and $\sigma$ stands for the Pauli
matrices.\\
The single-particle states of a supersymmetric theory fall into
irreducible representations of the supersymmetry algebra, called
{\it supermultiplets}. Each supermultiplet contains both fermion and
boson states, which are commonly known as {\it superpartners} of
each other. The squared-mass operator $P^2$ commutes with the
operators $Q$, $Q^\dagger$ and with all spacetime rotation and
translation operators, so it follows immediately that particles
inhabiting the same irreducible supermultiplet must have equal
eigenvalues of $P^2$, and therefore equal masses. The supersymmetry
generators $Q,Q^\dagger$ also commute with the generators of gauge
transformations. Therefore particles in the same supermultiplet must
also be in the same representation of the gauge group, and so must
have the same electric charges, weak isospin, and color degrees of
freedom.\\
Each supermultiplet contains an equal number of fermion and boson
degrees of freedom. There are in total two different types of
supermultiplets possible. The so called {\it chiral} supermultiplet
includes a two-component spin-$1/2$ Weyl fermion and a complex
spin-$0$ field, whereas the {\it gauge} supermultiplet combines a
spin-$1/2$ gaugino (the fermionic superpartner of a gauge boson) and
a spin-$1$ gauge boson.\\
In a supersymmetric extension of the SM, each of the known
fundamental particles is thus in either a chiral or a gauge
supermultiplet, and must have a superpartner with spin differing by
$1/2$ unit. What remains is to decide exactly how the known
particles fit into the supermultiplets, and to give them appropriate
names. This will be done in Chapter~\ref{chapter:mssm}, but first we
want to introduce the supersymmetric Lagrangian density and the
(soft) SUSY breaking terms.

\newpage
\chapter{Supersymmetric Lagrangians}

In this chapter we will describe the construction of supersymmetric
Lagrangians and soft supersymmetry breaking
terms~\cite{primer,bernhard-dipl}\footnote{In this work we use the
signature $(+,-,-,-)$ of the spacetime metric, contrary to the
convention found in~\cite{primer,bernhard-dipl}. On this account
some of the terms in our equations have different algebraic signs.}.
We can then apply these results to the special case of the MSSM.

\section{A Free Chiral Supermultiplet}

The simplest supersymmetric model we can build is the massless, non
interacting Wess-Zumino model~\cite{Wess1974}, which describes a
single chiral supermultiplet, consisting of a left-handed
two-component Weyl fermion $\psi$ and its superpartner, a complex
scalar field $\phi$. The simplest action we can write down is
\begin{eqnarray}
S & = & \int d^4x \> \left( \lagr_{\rm scalar} + \lagr_{\rm fermion} \right) \, , \\
\lagr_{\rm scalar} & = & \partial^\mu \phi^* \partial_\mu \phi \, , \\
\lagr_{\rm fermion} & = & i \psi^\dagger \bar{\sigma}^\mu \partial_\mu \psi \, .
\end{eqnarray}
A supersymmetry transformation should turn the scalar boson field $\phi$
into something involving the fermion field $\psi_\alpha$. The simplest
possibility for the transformation of the scalar field is
\begin{equation}
\delta \phi = \epsilon \psi \, , \qquad\qquad \delta \phi^* =
\epsilon^\dagger \psi^\dagger \, , \label{eq:delta_phi}
\end{equation}
where $\epsilon^\alpha$ is an infinitesimal, anticommuting,
two-component Weyl fermion object parameterizing the supersymmetry
transformation. As we only discuss global supersymmetry,
$\epsilon^\alpha$ is a constant, satisfying $\partial_\mu
\epsilon^\alpha=0$. The scalar part of the lagrangian now transforms
as
\begin{equation}
\delta \lagr_{\rm scalar} = \epsilon \partial^\mu \psi \> \partial_\mu \phi^* + \epsilon^\dagger \partial^\mu
\psi^\dagger \> \partial_\mu \phi \, .
\end{equation}
We would like for this to be canceled by $\delta \lagr_{\rm
fermion}$, at least up to a total derivative, so that the action
will be invariant under the supersymmetry transformation. This leads
to only one possibility (up to a multiplicative constant):
\begin{equation}
\delta \psi_\alpha = - i (\sigma^\mu \epsilon^\dagger)_\alpha \> \partial_\mu \phi \, , \qquad\qquad
\delta \psi^\dagger_{\dot{\alpha}} = i (\epsilon\sigma^\mu)_{\dot{\alpha}} \> \partial_\mu \phi^* \, .
\label{eq:delta_psi}
\end{equation}
With this guess, one immediately obtains
\begin{equation}
\delta \lagr_{\rm fermion} = - \epsilon \sigma^\mu \bar{\sigma}^\nu
\partial_\nu \psi \> \partial_\mu \phi^* + \psi^\dagger \bar{\sigma}^\nu
\sigma^\mu \epsilon^\dagger \> \partial_\mu \partial_\nu \phi \, .
\end{equation}
After applying the Pauli matrix identities
\begin{equation}
\bigl[ \sigma^\mu \bar{\sigma}^\nu + \sigma^\nu \bar{\sigma}^\mu \bigr]_\alpha{}^\beta =
2 \eta^{\mu\nu} \delta_\alpha^\beta \, , \qquad
\bigl[ \bar{\sigma}^\mu \sigma^\nu + \bar{\sigma}^\nu \sigma^\mu \bigr]^{\dot{\beta}}{}_{\dot{\alpha}} =
2 \eta^{\mu\nu} \delta_{\dot{\alpha}}^{\dot{\beta}} \, ,
\end{equation}
and using the fact that partial derivatives commute
$(\partial_\mu\partial_\nu = \partial_\nu\partial_\mu)$, it takes the form
\begin{eqnarray}
\delta \lagr_{\rm fermion} & = & - \epsilon\partial^\mu\psi \> \partial_\mu\phi^*
- \epsilon^\dagger \partial^\mu\psi^\dagger \> \partial_\mu \phi \nn \\
& & - \partial_\mu \left( \epsilon \sigma^\nu \bar{\sigma}^\mu \psi \> \partial_\nu \phi^*
- \epsilon \psi \> \partial^\mu \phi^* - \epsilon^\dagger \psi^\dagger \> \partial^\mu \phi \right) \, .
\end{eqnarray}
The first two terms here just cancel against $\delta \lagr_{\rm
scalar}$, while the remaining contribution is a total derivative. So
we arrive at
\begin{equation}
\delta S = \int d^4x \> \> \, (\delta \lagr_{\rm scalar} + \delta \lagr_{\rm fermion}) = 0 \, ,
\end{equation}
justifying our guess of the numerical multiplicative factor made in Eq.~(\ref{eq:delta_psi}).\\
We must also show that the supersymmetry algebra closes; in other
words, that the commutator of two supersymmetry transformations
parameterized by two different spinors $\epsilon_1$ and $\epsilon_2$
is another symmetry of the theory. Using Eq.~(\ref{eq:delta_psi}) in
Eq.~(\ref{eq:delta_phi}) one finds
\begin{equation}
(\delta_{\epsilon_2} \delta_{\epsilon_1} - \delta_{\epsilon_1}
\delta_{\epsilon_2}) \phi \, \equiv \, \delta_{\epsilon_2}
(\delta_{\epsilon_1} \phi) - \delta_{\epsilon_1}
(\delta_{\epsilon_2} \phi) = - i (\epsilon_1 \sigma^\mu
\epsilon_2^\dagger - \epsilon_2 \sigma^\mu \epsilon_1^\dagger) \>
\partial_\mu \phi \, .
\end{equation}
We have found that the commutator of two supersymmetry
transformations gives us back the derivative of the original field.
Since $\partial_\mu$ corresponds to the generator of spacetime
translations $P_\mu$, this equation implies the form of the
supersymmetry algebra that was foreshadowed in
Eq.~(\ref{eq:susy-algebra}).\\
For the fermion $\psi$ the commutator takes the form
\begin{equation}
(\delta_{\epsilon_2} \delta_{\epsilon_1} - \delta_{\epsilon_1}
\delta_{\epsilon_2}) \psi_\alpha = - i (\sigma^\mu\epsilon_1^\dagger)_\alpha \> \epsilon_2
\partial_\mu\psi + i (\sigma^\mu\epsilon_2^\dagger)_\alpha \>
\epsilon_1 \partial_\mu\psi \, .
\end{equation}
After applying the Fierz identity
\begin{equation}
\chi_{\alpha}\> (\xi\eta) = - \xi_{\alpha}\> (\eta\chi) - \eta_\alpha\> (\chi\xi)
\end{equation}
with $\chi = \sigma^\mu \epsilon_1^\dagger$, $\xi = \epsilon_{2}$,
$\eta = \partial_\mu \psi$, and again with $\chi = \sigma^\mu
\epsilon_2^\dagger$, $\xi = \epsilon_{1}$, $\eta = \partial_\mu
\psi$, followed in each case by an application of the identity
\begin{equation}
\xi^\dagger \bar{\sigma}^\mu \chi = -\chi \sigma^\mu \xi^\dagger =
(\chi^\dagger \bar{\sigma}^\mu \xi)^* =
-(\xi\sigma^\mu\chi^\dagger)^*
\end{equation}
we obtain
\begin{equation}
(\delta_{\epsilon_2} \delta_{\epsilon_1} - \delta_{\epsilon_1}
\delta_{\epsilon_2}) \psi_\alpha = - i (\epsilon_1 \sigma^\mu \epsilon_2^\dagger -
\epsilon_2 \sigma^\mu \epsilon_1^\dagger) \> \partial_\mu \psi_\alpha
+ i \epsilon_{1\alpha} \> \epsilon_2^\dagger \bar{\sigma}^\mu \partial_\mu\psi
- i \epsilon_{2\alpha} \> \epsilon_1^\dagger \bar{\sigma}^\mu \partial_\mu\psi \, .
\end{equation}
The last two terms vanish on-shell; that is, if the equation of
motion $\bar{\sigma}^\mu\partial_\mu \psi = 0$ following from the
action is enforced. The remaining piece is exactly the same
spacetime translation that we found for the scalar field. The fact
that the supersymmetry algebra only closes on-shell (when the
classical equations of motion are satisfied) might be somewhat
worrisome, since we would like the symmetry to hold even quantum
mechanically. This can be fixed by a trick. We invent a new complex
scalar field $F$, which does not have a kinetic term. Such fields
are called {\it auxiliary}, and they are really just book-keeping
devices that allow the symmetry algebra to close off-shell. The
Lagrangian density for $F$ and its complex conjugate is simply
\beq
\lagr_{\rm auxiliary} = F^* F
\eeq
which leads to the equations of motion $F = F^* = 0$. The new field is now included in the
supersymmetry transformation rules
\beq
\delta F = - i \epsilon^\dagger \sigmabar^\mu \partial_\mu \psi \, ,
\qquad\qquad
\delta F^* = i \partial_\mu \psi^\dagger \sigmabar^\mu \epsilon \, .
\eeq
Now the auxiliary part of the Lagrangian density
transforms as
\beq
\delta \lagr_{\rm auxiliary} = - i \epsilon^\dagger \sigmabar^\mu \partial_\mu \psi \> F^*
+ i \partial_\mu \psi^\dagger \sigmabar^\mu \epsilon \> F
\eeq
which vanishes on-shell, but not for arbitrary off-shell field
configurations. By adding an extra term to
the transformation law for $\psi$ and $\psi^\dagger$
\beq
\delta \psi_\alpha = - i (\sigma^\mu \epsilon^\dagger)_{\alpha}\> \partial_\mu\phi
+ \epsilon_\alpha F \, , \qquad \>\>
\delta \psi_{\dot{\alpha}}^\dagger = i (\epsilon\sigma^\mu)_{\dot{\alpha}}\> \partial_\mu \phi^*
+ \epsilon^\dagger_{\dot{\alpha}} F^* \, ,
\eeq
one obtains an additional contribution to $\delta \lagr_{\rm fermion}$,
which just cancels with $\delta \lagr_{\rm auxiliary}$, up to a total
derivative term. So our ``modified" theory with $\lagr = \lagr_{\rm
scalar} + \lagr_{\rm fermion} + \lagr_{\rm auxiliary}$ is still invariant
under supersymmetry transformations. Proceeding as before, one now obtains
for each of the fields $X=\phi,\phi^*,\psi,\psi^\dagger,F,F^*$ that the
supersymmetry algebra now closes also off-shell
\beq
(\delta_{\epsilon_2} \delta_{\epsilon_1} - \delta_{\epsilon_1} \delta_{\epsilon_2}) X =
- i (\epsilon_1 \sigma^\mu \epsilon_2^\dagger - \epsilon_2 \sigma^\mu \epsilon_1^\dagger) \> \partial_\mu X \, .
\eeq
The real reason for the necessity to introduce the auxiliary field $F$ is that the two fields $\phi$
and $\psi$ have a different number of degrees of freedom. While on-shell, the two real propagating
degrees of freedom of the complex scalar field $\phi$ match with the two spin polarization
states of $\psi$, there is a different situation off-shell. Off-shell, the Weyl fermion $\psi$ is a complex
two-component object and has four real degrees of freedom (half of the degrees of freedom
are eliminated by going on-shell). The difference of degrees of freedom off-shell is corrected
by the introduction of two more real scalar degrees of freedom in the complex field $F$, which
vanishes going on-shell.\\
Invariance of the action under a symmetry transformation always implies
the existence of a conserved current, and supersymmetry is no exception.
The {\it supercurrent} $J^\mu_\alpha$ is an anticommuting four-vector carrying a spinor index.
With the Noether procedure we receive
\beq
\epsilon J^\mu + \epsilon^\dagger J^{\dagger\mu} \equiv
\sum_X \, \delta X \> \frac{\delta\lagr}{\delta(\partial_\mu X)} - K^\mu
\eeq
where $X=\phi,\phi^*,\psi,\psi^\dagger,F,F^*$ and $K^\mu$ is an object
whose divergence is the variation of the Lagrangian density
under the supersymmetry transformation, $\delta \lagr = \partial_\mu K^\mu$.
Calculated explicitly, $J^\mu_\alpha$ and its hermitian conjugate become
\beq
J^\mu_\alpha = (\sigma^\nu\sigmabar^\mu\psi)_\alpha\> \partial_\nu \phi^* \, , \qquad\qquad
J^{\dagger\mu}_{\dot{\alpha}} = (\psi^\dagger \sigmabar^\mu \sigma^\nu)_{\dot{\alpha}} \> \partial_\nu \phi \, .
\eeq
Using the equations of motion, it can be shown that the supercurrent and its hermitian
conjugate are conserved separately:
\beq
\partial_\mu J^\mu_\alpha = 0 \, , \qquad\qquad \partial_\mu J^{\dagger\mu}_{\dot{\alpha}} = 0 \, .
\eeq
The corresponding conserved charges to these currents are
\beq
Q_\alpha = {\sqrt{2}}\int d^3 \vec{x} \> J^0_\alpha \, , \qquad\qquad
Q^\dagger_{\dot{\alpha}} = {\sqrt{2}} \int d^3\vec{x} \> J^{\dagger 0}_{\dot{\alpha}} \, .
\eeq
As quantum mechanical operators, they satisfy
\beq
\left[ \epsilon Q + \epsilon^\dagger Q^\dagger , X \right] = -i{\sqrt{2}} \> \delta X
\eeq
for any field $X$, up to terms that vanish on-shell. This further leads
(using the canonical equal-time commutation and anticommutation relations)
to the supersymmetry algebra
\beq
\{ Q_\alpha , Q^\dagger_{\dot{\alpha}} \} = 2 \sigma^\mu_{\alpha\dot{\alpha}} P_\mu \, ,
\qquad\qquad
\{ Q_\alpha, Q_\beta\} = \{ Q^\dagger_{\dot{\alpha}}, Q^\dagger_{\dot{\beta}} \} = 0 \, ,
\eeq
and since we only deal with global supersymmetry transformations we have
\beq
[Q_\alpha, P^\mu ] = 0 \, , \qquad\qquad
[Q^\dagger_{\dot{\alpha}}, P^\mu] = 0 \, .
\eeq
So, $Q_\alpha$ and $Q^\dagger_{\dot{\alpha}}$ finally can be really identified
as the generators of the supersymmetry transformation.

\section{Non-Gauge Interactions of Chiral Supermultiplets}

In a realistic theory like the MSSM, there are many chiral
supermultiplets, with both gauge and non-gauge interactions. In this
section, we will construct the most general possible theory of
masses and non-gauge interactions for particles that live in chiral
supermultiplets. We will find that the form of the non-gauge
couplings, including mass terms, is highly restricted by the
requirement that the action is invariant under supersymmetry
transformations.\\
We start with the Lagrangian density for a collection of free chiral
supermultiplets labeled by an index $i$, which runs over all gauge
and flavor degrees of freedom. Since we want to construct an
interacting theory with supersymmetry closing off-shell, each
supermultiplet contains a complex scalar $\phi_i$ and a left-handed
Weyl fermion $\psi_i$ as physical degrees of freedom, plus a complex
auxiliary field $F_i$ which does not propagate. The results of the
previous section tell us that the free part of the Lagrangian is
\beq
\lagr_{\rm free} = \partial^\mu \phi^{*i} \partial_\mu \phi_i
+ i {\psi}^{\dagger i} \sigmabar^\mu \partial_\mu \psi_i + F^{*i} F_i
\label{eq:lagr_free}
\eeq
where we sum over repeated indices $i$ (not to be confused with the
suppressed spinor indices), with the convention that fields $\phi_i$ and
$\psi_i$ always carry lowered indices, while their conjugates always carry
raised indices. It is invariant under the supersymmetry transformation
\beqa
&
\delta \phi_i = \epsilon\psi_i ,
\qquad\>\>\>\>\>\qquad\qquad\qquad
\phantom{xxxi}
&
\delta \phi^{*i} = \epsilon^\dagger {\psi}^{\dagger i} ,
\label{phitran}
\\
&
\delta (\psi_i)_\alpha =
- i (\sigma^\mu {\epsilon^\dagger})_{\alpha}\, \partial_\mu
\phi_i + \epsilon_\alpha F_i ,
\qquad
&
\delta ({\psi}^{\dagger i})_{\dot{\alpha}} =
i (\epsilon\sigma^\mu)_{\dot{\alpha}}\, \partial_\mu
\phi^{*i} + \epsilon^\dagger_{\dot{\alpha}} F^{*i} ,
\phantom{xxxx}
\\
&
\delta F_i = - i \epsilon^\dagger \sigmabar^\mu\partial_\mu \psi_i ,
\qquad\qquad\qquad
\phantom{xxi}
&
\delta F^{* i} =
i\partial_\mu {\psi}^{\dagger i} \sigmabar^\mu  \epsilon\> .
\phantom{xxx}
\eeqa
The most general renormalizable interaction Lagrangian
of these fields that is invariant under supersymmetry
transformation can be written as
\beq
\lagr_{\rm int} = \left (-{1 \over 2} W^{ij} \psi_i \psi_j + W^i F_i \right ) + {\rm c.c.}
\eeq
with the so called superpotential W (which is not a potential in the usual sense)
\beqa
W = {1 \over 2} M^{ij} \phi_i \phi_j + {1 \over 6} y^{ijk} \phi_i \phi_j \phi_k \, , \\
W^i = {\delta W\over \delta \phi_i} \, , \qquad W^{ij} = {\delta^2 \over \delta\phi_i\delta\phi_j} W \, .
\eeqa
$M^{ij}$ is a symmetric mass matrix for the fermion fields, and
$y^{ijk}$ is a Yukawa coupling of a scalar $\phi_k$ and two fermions
$\psi_i \psi_j$ that must be totally symmetric under interchange of
$i,j,k$. So we have found that the most general non-gauge interactions for
chiral supermultiplets are determined by a single analytic function of the
complex scalar fields, the superpotential $W$.\\
The auxiliary fields $F_i$ and $F^{*i}$ can be eliminated using their
classical equations of motion. The part of $\lagr_{\rm free} + \lagr_{\rm int}$
that contains the auxiliary fields is $ F_i F^{*i} + W^i F_{i} + W^{*}_i F^{*i}$,
leading to the equations of motion
\beq
F_i = -W_i^* \, , \qquad\qquad F^{*i} = -W^i \, ,
\eeq
so the auxiliary fields can be expressed by terms corresponding to the superpotential.
After making this replacement, the Lagrangian finally takes the form
\beq
\lagr_\mathrm{chiral} = \partial^\mu \phi^{*i} \partial_\mu \phi_i
+ i \psi^{\dagger i} \sigmabar^\mu \partial_\mu \psi_i
- {1 \over 2} \left( W^{ij} \psi_i \psi_j  + W^{*}_{ij} \psi^{\dagger i} \psi^{\dagger j} \right)
- W^i W^{*}_i \, .
\eeq
Another way of writing the Lagrangian is to divide it into kinetic-, mass- and Yukawa-coupling terms
and a scalar potential $V(\phi,\phi^*)$
\beqa
\lagr_\mathrm{chiral} &=&
\partial^\mu \phi^{*i} \partial_\mu \phi_i - V(\phi,\phi^*)
+ i \psi^{\dagger i} \sigmabar^\mu \partial_\mu \psi_i
- {1 \over 2} M^{ij} \psi_i\psi_j - {1 \over 2} M_{ij}^{*} \psi^{\dagger i} \psi^{\dagger j} \nn \\
&& - \half y^{ijk} \phi_i \psi_j \psi_k - \half y_{ijk}^{*} \phi^{*i} \psi^{\dagger j} \psi^{\dagger k}
\eeqa
with
\beqa
V(\phi,\phi^*) = W^k W_k^* = F^{*k} F_k =
\phantom{xxxxxxxxxxxxxxxxxxxxxxxxxxxxxxxxxxx}
&&
\nonumber
\\
M^*_{ik} M^{kj} \phi^{*i} \phi_{j}
+{1\over 2} M^{in} y_{jkn}^* \phi_i \phi^{*j} \phi^{*k}
+{1\over 2} M_{in}^{*} y^{jkn} \phi^{*i} \phi_j \phi_k
+{1\over 4} y^{ijn} y_{kln}^{*} \phi_i \phi_j \phi^{*k} \phi^{*l}
\, .
&&
\label{eq:pot_V}
\eeqa
Now we can compare the masses of the fermions and scalars by looking at
the linearized equations of motion:
\beq
\partial^\mu\partial_\mu \phi_i =
- M_{ik}^{*} M^{kj} \phi_j + \ldots \, , \qquad\qquad
\partial^\mu\partial_\mu \psi_i =
- M_{ik}^{*} M^{kj} \psi_j + \ldots \, .
\eeq
Therefore, the fermions and the bosons satisfy the same wave equation with
exactly the same squared-mass matrix with real non-negative eigenvalues,
namely ${(M^2)_i}^j = M_{ik}^{*} M^{kj}$. Since SUSY is not broken yet, we have
a collection of chiral supermultiplets, each of which contains a
mass-degenerate complex scalar and Weyl fermion.

\section{Lagrangians for Gauge Supermultiplets}

Besides the chiral supermultiplets there also exist gauge
supermultiplets that contain gauge bosons and their fermionic
superpartners. The gauge bosons are described (before spontaneous
symmetry breaking) by a massless gauge boson field $A_\mu^a$, the
superpartners by a two-component Weyl fermion gaugino $\lambda^a$.
The index $a$ runs over the adjoint representation of the gauge
group ($a=1,\ldots ,8$ for $SU(3)_C$ color gluons and gluinos;
$a=1,2,3$ for $SU(2)_L$ weak isospin; $a=1$ for $U(1)_Y$ weak
hypercharge). The gauge transformations of the vector supermultiplet
fields are
\beqa
&& \delta_{\rm gauge} A^a_\mu = - \partial_\mu \Lambda^a + g f^{abc} A^b_\mu \Lambda^c \, , \\
&& \delta_{\rm gauge} \lambda^a = g f^{abc} \lambda^b \Lambda^c \, ,
\eeqa
where $\Lambda^a$ is an infinitesimal gauge transformation parameter, $g$
is the gauge coupling, and $f^{abc}$ are the totally antisymmetric
structure constants that define the gauge group.\\
The on-shell degrees of freedom for $A^a_\mu$ and $\lambda^a_\alpha$
amount to two bosonic and two fermionic helicity states (for each $a$), as
required by supersymmetry. However, off-shell $\lambda^a_\alpha$ consists
of two complex, or four real, fermionic degrees of freedom, while
$A^a_\mu$ only has three real bosonic degrees of freedom. So, we will need
one real bosonic auxiliary field, traditionally called $D^a$, in order for
supersymmetry to be consistent off-shell. This field also transforms
as an adjoint of the gauge group and satisfies $(D^a)^* = D^a$.
Like the chiral auxiliary fields $F_i$, the
gauge auxiliary field $D^a$ has no kinetic
term, so it can be eliminated on-shell using its algebraic equation of
motion.\\
The Lagrangian density for a gauge supermultiplet is thus given by
\beq
\lagr_{\rm gauge} = -{1\over 4} F_{\mu\nu}^a F^{\mu\nu a}
+ i \lambda^{\dagger a} \sigmabar^\mu D_\mu \lambda^a
+ {1\over 2} D^a D^a \, ,
\eeq
where
\beq
F^a_{\mu\nu} = \partial_\mu A^a_\nu - \partial_\nu A^a_\mu
               - g f^{abc} A^b_\mu A^c_\nu
\eeq
is the usual Yang-Mills field strength, and
\beq
D_\mu \lambda^a = \partial_\mu \lambda^a
                  - g f^{abc} A^b_\mu \lambda^c
\eeq
is the covariant derivative of the gaugino field. Up to multiplicative
factors, the supersymmetry transformations of the fields are
\beqa
&& \delta A_\mu^a =
- {1\over \sqrt{2}} \left (\epsilon^\dagger \sigmabar_\mu
\lambda^a + \lambda^{\dagger a} \sigmabar_\mu \epsilon \right )
\, , \\
&& \delta \lambda^a_\alpha =
- {i\over 2\sqrt{2}} (\sigma^\mu \sigmabar^\nu \epsilon)_\alpha
\> F^a_{\mu\nu} + {1\over \sqrt{2}} \epsilon_\alpha\> D^a
\, , \\
&& \delta D^a =
- {i\over \sqrt{2}} \left (
\epsilon^\dagger \sigmabar^\mu D_\mu \lambda^a -
D_\mu \lambda^{\dagger a} \sigmabar^\mu \epsilon \right ) \, .
\eeqa
Like the transformations of the chiral fields, they satisfy
the commutator relation
\beq
(\delta_{\epsilon_2} \delta_{\epsilon_1} - \delta_{\epsilon_1}
\delta_{\epsilon_2} ) X =
- i (\epsilon_1\sigma^\mu \epsilon_2^\dagger
- \epsilon_2\sigma^\mu \epsilon_1^\dagger) D_\mu X
\eeq
for $X$ equal to any of the gauge-covariant fields $F_{\mu\nu}^a$,
$\lambda^a$, $\lambda^{\dagger a}$, $D^a$, as well as for arbitrary
covariant derivatives acting on them. This ensures that the supersymmetry
algebra is realized on gauge-invariant combinations of fields
in gauge supermultiplets, as they were on the chiral supermultiplets.

\section{Supersymmetric Gauge Interactions}

Finally we are ready to consider a general Lagrangian density for a
supersymmetric theory with both chiral and gauge supermultiplets. Suppose
that the chiral supermultiplets transform under the gauge group in a
representation with hermitian matrices ${(T^a)_i}^j$ satisfying $[T^a,T^b]
=i f^{abc} T^c$. [For example, if the gauge group is $SU(2)$, then
$f^{abc} = \epsilon^{abc}$, and the $T^a$ are $1/2$ times the Pauli
matrices for a chiral supermultiplet transforming in the fundamental
representation.] Since supersymmetry and gauge transformations commute,
the scalar, fermion, and auxiliary fields must be in the same
representation of the gauge group, so
\beq
\delta_{\rm gauge}X_i = i g \Lambda^a (T^a X)_i
\eeq
for $X_i = \phi_i,\psi_i,F_i$. To have a gauge-invariant Lagrangian, we
now need to replace the ordinary derivatives in Eq.~(\ref{eq:lagr_free}) with
covariant derivatives:
\beqa
\partial_\mu \phi_i &\rightarrow& D_\mu \phi_i =
\partial_\mu \phi_i + i g A^a_\mu (T^a\phi)_i \\
\partial_\mu \phi^{*i} &\rightarrow& D_\mu \phi^{*i} =
\partial_\mu \phi^{*i} - i g A^a_\mu (\phi^* T^a)^i \\
\partial_\mu \psi_i &\rightarrow& D_\mu \psi_i =
\partial_\mu \psi_i + i g A^a_\mu (T^a\psi)_i \, .
\eeqa
This simple procedure achieves the goal of coupling the vector
bosons in the gauge supermultiplet to the scalars and fermions in the
chiral supermultiplets. Nevertheless, there are three additional
possible interactions that are gauge invariant and
renormalizable and can be included in the Lagrangian, namely
\beq
(\phi^* T^a \psi)\lambda^a,\qquad
\lambda^{\dagger a} (\psi^\dagger T^a
\phi),
\qquad {\rm and} \qquad
(\phi^* T^a \phi) D^a .
\eeq
It is only possible to add these terms to the Lagrangians for
the chiral and gauge supermultiplets, if the
supersymmetry transformation laws for the matter fields are modified to
include gauge-covariant rather than ordinary derivatives. Also, it is
necessary to include one strategically chosen extra term in $\delta F_i$, so
\beqa
\delta \phi_i &=& \epsilon\psi_i \\
\delta \psi_{i\alpha} &=& - i (\sigma^\mu \epsilon^\dagger)_{\alpha}\> D_\mu \phi_i + \epsilon_\alpha F_i \\
\delta F_i &=& - i \epsilon^\dagger \sigmabar^\mu D_\mu \psi_i \>
+ \> \sqrt{2} g (T^a \phi)_i\> \epsilon^\dagger \lambda^{\dagger a} \, .
\eeqa
Like the auxiliary fields $F_i$ and $F^{*i}$, the $D^a$ are
expressible in terms of the scalar fields, using an equation of motion
\beq
D^a = -g ( \phi^* T^a \phi ) \, .
\eeq
One thus finds that the complete scalar potential (see Eq.~(\ref{eq:pot_V})) becomes
\beq
V(\phi,\phi^*) = F^{*i} F_i + \half \sum_a D^a D^a = W_i^* W^i +
\half \sum_a g_a^2 (\phi^* T^a \phi)^2 \, .
\eeq
Summing up all the previous results we obtain as the full Lagrangian density
for a renormalizable supersymmetric theory
\beqa
\lagr_\mathrm{SUSY} & = & \lagr_{\rm chiral} + \lagr_{\rm gauge} \nn \\
&& - \sqrt{2} g (\phi^* T^a \psi)\lambda^a - \sqrt{2} g\lambda^{\dagger a} (\psi^\dagger T^a \phi)
+ g (\phi^* T^a \phi) D^a \\
& = & i \psi^{\dagger i} \sigmabar^\mu D_\mu \psi_i + D^\mu \phi^{*i} D_\mu \phi_i
- {1\over 4} F_{\mu\nu}^a F^{\mu\nu a} + i \lambda^{\dagger a} \sigmabar^\mu D_\mu \lambda^a \nn \\
&& - V(\phi,\phi^*) - {1 \over 2} M^{ij} \psi_i\psi_j - {1 \over 2} M_{ij}^{*} \psi^{\dagger i} \psi^{\dagger j}
- \half y^{ijk} \phi_i \psi_j \psi_k - \half y_{ijk}^{*} \phi^{*i} \psi^{\dagger j} \psi^{\dagger k} \nn \\
&& - \sqrt{2} g \left( (\phi^* T^a \psi)\lambda^a + \lambda^{\dagger a} (\psi^\dagger T^a \phi) \right) \, .
\label{eq:lagr_susy}
\eeqa
The first line of Eq.~(\ref{eq:lagr_susy}) shows the kinetic terms of fermions and scalars,
the self-interaction of gauge-fields and the kinetic term of gauginos.
In the second line, the first term is the scalar-potential while
the remaining terms come from the superpotential and include
Yukawa-couplings and fermion-mass terms. The last line consists
of the additional supersymmetric couplings.

\section{Soft Supersymmetry Breaking Interactions}

Since no supersymmetric particles are discovered yet, they must have
significantly higher masses than the standard model particles. In
other words, supersymmetry is broken. Like electroweak symmetry
breaking, supersymmetry should be broken spontaneously. Although
this mechanism of producing the symmetry breaking at high energies
is not totally understood yet --- and there is no consensus on which
of the several models is the right choice --- it is possible to
write down the general form of possible supersymmetry breaking terms
in the Lagrangian at low energies
\beq
\lagr_{\rm soft} = - \left( \half M_a \, \lambda^a \lambda^a
+ {1\over 6} a^{ijk} \phi_i\phi_j\phi_k
+ \half b^{ij} \phi_i\phi_j \right)
+ \mathrm{c.c.}
- (m^2)_j^i \phi^{j*} \phi_i
\label{eq:lagr_soft}
\eeq
with gaugino masses $M_a$ for each gauge group, scalar
squared-mass terms $(m^2)_i^j$ and $b^{ij}$, and (scalar)$^3$ couplings
$a^{ijk}$.\\
Because the effective Lagrangian can be written in the form $\lagr =
\lagr_{\rm SUSY} + \lagr_{\rm soft}$~(see the next section) and
because $\lagr_{\rm soft}$ contains only mass terms and coupling
parameters with {\it positive} mass dimension, broken supersymmetry
is still providing a solution to the hierarchy problem. It is
therefore named ``soft"
supersymmetry breaking.\\
Furthermore, supersymmetry is indeed broken by $\lagr_{\rm soft}$,
because it involves only scalars and gauginos and not their
respective superpartners.

\section{The Complete Lagrangian}

Finally, after all the efforts made in this chapter, we can present the
full Lagrangian of a supersymmetric theory, consisting of the full
supersymmetric Lagrangian from Eq.~(\ref{eq:lagr_susy}) and the soft
supersymmetry breaking Lagrangian from Eq.~(\ref{eq:lagr_soft}) resulting in
\beqa
\lagr & = & \lagr_\mathrm{SUSY} + \lagr_{\rm soft} \nn \\
& = & i \psi^{\dagger i} \sigmabar^\mu D_\mu \psi_i + D^\mu \phi^{*i} D_\mu \phi_i
- {1\over 4} F_{\mu\nu}^a F^{\mu\nu a} + i \lambda^{\dagger a} \sigmabar^\mu D_\mu \lambda^a \nn \\
&& - V(\phi,\phi^*) - {1 \over 2} M^{ij} \psi_i\psi_j - {1 \over 2} M_{ij}^{*} \psi^{\dagger i} \psi^{\dagger j}
- \half y^{ijk} \phi_i \psi_j \psi_k - \half y_{ijk}^{*} \phi^{*i} \psi^{\dagger j} \psi^{\dagger k} \nn \\
&& - \sqrt{2} g \left( (\phi^* T^a \psi)\lambda^a + \lambda^{\dagger a} (\psi^\dagger T^a \phi) \right) \nn \\
&& - \left( \half M_a \, \lambda^a \lambda^a
+ {1\over 6} a^{ijk} \phi_i\phi_j\phi_k
+ \half b^{ij} \phi_i\phi_j \right)
+ \mathrm{c.c.}
- (m^2)_j^i \phi^{j*} \phi_i \, .
\eeqa

\newpage
\chapter{The Minimal Supersymmetric Standard Model (MSSM)}
\label{chapter:mssm}

The MSSM extends the particle content of the Standard Model in two
different ways~\cite{primer,bernhard-dipl}. On the one hand, all
particles get a superpartner, on the other hand there is a larger
Higgs sector with two complex Higgs doublets. The superpartners get
the same names as their corresponding SM-particles, just with the
prefix ``s" (scalar) for spin~$=0$ superpartners and the suffix
``-ino" for spin~$=1/2$ superpartners. That leads to names like
sfermions on the one, and gauginos and higgsinos on the other hand.
Particles and superpartners are placed in the chiral and gauge
supermultiplets as shown in Table~\ref{table:chiral_supermultiplets}
and~\ref{table:gauge_supermultiplets}.
\renewcommand{\arraystretch}{1.4}
\begin{table}[htbp]
\begin{center}
\begin{tabular}{|c|c|c|c|c|}
\hline \multicolumn{2}{|c|}{Names} & spin 0 & spin 1/2 & $SU(3)_C
,\, SU(2)_L ,\, U(1)_Y$
\\  \hline\hline
squarks, quarks & $\tilde Q$ & $({\stilde u}_L\>\>\>{\stilde d}_L
)$&
 $(u_L\>\>\>d_L)$ & $(\>{\bf 3},\>{\bf 2}\>,\>{1\over 6})$
\\
($\times 3$ families) & $\sbar u$ &${\stilde u}^*_R$ & $u^\dagger_R$
& $(\>{\bf \overline 3},\> {\bf 1},\> -{2\over 3})$
\\ & $\sbar d$ &${\stilde d}^*_R$ & $d^\dagger_R$ &
$(\>{\bf \overline 3},\> {\bf 1},\> {1\over 3})$
\\  \hline
sleptons, leptons & $\tilde L$ &$({\stilde \nu}\>\>{\stilde e}_L )$&
 $(\nu\>\>\>e_L)$ & $(\>{\bf 1},\>{\bf 2}\>,\>-{1\over 2})$
\\
($\times 3$ families) & $\sbar e$ &${\stilde e}^*_R$ & $e^\dagger_R$
& $(\>{\bf 1},\> {\bf 1},\>1)$
\\  \hline
Higgs, higgsinos &$H_u$ &$(H_2^1\>\>\>H_2^2 )$& $(\stilde H_2^1
\>\>\> \stilde H_2^2)$& $(\>{\bf 1},\>{\bf 2}\>,\>+{1\over 2})$
\\ &$H_d$ & $(H_1^1 \>\>\> H_1^2)$ & $(\stilde H_1^1 \>\>\> \stilde H_1^2)$&
$(\>{\bf 1},\>{\bf 2}\>,\>-{1\over 2})$
\\  \hline
\end{tabular}
\end{center}
\caption{Chiral supermultiplets in the MSSM. The spin~$=0$ fields
are complex scalars and the spin~$=1/2$ fields are left-handed
two-component Weyl fermions.}
\label{table:chiral_supermultiplets}
\renewcommand{\arraystretch}{1.4}
\end{table}
\begin{table}[htbp]
\begin{center}
\begin{tabular}{|c|c|c|c|}
\hline
Names & spin 1/2 & spin 1 & $SU(3)_C, \> SU(2)_L,\> U(1)_Y$\\
\hline\hline gluino, gluon &$ \stilde g$& $g$ & $(\>{\bf 8},\>{\bf
1}\>,\> 0)$
\\
\hline winos, W bosons & $ \stilde \lambda^\pm\>\>\> \stilde
\lambda^3 $&
 $W^\pm\>\>\> W^0$ & $(\>{\bf 1},\>{\bf 3}\>,\> 0)$
\\
\hline bino, B boson &$\stilde \lambda'$&
 $B^0$ & $(\>{\bf 1},\>{\bf 1}\>,\> 0)$
\\
\hline
\end{tabular}
\end{center}
\caption{Gauge supermultiplets in the MSSM.}
\label{table:gauge_supermultiplets}
\end{table}

\noindent The left-handed and right-handed pieces of the quarks and
leptons are separate two-component Weyl fermions with different
gauge transformation properties in the SM, so each must have its own
complex scalar partner. It is important to keep in mind that the
``handedness" here does not refer to the helicity of the sfermions
(they are spin-0 particles) but to that of their superpartners.\\
The symbols for the superpartners carry a tilde
($\phantom{.}\stilde{\phantom{.}}\phantom{.}$) for distinction.\\
Due to the spontaneous electroweak symmetry breaking, the
interaction eigenstates (winos, bino) are no longer mass eigenstates
and thus no physical particles. Therefore one needs to consider the
mixing of interaction eigenstates to mass eigenstates as we will do
in detail in Chapter~\ref{chapter:mass-matrices}.
Table~\ref{table:mixing_mass_gauge} shows the mass eigenstates and
corresponding interaction eigenstates.
\begin{table}[htbp]
\begin{center}
\begin{tabular}{|c|c|c|c|c|}
\hline
Names & Spin & $P_R$ & Gauge Eigenstates & Mass Eigenstates \\
\hline\hline Higgs bosons & 0 & $+1$ & $H_2^0\>\> H_1^0\>\> H_2^+
\>\> H_1^-$ & $h^0\>\> H^0\>\> A^0 \>\> H^\pm$
\\ \hline
& & &${\stilde u}_L\>\> {\stilde u}_R\>\> \stilde d_L\>\> \stilde
d_R$&${\stilde u}_1\>\> {\stilde u}_2\>\> \stilde d_1\>\> \stilde
d_2$
\\
squarks& 0&$-1$& ${\stilde c}_L\>\> {\stilde c}_R\>\> \stilde
s_L\>\>
\stilde s_R$& ${\stilde c}_1\>\> {\stilde c}_2\>\> \stilde s_1\>\> \stilde s_2$ \\
& & & $\stilde t_L \>\>\stilde t_R \>\>\stilde b_L\>\> \stilde b_R$
& ${\stilde t}_1\>\> {\stilde t}_2\>\> \stilde b_1\>\> \stilde b_2$
\\ \hline
& & &${\stilde e}_L\>\> {\stilde e}_R \>\>\stilde \nu_e$&${\stilde
e}_1 \>\>{\stilde e}_2 \>\>\stilde \nu_e$
\\
sleptons& 0&$-1$&${\stilde \mu}_L\>\>{\stilde
\mu}_R\>\>\stilde\nu_\mu$&${\stilde \mu}_1 \>\>{\stilde \mu}_2
\>\>\stilde \nu_\mu$
\\
& & & $\stilde \tau_L\>\> \stilde \tau_R \>\>\stilde \nu_\tau$ &
${\stilde \tau}_1 \>\>{\stilde \tau}_2 \>\>\stilde \nu_\tau$
\\
\hline neutralinos & $1/2$&$-1$ & $\stilde \lambda' \>\>\>\stilde
\lambda^3\>\>\> \stilde H_2^2\>\>\> \stilde H_1^1$ & $\stilde
\chi_1\>\> \stilde \chi_2 \>\>\stilde \chi_3\>\> \stilde \chi_4$
\\
\hline charginos & $1/2$&$-1$ & $\stilde \lambda^\pm\>\>\> \stilde
H_2^1 \>\>\>\stilde H_1^2$ & $\stilde \chi_1^\pm\>\>\>\stilde
\chi_2^\pm $
\\
\hline
gluino & $1/2$&$-1$ &$\stilde g$  &$\stilde g$ \\
\hline
% ${\rm goldstino}\atop{\rm (gravitino)}$ &
% ${1/2}\atop{(3/2)}$&$-1$&$\stilde
% G$  &(same) \\
% \hline
\end{tabular}
\end{center}
\caption{Mass eigenstates and corresponding interaction eigenstates
of the particles in the MSSM.} \label{table:mixing_mass_gauge}
\end{table}

\newpage
\chapter{Mass Matrices}
\label{chapter:mass-matrices}

Due to the spontaneous electroweak symmetry breaking, the mass
eigenstates (and therefore the physical particles) of the
superpartners are not identical with the gauge eigenstates of the
interaction, but mixtures of them. Thus one has to deal with the
mixing of the fields and derive mass matrices, which can then be
diagonalized using rotation matrices in order to obtain mass
eigenstates. Since these rotation matrices include in general
complex values, they are an important source of CP violation.

\section{Sfermion Sector}

The relevant terms for the sfermion mass matrix are derived from the
soft SUSY breaking terms and the auxiliary field terms (for a
detailed derivation see \cite{theoretikum}). For the mass matrix in
the basis $\psi = ( \tilde f_L \; \tilde f_R )^{\top}$ ($\tilde f =
\lbrace \tilde t, \tilde b, \tilde \tau, \dots \rbrace$) we get
\begin{equation}
M_{\tilde f}^2 = \left( \begin{array}{cc}
m_{LL}^2 & m_{LR}^2 \\
m_{RL}^2 & m_{RR}^2
\end{array} \right)
= \left( \begin{array}{cc}
m_{\tilde f_L}^2 & m_f a_f \\
m_f a_f^* & m_{\tilde f_R}^2
\end{array} \right) \label{eq:sfermion_mass}
\end{equation}
with the following entries (see also \cite{bernhard-dipl}):
\begin{eqnarray}
m_{\tilde f_L}^2 & = & M_{\lbrace Q; L \rbrace}^2 + m_f^2 + m_Z^2 \cos 2 \beta ( I_f^{3 L} - e_f \sin^2 \theta_W ) \, , \\
m_{\tilde f_R}^2 & = & M_{\lbrace \bar u; \bar d; \bar e \rbrace}^2 + m_f^2 + m_Z^2 \cos 2 \beta e_f \sin^2 \theta_W \, , \\
m_f a_f & = & \biggl\{ \begin{array}{l}
m_u ( A_u^* - \mu \cot \beta ) \; \dots \; \textnormal{up-type sfermions} \\
m_d ( A_d^* - \mu \tan \beta ) \; \dots \; \textnormal{down-type sfermions} \, .
\end{array} \label{eq:sfermion_mass_offdiag}
\end{eqnarray}
$M_{\lbrace Q; L \rbrace}$, $M_{\lbrace \bar u; \bar d; \bar e
\rbrace}$ are real soft SUSY breaking masses, $A_{u}$, $A_{d}$ are
complex trilinear breaking parameters and $\mu$ is the complex
higgsino mass. $I_f^{3 L}$ is the third component of the weak
isospin, $e_{f}$ is the electric charge in terms of the elementary
charge, $m_{f}$, $m_{u}$, $m_{d}$ are the masses of the fermionic
superpartners and $\tan \beta = \frac{v_{2}}{v_{1}}$ is the ratio of
the two VEVs of the Higgs fields.\\
The mass matrix can now be diagonalized with a rotation matrix $R^{\tilde f}$
\begin{equation}
M_{\tilde f}^2 = \left( \begin{array}{cc}
m_{\tilde f_L}^2 & m_f a_f \\
m_f a_f^* & m_{\tilde f_R}^2
\end{array} \right)
= ( R^{\tilde f} )^{\dagger} \left( \begin{array}{cc}
m_{\tilde f_1}^2 & 0 \\
0 & m_{\tilde f_2}^2
\end{array} \right) R^{\tilde f} \, .
\end{equation}
We parameterize the unitary matrix as
\begin{equation}
R^{\tilde f} = \left( \begin{array}{cc}
R^{\tilde f}_{1 L} & R^{\tilde f}_{1 R} \\
R^{\tilde f}_{2 L} & R^{\tilde f}_{2 R}
\end{array} \right)
= \left( \begin{array}{cc}
\cos \theta_{\tilde f} & e^{i \varphi_{\tilde f}} \sin \theta_{\tilde f} \\
- e^{- i \varphi_{\tilde f}} \sin \theta_{\tilde f} & \cos \theta_{\tilde f}
\end{array} \right)
\end{equation}
with the mixing angle $\theta_{\tilde f}$ and the complex phase
$\varphi_{\tilde f}$ which is a new source for CP violation.\\
Applying the rotation matrix $R^{\tilde f}_{i \alpha}$ to the gauge
eigenstates $\tilde f_{\alpha} \, ( \alpha = L, R )$ one obtains the
mass eigenstates $\tilde f_i \, ( i = 1, 2 )$ and vice versa:
\begin{eqnarray}
{ \tilde f_1 \choose \tilde f_2 } = R^{\tilde f} { \tilde f_L \choose \tilde f_R } & \Leftrightarrow & \tilde f_i = R^{\tilde f}_{i \alpha} \tilde f_{\alpha} \, , \\
{ \tilde f_L \choose \tilde f_R } = ( R^{\tilde f} )^{\dagger} {
\tilde f_1 \choose \tilde f_2 } & \Leftrightarrow & \tilde
f_{\alpha} = R^{\tilde f *}_{i \alpha} \tilde f_i \, ,
\label{eq:sfermion-gauge2mass}
\end{eqnarray}
with the convention $m_{\tilde f_{1}} \leq m_{\tilde f_{2}}$. We
assume that the super CKM matrix is diagonal, so there is no mixing
between the three generations of sfermions.

\section{Chargino Sector}

The charged fermionic superpartners $\lambda^{\pm}$ and
$\psi^1_{H_2}$, $\psi^2_{H_1}$ of the $W$ and Higgs bosons mix to
two mass eigenstates with charge $\pm 1$, the so called charginos.
We identify these eigenstates in the Dirac spinor representation
with $\tilde \chi^{\pm}_i \, (i = 1,2)$ using the convention
$m_{\tilde \chi^{\pm}_1} \leq m_{\tilde \chi^{\pm}_2}$ (see
\cite{primer}, \cite{helmut-diss}, \cite{bernhard-dipl}).\\
The relevant terms for the chargino mass matrix are derived from the
soft SUSY breaking terms, the SUSY gauge interaction terms and the
Yukawa interaction terms, which can be deduced from the
superpotential (for a detailed derivation see \cite{theoretikum}).\\
The mass term of the lagrangian in the basis $\psi^+ = ( - i
\lambda^+ \; \psi^1_{H_2} )^{\top}$, $\psi^- = ( - i \lambda^- \;
\psi^2_{H_1} )^{\top}$ is given by
\begin{equation}
{\cal L}_{\tilde \chi^{\pm}} = - \frac{1}{2} ( (\psi^+)^{\top} (\psi^-)^{\top} ) \cdot \left( \begin{array}{cc}
0 & X^{\top} \\
X & 0
\end{array} \right) \cdot { \psi^+ \choose \psi^- } + h.c.
\end{equation}
with the mass matrix
\begin{equation}
X = \left( \begin{array}{cc}
M_2 & \sqrt{2} m_W \sin \beta \\
\sqrt{2} m_W \cos \beta & \mu
\end{array} \right) \, .
\end{equation}
$M_{2}$ is the real SUSY breaking mass of the gauginos $\lambda^{1}$
and $\lambda^{2}$, the superpartners of the $SU(2)_L$ bosons $A^1$
and $A^2$. The parameter $\mu$ is the complex higgsino mass.\\
This matrix can be diagonalized with two $2 \times 2$ rotation
matrices $U$ and $V$:
\begin{equation}
X_D = U^* X V^{-1} = \left( \begin{array}{cc}
m_{\tilde \chi^{\pm}_1} & 0 \\
0 & m_{\tilde \chi^{\pm}_2}
\end{array} \right).
\end{equation}
Applying the unitary matrices $U$ and $V$ to the gauge eigenstates
$\psi^{\pm}_j$ one obtains the mass eigenstates $\chi^{\pm}_i$ and
vice versa:
\begin{eqnarray}
\chi^+_i = V_{i j} \psi^+_j & & \chi^-_i = U_{i j} \psi^-_j \\
\psi^+_j = V_{i j}^* \chi^+_i & & \psi^-_j = U_{i j}^* \chi^-_i \, .
\label{eq:chargino-gauge2mass}
\end{eqnarray}
Finally we can define the charginos as Dirac spinors $(i = 1,2)$:
\begin{equation}
\tilde \chi^+_i = { \chi^+_i \choose \bar \chi^-_i } \qquad \tilde
\chi^-_i = { \chi^-_i \choose \bar \chi^+_i } \, .
\end{equation}

\section{Neutralino Sector}

The uncharged fermionic superpartners of the $U(1)_Y$ interaction
field $B^{\mu}$, the third component of the $SU(2)_L$ interaction
field $A^{3 \mu}$ and the Higgs bosons $H_1^1$ and $H_2^2$ mix to
four mass eigenstates called neutralinos. We specify them in the
Majorana representation with $\tilde \chi^0_i \, (i = 1,2,3,4)$
using the convention $m_{\tilde \chi^0_1} \leq m_{\tilde \chi^0_2}
\leq m_{\tilde \chi^0_3} \leq m_{\tilde \chi^0_4}$ (see
\cite{primer}, \cite{helmut-diss}, \cite{bernhard-dipl}).\\
The relevant terms for the neutralino mass matrix are again derived
from the soft SUSY breaking terms, the SUSY gauge interaction terms
and the Yukawa interaction terms, which can be deduced from the
superpotential (for a detailed derivation see \cite{theoretikum}).\\
The mass term of the lagrangian in the basis $\psi^0 = ( - i
\lambda', - i \lambda^3, \psi^1_{H_1}, \psi^2_{H_2} )$ is given by
\begin{equation}
{\cal L}_{\tilde \chi^0} = - \frac{1}{2} ( \psi^0 )^{\top} Y \psi^0 + h.c.
\end{equation}
with the mass matrix
\begin{equation}
Y = \left( \begin{array}{cccc}
M_1 & 0 & - m_Z s_{\theta_W} c_{\beta} & m_Z s_{\theta_W} s_{\beta} \\
0 & M_2 & m_Z c_{\theta_W} c_{\beta} & - m_Z c_{\theta_W} s_{\beta} \\
- m_Z s_{\theta_W} c_{\beta} & m_Z c_{\theta_W} c_{\beta} & 0 & - \mu \\
m_Z s_{\theta_W} s_{\beta} & - m_Z c_{\theta_W} s_{\beta} & - \mu & 0 \\
\end{array} \right),
\end{equation}
with the abbreviations $s_{\alpha} := \sin \alpha$ and $c_{\alpha} := \cos \alpha$.\\
The matrix can be diagonalized with the unitary rotation matrix $Z$:
\begin{equation}
Y_D = Z^* Y Z^{-1} = {\rm diag} ( m_{\tilde \chi^0_1}, m_{\tilde \chi^0_2}, m_{\tilde \chi^0_3}, m_{\tilde \chi^0_4} ) \, .
\end{equation}
Applying the matrix $Z$ to the gauge eigenstates $\psi^0_j$ one
receives the mass eigenstates $\chi^0_i$ and vice versa:
\begin{equation}
\chi^0_i = Z_{i j} \psi^0_j \qquad \psi^0_j = Z_{i j}^* \chi^0_i \, .
\end{equation}
At last the neutralinos are defined as Majorana spinors with mass
index $(i = 1,2,3,4)$:
\begin{equation}
\tilde \chi^0_i = { \chi^0_i \choose {\bar \chi^0_i} } \, .
\end{equation}

\newpage
\chapter{Couplings}
\label{chapter:couplings}

In this chapter we provide the relevant terms of the interaction
Lagrangian (and thus the couplings) for the leading contributions
mentioned in Chapter~\ref{chapter:contributions} and some more. For
a listing of more couplings see \cite{bernhard-dipl,helmut-diss,kuroda}.\\
Note that although we have written the fermions and sfermions only in the
third generation we sum over all generations nevertheless. We
apply the Einstein summation convention, repeated indices are
summed. We use the same convention as in \cite{bernhard-dipl}.\\
The chargino-squark-quark $(\tilde \chi^+_k q \tilde q'_i)$ and
the chargino-slepton-lepton $(\tilde \chi^+_k l \tilde l'_i)$
interaction Lagrangian can be written as
\begin{eqnarray}
{\cal L}_{\tilde \chi^+ q \tilde q'} & = & \bar t ( A^R_{k i} P_R + A^L_{k i} P_L ) \tilde \chi^+_k \tilde b_i + \bar b (B^R_{k i} P_R + B^L_{k i} P_L ) \tilde \chi^{+ c}_k \, \tilde t_i \nonumber \\
& & + \overline{\tilde \chi^+_k} ( A^{L *}_{k i} P_R + A^{R *}_{k i}
P_L ) t \, \tilde b^*_i + \overline{\tilde \chi^{+ c}_k} ( B^{L
*}_{k i} P_R + B^{R *}_{k i} P_L ) b \, \tilde t^*_i \\
{\cal L}_{\tilde \chi^+ l \tilde l'} & = & \bar \nu_{\tau} ( A'^R_{k i} P_R + A'^L_{k i} P_L ) \tilde \chi^+_k \tilde \tau_i + \bar \tau (B'^R_{k} P_R + B'^L_{k} P_L ) \tilde \chi^{+ c}_k \, \tilde \nu_{\tau} \nonumber \\
& & + \overline{\tilde \chi^+_k} ( A'^{L *}_{k i} P_R + A'^{R *}_{k
i} P_L ) \nu_{\tau} \tilde \tau^*_i + \overline{\tilde \chi^{+ c}_k}
( B'^{L *}_{k} P_R + B'^{R *}_{k} P_L ) \tau \tilde \nu^*_{\tau}
\end{eqnarray}
with the projection operators $P_{R,L} = (1 \pm \gamma^5)/2$. The
abbreviated coupling matrices are
\begin{eqnarray}
A^R_{k i} & = & \frac{g}{\sqrt 2} \left( \frac{m_b}{m_W \cos \beta} U_{k 2} R^{\tilde b *}_{i 2} - \sqrt 2 U_{k 1} R^{\tilde b *}_{i 1} \right) \nonumber \\
A^L_{k i} & = & \frac{g m_t}{\sqrt 2 m_W \sin \beta} V^*_{k 2} R^{\tilde b *}_{i 1} \nonumber \\
B^R_{k i} & = & \frac{g}{\sqrt 2} \left( \frac{m_t}{m_W \sin \beta} V_{k 2} R^{\tilde t *}_{i 2} - \sqrt 2 V_{k 1} R^{\tilde t *}_{i 1} \right) \nonumber \\
B^L_{k i} & = & \frac{g m_b}{\sqrt 2 m_W \cos \beta} U^*_{k 2} R^{\tilde t *}_{i 1} \label{eq:cha-sq-q-couplings} \\
A'^R_{k i} & = & \frac{g}{\sqrt 2} \left( \frac{m_{\tau}}{m_W \cos \beta} U_{k 2} R^{\tilde \tau *}_{i 2} - \sqrt 2 U_{k 1} R^{\tilde \tau *}_{i 1} \right) \nn \\
{A'}^L_{k i} & = & 0 \nn \\
B'^R_{k} & = & - g V_{k 1} \nn \\
B'^L_{k} & = & \frac{g m_{\tau}}{\sqrt 2 m_W \cos \beta} U^*_{k 2}
\, . \label{eq:cha-sl-l-couplings}
\end{eqnarray}
The neutralino-squark-quark $(\tilde \chi^0_k q \tilde q_i)$ and the
neutralino-slepton-lepton $(\tilde \chi^0_k l \tilde l_i)$
interaction Lagrangian can be written as
\begin{eqnarray}
{\cal L}_{\tilde \chi^0 q \tilde q} & = & \bar t ( C^R_{k i} P_R + C^L_{k i} P_L ) \tilde \chi^0_k \tilde t_i + \bar b ( D^R_{k i} P_R + D^L_{k i} P_L ) \tilde \chi^0_k \tilde b_i \nn \\
& & + \overline{\tilde \chi^0_k} ( C^{L *}_{k i} P_R + C^{R *}_{k i}
P_L ) t \tilde t^*_i + \overline{\tilde \chi^0_k} ( D^{L *}_{k i}
P_R + D^{R *}_{k i} P_L ) b \tilde b^*_i \\
{\cal L}_{\tilde \chi^0 l \tilde l} & = & \bar \nutau ( C'^R_{k} P_R + C'^L_{k} P_L ) \tilde \chi^0_k \tilde \nu_{\tau} + \bar \tau ( D'^R_{k i} P_R + D'^L_{k i} P_L ) \tilde \chi^0_k \tilde \tau_i \nn \\
& & + \overline{\tilde \chi^0_k} ( C'^{L *}_{k} P_R + C'^{R *}_{k}
P_L ) \nutau \tilde \nu_{\tau}^* + \overline{\tilde \chi^0_k} (
D'^{L *}_{k i} P_R + D'^{R *}_{k i} P_L ) \tau \tilde \tau^*_i \, .
\end{eqnarray}
The abbreviated coupling matrices are
\begin{eqnarray}
C^R_{k i} & = & \frac{g}{\sqrt{2}} \left( - \frac{m_t}{m_W \sin \beta} Z_{k 4} R^{\tilde t *}_{i 2} - ( Z_{k 2} + \frac{1}{3} \tan \theta_W Z_{k 1} ) R^{\tilde t *}_{i 1} \right) \nn \\
C^L_{k i} & = & \frac{g}{\sqrt{2}} \left( - \frac{m_t}{m_W \sin \beta} Z_{k 4}^* R^{\tilde t *}_{i 1} + \frac{4}{3} \tan \theta_W Z^*_{k 1} R^{\tilde t *}_{i 2} \right) \nn \\
D^R_{k i} & = & \frac{g}{\sqrt{2}} \left( - \frac{m_b}{m_W \cos \beta} Z_{k 3} R^{\tilde b *}_{i 2} + ( Z_{k 2} - \frac{1}{3} \tan \theta_W Z_{k 1} ) R^{\tilde b *}_{i 1} \right) \nn \\
D^L_{k i} & = & \frac{g}{\sqrt{2}} \left( - \frac{m_b}{m_W \cos
\beta} Z^*_{k 3} R^{\tilde b *}_{i 1} - \frac{2}{3} \tan \theta_W
Z^*_{k 1} R^{\tilde b *}_{i 2} \right) \\
C'^R_{k} & = & \frac{g}{\sqrt{2}} \left( - Z_{k 2} + \tan \theta_W Z_{k 1} \right) \nn \\
C'^L_{k} & = & 0 \nn \\
D'^R_{k i} & = & \frac{g}{\sqrt{2}} \left( - \frac{m_{\tau}}{m_W \cos \beta} Z_{k 3} R^{\tilde \tau *}_{i 2} + ( Z_{k 2} + \tan \theta_W Z_{k 1} ) R^{\tilde \tau *}_{i 1} \right) \nn \\
D'^L_{k i} & = & \frac{g}{\sqrt{2}} \left( - \frac{m_{\tau}}{m_W
\cos \beta} Z^*_{k 3} R^{\tilde \tau *}_{i 1} - 2 \tan \theta_W
Z^*_{k 1} R^{\tilde \tau *}_{i 2} \right) \, .
\end{eqnarray}
The gluino-squark-quark $(\tilde g q \tilde q_i)$ interaction
Lagrangian is
\begin{equation}
{\cal L}_{\tilde g q \tilde q} = \overline{\tilde g} ( G^{R}_{i}
P_R + G^{L}_{i} P_L ) q \tilde q^{*}_i + \overline{q} ( G^{L
*}_{i} P_R + G^{R *}_{i} P_L ) \tilde g \, \tilde q_i
\end{equation}
with the coupling matrices
\begin{eqnarray}
G^{R}_{i} & = & \sqrt{2} \, g_s T R^{\tilde q}_{i 2} \nn \\
G^{L}_{i} & = & - \sqrt{2} \, g_s T R^{\tilde q}_{i 1} \, .
\end{eqnarray}
$g_s$ is the coupling constant of the strong interaction and $T$ is
the generator of the $SU(3)_C$ group. We omitted the colour indices
of the quark and squark and the gluino index for the sake of
simplicity.\\
The photon-chargino-chargino $(\gamma \tilde \chi^+_i \tilde
\chi^+_i)$ interaction Lagrangian is
\begin{equation}
\mathcal{L}_{\gamma \tilde \chi^+ \tilde \chi^+} = e A^\mu
\overline{\tilde \chi^+_i} \gamma_\mu \tilde \chi^+_i
\end{equation}
with the electric positron charge $e$.\\
The Lagrangian of the photon-sfermion-sfermion $(\gamma \tilde f_i
\tilde f_i)$ interaction can be written as
\begin{equation}
\mathcal{L}_{\gamma \tilde f \tilde f} = - i e e_f A^\mu \tilde
f^*_i \stackrel{\leftrightarrow}{\partial_\mu} \tilde f_i
\end{equation}
where $e_f$ is the charge of the particle in $e$ (e.g. for the
electron $e_f = -1$). If we define the momenta $\tilde f_i (k_1)$
and $\tilde f^*_i (k_2)$ we obtain as the Feynman rule $- i e e_f
(k_1 - k_2)^\mu$.

\newpage
\chapter{Definitions}
\label{chapter:definitions}

In this chapter we give some definitions which are important for the
later study of CP violating effects. After a brief information about
the CP transformation (see~\cite{garfield-diss}) we define the CP
violating decay rate asymmetry $\delta^{CP}$
(see~\cite{bernhard-dipl}, for more details see
Chapter~\ref{chapter:cp-violation}). Further we define the branching
ratio $BR$ and explain why it is important to observe both
$\delta^{CP}$ and $BR$ simultaneously.

\section{CP Transformation}

The C transformation (charge transformation) changes particles to
its anti-particles and vice versa. It changes the sign of the
electric charge and inverts the colour charge of a particle. For
spinors we have the relations
\begin{eqnarray}
u(k) = C \bar v^\top(k) & & \bar u(k) = v^\top(k) C \nn \\
v(k) = C \bar u^\top(k) & & \bar v(k) = u^\top(k) C
\end{eqnarray}
and for the vector bosons
\begin{equation}
\epsilon^*_\mu = C \epsilon_\mu \, .
\end{equation}
The P transformation (parity transformation) is equivalent to a
spatial point reflection. It changes the sign of the coordinate
system and thus a right-handed coordinate system into a left-handed
one or vice versa. For spinors we use the definition
\begin{equation}
P \psi(t,\vec x) = \gamma^0 \psi(t,-\vec x) \, .
\end{equation}
Together they form the CP transformation. Applied to the Lagrangian,
it changes signs of four momenta, left- and right-handed parts and
spinors and therefor it also conjugates the tree-level coupling
matrices. Since these are now complex (due to the complex chargino
rotation matrices U and V, the complex neutralino rotation matrix Z
and the complex sfermion rotation matrix R) we have a violation of
CP symmetry. As we assume the Cabibbo-Kobayashi-Maskawa (CKM) matrix
(and the super CKM matrix as well) to be diagonal we neglect the
small CP violating effects coming from flavour mixing.

\section{Decay Rate Asymmetry $\delta^{CP}$}

The CP violating decay rate asymmetry $\delta^{CP}$ for our decay is
defined by
\begin{equation}
\delta^{CP} = \frac{\Gamma^+(\tilde t_i \to b \, \tilde \chi^+_k) -
\Gamma^-(\tilde t^*_i \to \bar b \, \tilde \chi^{+
c}_k)}{\Gamma^+(\tilde t_i \to b \, \tilde \chi^+_k) +
\Gamma^-(\tilde t^*_i \to \bar b \, \tilde \chi^{+ c}_k)}
\label{eq:deltaCP}
\end{equation}
with the decay widths
\begin{eqnarray}
\Gamma^+ & \propto & \sum_s |\mathcal{M}^+_\mathrm{tree}|^2 + 2 \mathrm{Re} \Big( \sum_s ( \mathcal{M}^+_\mathrm{tree} )^\dagger \mathcal{M}^+_\mathrm{loop} \Big) \\
\Gamma^- & \propto & \sum_s |\mathcal{M}^-_\mathrm{tree}|^2 + 2 \mathrm{Re} \Big( \sum_s ( \mathcal{M}^-_\mathrm{tree} )^\dagger \mathcal{M}^-_\mathrm{loop} \Big) \, .
\end{eqnarray}
The matrix elements at tree- and one-loop-level are given by
\begin{eqnarray}
\mathcal{M}^+_\mathrm{tree} & = & i \, \bar u(k_1) ( B^R P_R + B^L P_L ) v(-k_2) \nn \\
\mathcal{M}^-_\mathrm{tree} & = & i \, \bar u(k_2) ( B^{R*} P_R + B^{L*} P_L ) v(-k_1) \nn \\
\mathcal{M}^+_\mathrm{loop} & = & i \, \bar u(k_1) ( \delta B^R_+ P_R + \delta B^L_+ P_L ) v(-k_2) \nn \\
\mathcal{M}^-_\mathrm{loop} & = & i \, \bar u(k_2) ( \delta B^R_- P_R + \delta B^L_- P_L ) v(-k_1) \, .
\end{eqnarray}
Since there is no CP violation at tree-level, we set
$|\mathcal{M}_\mathrm{tree}|^2 := |\mathcal{M}^+_\mathrm{tree}|^2 =
|\mathcal{M}^-_\mathrm{tree}|^2$. The decay rate asymmetry can then
be approximated to
\begin{eqnarray}
\delta^{CP} & = & \frac{\Gamma^+ - \Gamma^-}{2 \Gamma_\mathrm{tree}} \nn \\
& = & \frac{2 \mathrm{Re} \Big( \sum_s ( \mathcal{M}^+_\mathrm{tree}
)^\dagger \mathcal{M}^+_\mathrm{loop} \Big) - 2 \mathrm{Re} \Big(
\sum_s ( \mathcal{M}^-_\mathrm{tree} )^\dagger
\mathcal{M}^-_\mathrm{loop} \Big)}{2 \sum_s
|\mathcal{M}_\mathrm{tree}|^2} \label{eq:deltaCP_approx}
\end{eqnarray}
with
\begin{eqnarray}
2 \mathrm{Re} \Big( \sum_s ( \mathcal{M}^+_\mathrm{tree} )^\dagger
\mathcal{M}^+_\mathrm{loop} \Big) & = & 2 (m_{\tilde t_i}^2 - m_b^2
- m_{\tilde \chi^+_k}^2) \mathrm{Re} ( B^{R*} \delta B^R_+ + B^{L*}
\delta B^L_+ ) \nn \\
& & - 4 m_b m_{\tilde \chi^+_k} \mathrm{Re} ( B^{R*} \delta B^L_+ +
B^{L*} \delta B^R_+ ) \label{eq:2_Re_M+M+} \\
2 \mathrm{Re} \Big( \sum_s ( \mathcal{M}^-_\mathrm{tree} )^\dagger
\mathcal{M}^-_\mathrm{loop} \Big) & = & 2 (m_{\tilde t_i}^2 - m_b^2
- m_{\tilde \chi^+_k}^2) \mathrm{Re} ( B^{R} \delta B^R_- + B^{L}
\delta B^L_- ) \nn \\
& & - 4 m_b m_{\tilde \chi^+_k} \mathrm{Re} ( B^{R} \delta B^L_- +
B^{L} \delta B^R_- ) \label{eq:2_Re_M-M-}
\end{eqnarray}
and
\begin{equation}
2 \sum_s |\mathcal{M}_\mathrm{tree}|^2 = 2 (m_{\tilde t_i}^2 - m_b^2
- m_{\tilde \chi^+_k}^2) ( |B^R|^2 + |B^L|^2 ) - 8 m_b m_{\tilde
\chi^+_k} \mathrm{Re} ( B^{R*} B^L ) \, . \label{eq:M-tree-generic}
\end{equation}
In Chapter~\ref{chapter:contributions} we will only provide the form
factors of the graph but not the anti-graph, so $\Lambda_{R,L} = (4
\pi)^2 \delta B^{R,L}_+$. The form factors $\delta B^{R,L}_-$ of the
anti-graph can be easily obtained by conjugating all the couplings
involved.\\
We can refine our result by deriving more specific expressions for
Eq.~(\ref{eq:2_Re_M+M+}) and (\ref{eq:2_Re_M-M-}). We rewrite these
equations to
\begin{eqnarray}
\delta \Gamma^\pm \propto 2 \mathrm{Re} \Big( \sum_s (
\mathcal{M}^\pm_\mathrm{tree} )^\dagger
\mathcal{M}^\pm_\mathrm{loop} \Big) & = & 2 \Delta \mathrm{Re} ( B^R_\mp \delta B^R_\pm + B^L_\mp \delta B^L_\pm ) \nn \\
& & - 4 m_b m_{\tilde \chi^+_k} \mathrm{Re} ( B^R_\mp \delta B^L_\pm
+ B^L_\mp \delta B^R_\pm )
\end{eqnarray}
using $\Delta = (m_{\tilde t_i}^2 - m_b^2 - m_{\tilde \chi^+_k}^2)$,
$B^{R,L}_+ = B^{R,L}$ and $B^{R,L}_- = B^{R,L *}$. Then we define
the combined coupling matrices
\begin{equation}
C^{i j}_+ = B^i_- \delta B^j_+ \qquad C^{i j}_- = B^i_+ \delta B^j_-
\end{equation}
with $i,j \in \{R,L\}$ and we get
\begin{equation}
\delta \Gamma^\pm \propto 2 \Delta \left( \mathrm{Re} (C^{RR}_\pm) +
\mathrm{Re} (C^{LL}_\pm) \right) - 4 m_b m_{\tilde \chi^+_k} \left(
\mathrm{Re} (C^{RL}_\pm) + \mathrm{Re} (C^{LR}_\pm) \right) \, .
\end{equation}
The coupling matrix $C^{i j}_\pm$ can be generally expressed by (see
Eq.~(\ref{eq:A_R-generic-scalar-selfenergy}) and
(\ref{eq:A_R-generic-vertex-correction}) for specific examples of
$\delta B_\pm$)
\begin{equation}
C^{i j}_+ \propto b \times g_0 g_1 g_2 \times \mathrm{PaVe} \qquad
C^{i j}_- \propto b^* \times (g_0 g_1 g_2)^* \times \mathrm{PaVe}
\end{equation}
where $b=B_-, b^*=B_+$ is the coupling at tree-level, $g_0 g_1 g_2$
are the couplings of the three vertices and PaVe are the
Passarino-Veltman-Integrals (see Chapter~\ref{chapter:contributions}
and Appendix~\ref{chapter:pave}). Taking the real part of the
coupling matrix $C^{i j}_\pm$ and keeping in mind that both the
couplings and the PaVe's are in general complex we get
\begin{eqnarray}
\mathrm{Re} (b g_0 g_1 g_2 \times \mathrm{PaVe}) & = & \mathrm{Re}
(b g_0 g_1 g_2) \mathrm{Re} (\mathrm{PaVe}) - \mathrm{Im} (b g_0 g_1
g_2) \mathrm{Im} (\mathrm{PaVe}) \\
\mathrm{Re} ((b g_0 g_1 g_2)^* \times \mathrm{PaVe}) & = &
\mathrm{Re} (b g_0 g_1 g_2) \mathrm{Re} (\mathrm{PaVe}) +
\mathrm{Im} (b g_0 g_1 g_2) \mathrm{Im} (\mathrm{PaVe}) \, .
\end{eqnarray}
This finally leads to the important decomposition into both CP
invariant and CP violating parts
\begin{equation}
\mathrm{Re} (C^{i j}_\pm) = C^{i j}_\mathrm{inv} \pm \frac{1}{2}
C^{i j}_\mathrm{CP} \label{eq:C_inv&C_CP}
\end{equation}
with the definitions
\begin{eqnarray}
C^{i j}_\mathrm{inv} & \propto & \mathrm{Re} (b g_0 g_1 g_2)
\mathrm{Re}
(\mathrm{PaVe}) \\
C^{i j}_\mathrm{CP} & \propto & - 2 \mathrm{Im} (b g_0 g_1 g_2)
\mathrm{Im} (\mathrm{PaVe}) \, . \label{eq:C_CP}
\end{eqnarray}
To show that the decomposition in Eq.~(\ref{eq:C_inv&C_CP}) is
correct one simply has to calculate $\delta^{CP} \propto \delta
\Gamma^+ - \delta \Gamma^-$ using the rewritten equation
\begin{eqnarray}
\delta \Gamma^\pm & \propto & 2 \Delta \left( C^{RR}_\mathrm{inv} +
C^{LL}_\mathrm{inv} \pm \frac{1}{2} (C^{RR}_\mathrm{CP} +
C^{LL}_\mathrm{CP}) \right) \nn \\
& & - 4 m_b m_{\tilde \chi^+_k} \left( C^{RL}_\mathrm{inv} +
C^{LR}_\mathrm{inv} \pm \frac{1}{2} (C^{RL}_\mathrm{CP} +
C^{LR}_\mathrm{CP}) \right)
\end{eqnarray}
resulting in
\begin{equation}
\delta^{CP} \propto 2 \Delta (C^{RR}_\mathrm{CP} +
C^{LL}_\mathrm{CP}) - 4 m_b m_{\tilde \chi^+_k} (C^{RL}_\mathrm{CP}
+ C^{LR}_\mathrm{CP}) \, . \label{eq:deltaCP-C_P}
\end{equation}
From Eq.~(\ref{eq:C_CP}) we can see that we need not only the
couplings but also the PaVe's to be complex in order to obtain a
non-zero $\delta^{CP}$.

\section{Branching Ratio $BR$}

The branching ratio $BR$ is simply defined as the ratio between the
tree-level decay width of a certain decay and the total tree-level
decay width. We write
\begin{equation}
BR =
\frac{\Gamma_\mathrm{tree}}{\Gamma_\mathrm{tree}^\mathrm{total}} \, .
\end{equation}
An example would be
\begin{equation}
BR (\tilde t_i \to b \tilde \chi^+_k) = \frac{ \Gamma_{\tilde t_i} (
b \tilde \chi^+_k ) } { \Gamma_{\tilde t_i}^\mathrm{total} } \, .
\label{eq:BR}
\end{equation}
This quantity gives you the probability how often the certain decay
occurs compared to all possible decay channels. Combined with the
decay rate asymmetry to $\delta^{CP} \times BR$ one obtains the
overall probability how often the certain decay channel is CP
violated compared to the rest. As $\delta^{CP}$ usually gets high
when $BR$ gets low and vice versa one has to find an optimum in
$\delta^{CP} \times BR$ for a good measurability at colliders.

\newpage
\chapter{CP Violation}
\label{chapter:cp-violation}

In this chapter we provide a deeper insight into CP violation by
stating more precisely how and when the CP violating decay rate
asymmetry $\delta^{CP}$ defined in Chapter~\ref{chapter:definitions}
occurs.

As we already found out, the tree-level couplings introduced in
Chapter~\ref{chapter:couplings} being complex is just a necessary
but not sufficient condition for CP violation. There are in total
two conditions which have to be fulfilled in order to obtain a
non-zero decay rate asymmetry $\delta^{CP}$ at one-loop level:
\begin{enumerate}
\item Complex tree-level couplings
\item
\begin{enumerate}
\item At least two open decay channels\\
or equivalently
\item At least one Passarino-Veltman-Integral has to become complex\\
or equivalently
\item At least two particles in the loop have to become on-shell.
\end{enumerate}
\end{enumerate}
The first condition can be easily understood by remembering that a
CP transformation of a Lagrangian results in the conjugation of the
tree-level couplings involved (see
Chapter~\ref{chapter:definitions}). Real tree-level couplings leave
the CP symmetry intact. Pragmatically one can also see this
condition in Eq.~(\ref{eq:C_CP}).\\
Furthermore, the reason why $\delta^{CP}$ does not occur at
tree-level can be seen by looking at Eq.~(\ref{eq:M-tree-generic}),
keeping in mind that $\Gamma \propto \sum_s |\mathcal{M}|^2 = \sum_s
\mathcal{M}^\dagger \mathcal{M}$. The tree-level couplings $B^{R,L}$
and their complex conjugates $B^{R,L *}$ appear symmetric in the
equation, so a CP transformation (i.e. conjugating the couplings)
keeps the decay rate CP invariant. An other way to see this is the
fact that both CP transformation and calculating an adjoint matrix
$\mathcal{M}^\dagger$ conjugate the couplings so the net effect is
zero. The decay rate asymmetry is thus a pure loop-effect starting
at one-loop level.\\
For the second condition we need to expatiate on the subject in
order to understand the three conditions and their equivalency. We
start with the requirement~2.(a). Following from the CPT Theorem one
can prove that the total decay width $\Gamma_\mathrm{total}$ of a
certain particle is invariant under CP transformation. That means if
we we assume at least two different kinematically possible decays we
have
\begin{equation}
\Gamma^+_\mathrm{total} = \Gamma^+_1 + \Gamma^+_2 = \Gamma^-_1 +
\Gamma^-_2 = \Gamma^-_\mathrm{total} \, .
\end{equation}
Note that only the total decay width keeps CP invariant but the
partial decay widths are not necessarily CP invariant! So in our
example we have in general $\Gamma^+_1 \not= \Gamma^-_1$,
$\Gamma^+_2 \not= \Gamma^-_2$ and we can write $\delta^{CP} \propto
\Gamma^+_{1,2} - \Gamma^-_{1,2} \not= 0$. Only in the case of just
one open decay channel the partial decay rate (being now the total
decay rate) is becoming always CP invariant.

The requirement~2.(b) can be directly seen in Eq.~(\ref{eq:C_CP}).
If $\mathrm{Im}(\mathrm{PaVe})=0$ the CP violating part becomes zero
thus rendering $\delta^{CP}$ in Eq.~(\ref{eq:deltaCP-C_P}) zero as
well.\\
The last requirement~2.(c) can be understood in two ways. First one
can calculate the discontinuity (and hence the imaginary part) of a
PaVe using Cutkosky Rules~\cite{Cutkosky1960,Hooft1973,Peskin1995}.
After applying these rules one directly obtains the condition that
the particles in the loop have to become on-shell for a non-zero
$\mathrm{Im}(\mathrm{PaVe})$. For example, if we apply the rules and
cut through the loop of the Generic Structure~I in
Chapter~\ref{chapter:contributions}, we receive the condition $m_0
\geq M_1 + M_2$ for $\mathrm{Im}(\mathrm{PaVe}) \not= 0$. If we take
the Generic Structure~II we can cut in three ways (vertically and
two times horizontally) and get the conditions $m_0 \geq M_1 + M_2$,
$m_1 \geq M_0 + M_1$ or $m_2 \geq M_0 + M_2$ for
$\mathrm{Im}(\mathrm{PaVe}) \not= 0$. In both examples we need at
least two loop-particles who have to be on-shell, which is exactly
the requirement~2.(c). Note that the condition $m_i \geq \sum_j M_j$
implies that the decay of a particle $i$ into other particles $j$ is
kinematically possible, i.e. $\Gamma (i \to j) \not= 0$.\\
The second way to understand the last requirement~2.(c) is by using
the Optical Theorem~\cite{Peskin1995} directly, which we will derive
very briefly. From unitarity of the S-matrix $\sum_k S_{f k}^* S_{k
i} = \delta_{f i}$ and the relation $S_{f i} = \delta_{f i} + i
\mathcal{M}_{f i}$ to the matrix element $\mathcal{M}$ we can write
\begin{eqnarray}
\delta_{fi} & = & \sum_k (\delta_{fk} - i \mathcal{M}_{fk}^*)
(\delta_{ki} + i \mathcal{M}_{ki}) \nn \\
& = & \delta_{fi} \underbrace{- i \mathcal{M}_{fi}^* + i
\mathcal{M}_{fi} + \sum_k (\mathcal{M}_{fk}^* \mathcal{M}_{ki})}_{0}
\end{eqnarray}
with the indices $i,f,k$ denoting initial-, final- and
intermediate-states, respectively. Rewriting this condition directly
results in the Optical Theorem
\begin{equation}
2 \mathrm{Im} (\mathcal{M}_{fi}) = \sum_k (\mathcal{M}_{fk}^*
\mathcal{M}_{ki}) \propto \sigma_\mathrm{total}
\end{equation}
which relates the scattering amplitude $\mathcal{M}_{fi}$ to the
total cross section $\sigma_\mathrm{total}$ of the process. Note
that if we set $i=f$ the theorem relates $\mathcal{M}_{ii}$ with the
total decay width $\Gamma_\mathrm{total} \propto \sum_k
|\mathcal{M}_{ki}|^2$ of the process $i \to k$. Without loss of
generality, we choose the $i \to k$~process to be a tree-level decay
$\Gamma_\mathrm{tree} (i \to k)$ of one particle~$i$ into two
others. The amplitude $\mathcal{M}_{ii}$ then represents a one-loop
selfenergy of the particle~$i$. Now the theorem states that the
imaginary part of the scattering amplitude $\mathcal{M}_{ii}$ (and
hence the imaginary part of the PaVe inside of $\mathcal{M}_{ii}$)
becomes non-zero only in the case of a kinematically possible decay
$\Gamma_\mathrm{tree} (i \to k)$, i.e. $m_i \geq \sum_k M_k$.
Pictorially speaking, the one-loop selfenergy of particle $i$ can
then be ``assembled'' by two tree-level processes (or equivalently
``cut'' through to get two tree-level processes) as shown below:
\begin{center}
\includegraphics[width=0.5\paperwidth]{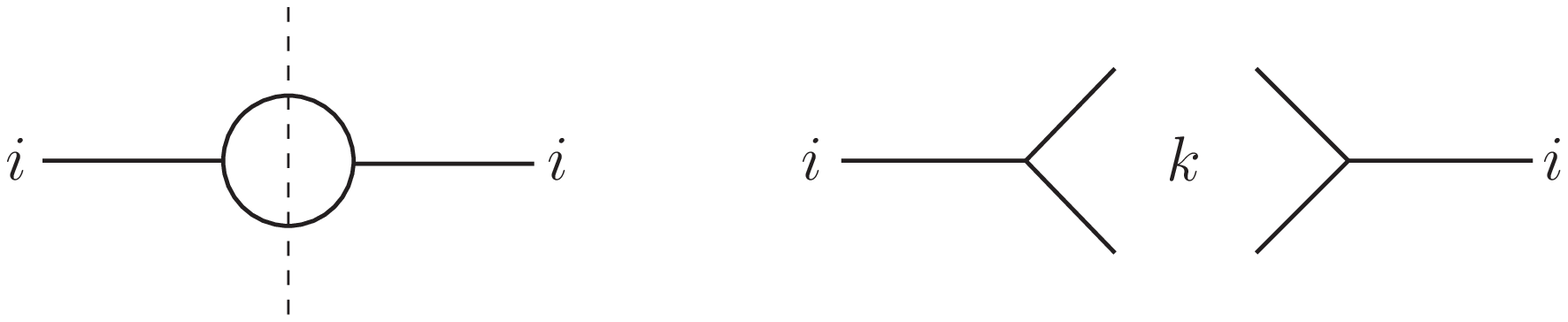}
\end{center}
This is exactly what Cutkosky Rules are doing and so these two ways
to understand the requirement~2.(c) are equivalent.\\
Because of the connection between $\mathrm{Im}(\mathrm{PaVe})$ and
$m_i \geq \sum_j M_j$ the requirement~2.(c) is equivalent to
requirement~2.(b). To show the equivalency with the first
requirement~2.(a) we take the Graph~1 in
Chapter~\ref{chapter:contributions} as example. Let's assume that
only two decay channels are open: $\tilde t_i \to b \tilde \chi^+_k$
and $\tilde t_i \to t \tilde g$. Condition~2.(a) as well as the
conditions~2.(b-c) state that CP violation is possible. Now let's
assume that only one decay is kinematically allowed: $\tilde t_i \to
b \tilde \chi^+_k$. Condition~2.(a) states that CP is not violated.
Because of $m_{\tilde t_i} < m_t + m_{\tilde g}$ (and hence
$\mathrm{Im}(\mathrm{PaVe})=0$) the conditions~2.(b-c) predict no CP
violation as well. So we demonstrated that all three requirements
(a-c) of condition~2. are really actually equivalent.\\
Concluding we want to mention that selfenergy-loops like the one in
Graph~1 can only contribute to $\delta^{CP}$ in the case of
different particles, i.e. $i \not= j$. If the particles are the
same, the couplings on the two vertices of the loop are complex
conjugates of each other ($g_1^R=g_2^{L*}$, $g_1^L=g_2^{R*}$)
resulting in an CP invariance of the form factors of the loop. This
is also the reason why there are only three different topologies in
Chapter~\ref{chapter:contributions} who can contribute. The possible
fourth topology, a selfenergy-loop for the particle with momentum
$k_1$, cannot contribute, because this particle is associated with a
bottom-quark. As we have set the CKM matrix to be diagonal, the
b-quark cannot change into another quark of the first or second
generation, thus it has to stay the same and cannot fulfill the
requirement.

\newpage
\chapter{Contributions}
\label{chapter:contributions}

In this chapter we list the most important processes at one-loop
level who are expected to yield the highest CP violating
asymmetries. We start by specifying all possible topologies and the
convention of the four momenta and the masses. Then we insert scalar
and fermionic fields in order to obtain the most important generic
contributions and give the matrix element and the corresponding form
factors. In the last step we list the most important processes with
their specific form factors and mention the remaining contributions.

\section{Topologies}

There are in total three possible topologies who can contribute to a
CP violating asymmetry: two different topologies with a selfenergy
contribution and one vertex correction.\\
The masses of the external ($m_{0}$, $m_{1}$, $m_{2}$) and internal
particles ($M_{0}$, $M_{1}$, $M_{2}$) and the four momenta of the
external particles ($p$, $k_{1}$, $k_{2}$) are defined as shown
below. The on-shell relations $p^2=m_{0}^2$, $k_{1}^2=m_{1}^2$,
$k_{2}^2=m_{2}^2$ as well as the relations $p=k_{1}-k_{2}$ and
$k=k_{1}+k_{2}$ are used.\\
For the calculation of the loop integrals we use the formalism of
Passarino-Veltman-Integrals defined in the
Appendix~\ref{chapter:pave}. The convention of the arguments of the
B- and C-functions are shown below. All the above conventions for
momenta and masses are given in \cite{helmut-diss}.

\begin{center}
\begin{tabular}{rl}
  \begin{tabular}{c}
  \includegraphics[width=0.4\paperwidth]{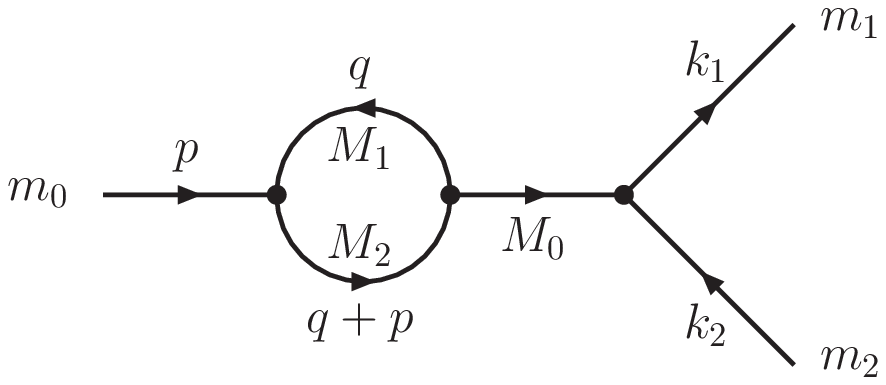}
  \end{tabular}
&
  \begin{tabular}{c}
  $\equiv B(m_{0}^2,M_{1}^2,M_{2}^2)$
  \end{tabular}
\end{tabular}
\end{center}

\begin{center}
\begin{tabular}{rl}
  \begin{tabular}{c}
  \includegraphics[width=0.4\paperwidth]{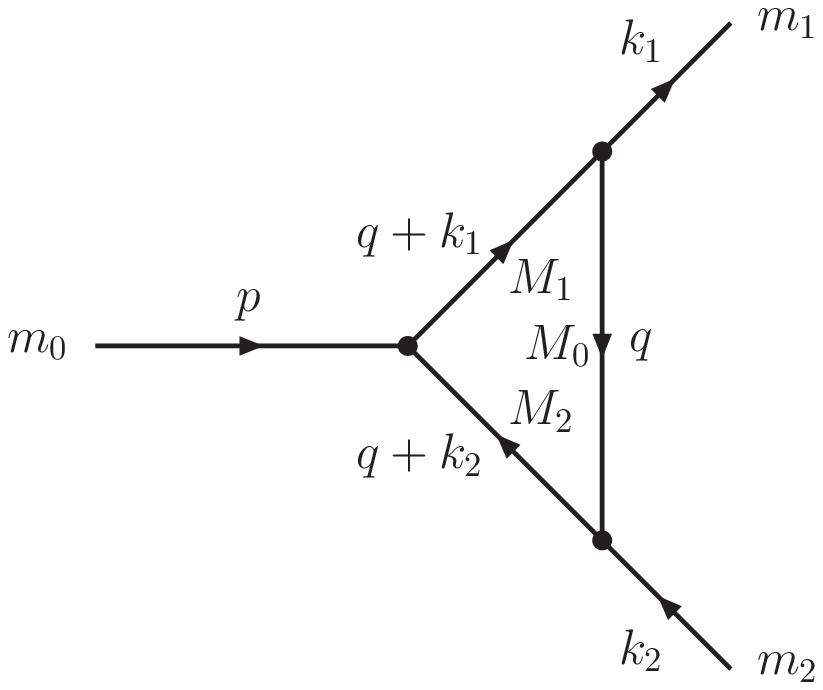}
  \end{tabular}
&
  \begin{tabular}{c}
  $\equiv C(m_{1}^2,m_{0}^2,m_{2}^2,M_{0}^2,M_{1}^2,M_{2}^2)$
  \end{tabular}
\end{tabular}
\end{center}

\begin{center}
\begin{tabular}{rl}
  \begin{tabular}{c}
  \includegraphics[width=0.4\paperwidth]{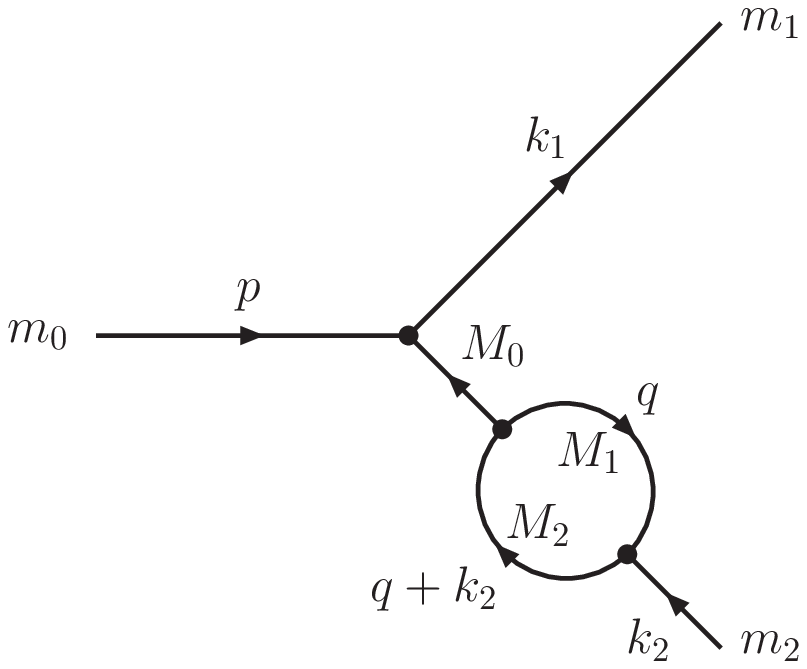}
  \end{tabular}
&
  \begin{tabular}{c}
  $\equiv B(m_{2}^2,M_{1}^2,M_{2}^2)$
  \end{tabular}
\end{tabular}
\end{center}

\section{Most Important Generic Structures}
\label{section:most-important-generic-contributions}

Now we insert scalar and fermionic fields into the topologies in
order to obtain the leading generic contributions. We list the
matrix element and the corresponding form factors which can then be
used for the calculation of specific processes. All form factors are
calculated in the SUSY invariant Dimensional Reduction
regularization scheme $(\overline{\mathrm{DR}})$ (see~\cite{Siegel1979,Capper1980}).\\
The general matrix element can be written as
(see~\cite{helmut-diss})
\begin{equation}
{\cal M} = \frac{i}{(4 \pi)^2} \bar u(k_1) ( \Lambda_R P_R +
\Lambda_L P_L ) v(-k_2) \, .
\end{equation}
We only give the form factor $\Lambda_R$ since $\Lambda_L$ can be
easily obtained by exchanging right- and left-handed couplings ($g^R
\leftrightarrow g^L$). The argument set of the form factors is
always $f =
(m_0,m_1,m_2,M_0,M_1,M_2,g_0^R,g_0^L,g_1^R,g_1^L,g_2^R,g_2^L)$.

\subsection{Generic Structure I}

The first leading generic structure is a scalar selfenergy-loop with
a fermion-fermion pair shown below.
\begin{center}
\begin{tabular}{rl}
  \begin{tabular}{c}
  \includegraphics[width=0.4\paperwidth]{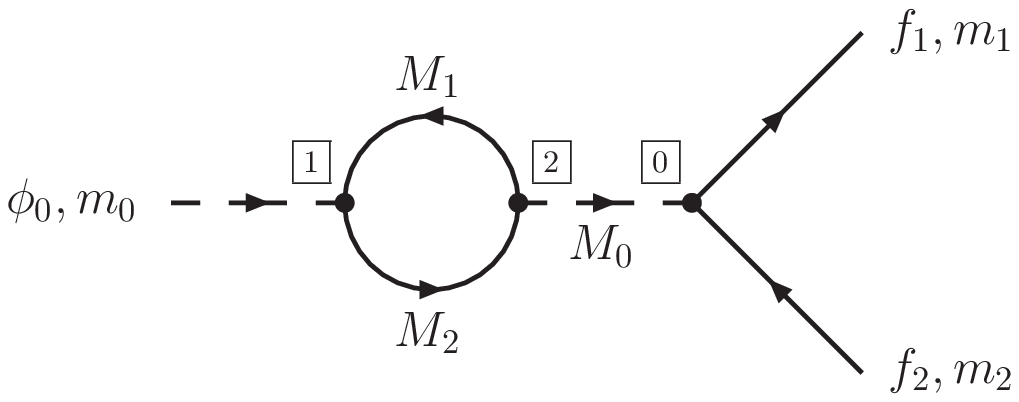}
  \end{tabular}
&
  \begin{tabular}{l}
  \framebox{0} : $i (g_0^R P_R + g_0^L P_L)$\\
  \framebox{1} : $i (g_1^R P_R + g_1^L P_L)$\\
  \framebox{2} : $i (g_2^R P_R + g_2^L P_L)$
  \end{tabular}
\end{tabular}
\end{center}
The appropriate form factor is
\begin{eqnarray}
\Lambda_R^{I}(f) & = & \frac{1}{m_0^2 - M_0^2} g_0^R \bigg[ 2 M_1 M_2 \Big( g_1^L g_2^L + g_1^R g_2^R \Big) B_0 \nn \\
& & + ( g_1^L g_2^R + g_1^R g_2^L ) \Big( A_0(M_1^2) + A_0(M_2^2) +
( M_1^2 + M_2^2 - m_0^2 ) B_0 \Big) \bigg] \, .
\label{eq:A_R-generic-scalar-selfenergy}
\end{eqnarray}

\subsection{Generic Structure II}

The second leading generic structure is a scalar-fermion-fermion
vertex correction.
\begin{center}
\begin{tabular}{rl}
  \begin{tabular}{c}
  \includegraphics[width=0.4\paperwidth]{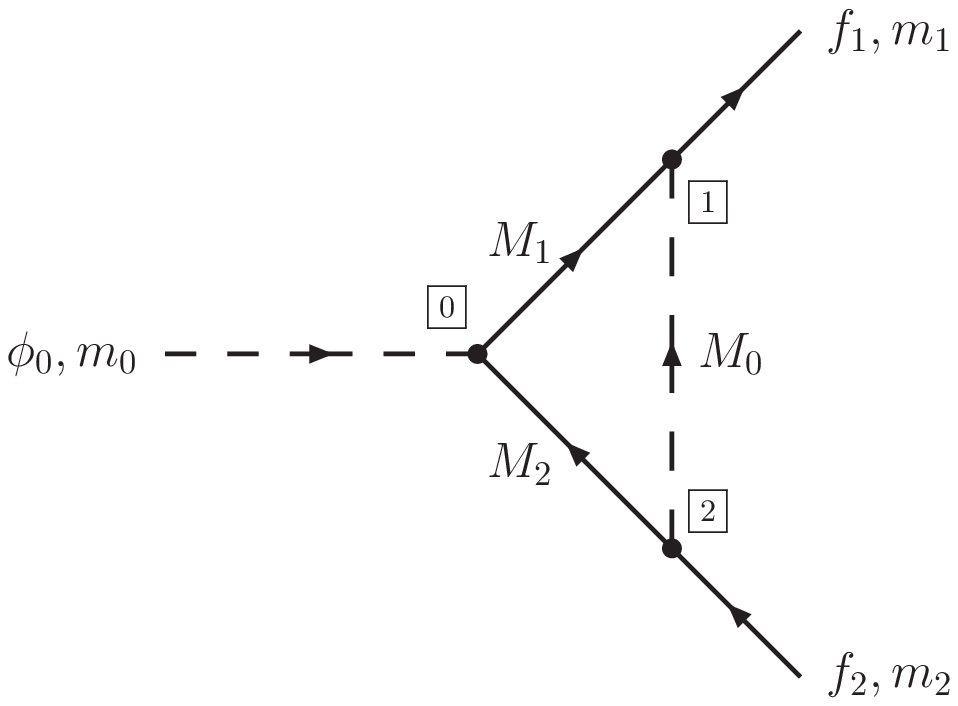}
  \end{tabular}
&
  \begin{tabular}{l}
  \framebox{0} : $i (g_0^R P_R + g_0^L P_L)$\\
  \framebox{1} : $i (g_1^R P_R + g_1^L P_L)$\\
  \framebox{2} : $i (g_2^R P_R + g_2^L P_L)$
  \end{tabular}
\end{tabular}
\end{center}
Here the respective form factor is
\begin{eqnarray}
\Lambda_R^{II}(f) & = & - \bigg[ \Big( g_0^L g_2^R ( g_1^R m_1 + g_1^L M_1 ) + g_0^R g_1^L ( g_2^L m_2 + g_2^R M_2 ) \Big) m_1 C_1 \nn \\
& & + \Big( g_0^R g_2^L ( g_1^L m_1 + g_1^R M_1 ) + g_0^L g_1^R ( g_2^R m_2 + g_2^L M_2 ) \Big) m_2 C_2 \nn \\
& & + g_0^R \Big( g_1^L g_2^L m_1 m_2 + g_1^R g_2^R M_1 M_2 + g_1^L g_2^R m_1 M_2 + g_1^R g_2^L m_2 M_1 \Big) C_0 \nn \\
& & + g_0^L g_1^R g_2^R \Big( B_0(m_0^2,M_1^2,M_2^2) + M_0^2 C_0
\Big) \bigg] \, . \label{eq:A_R-generic-vertex-correction}
\end{eqnarray}

\section{Most Important Processes}

Finally we list the most important processes with their specific
form factors. They are the two possible processes involving a gluino
$\tilde g$. Because the gluino couples like its superpartner the
gluon with the strong interaction force, these contributions are
expected to dominate over all others, if the decay channel $\tilde
t_i \to t \, \tilde g$ is open ($m_{\tilde t_i} \geq m_t + m_{\tilde
g}$). For a listing of all contributions at full one-loop level
(including these leading contributions) which we calculated see
Appendix~\ref{chapter:all-contributions}. The coupling matrices we
used are defined in Chapter~\ref{chapter:couplings}.

\subsection{Graph 1}

The first leading process is a stop selfenergy with a gluino-top in
the loop. Note that $i \neq j$ in order to contribute to CP
violation.
\begin{center}
  \includegraphics[width=0.4\paperwidth]{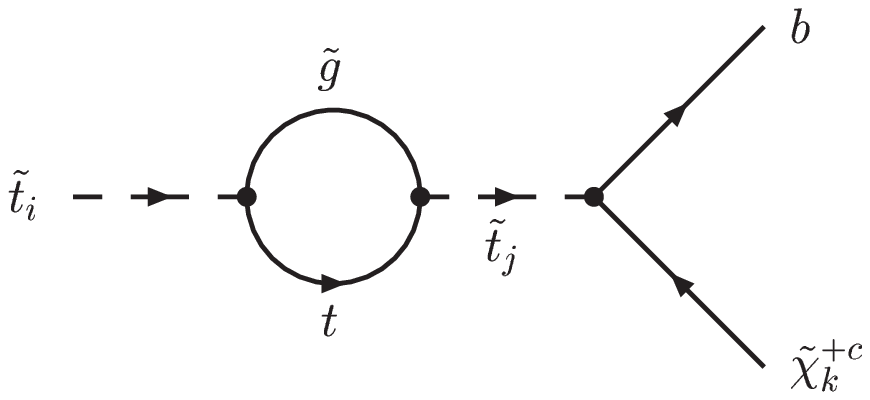}
\end{center}
The form factor has the following arguments:
% $f = (m_0,m_1,m_2,M_0,M_1,M_2,g_0^R,g_0^L,g_1^R,g_1^L,g_2^R,g_2^L)$
\begin{equation}
\Lambda^1_{R,L} = \Lambda_{R,L}^{I}(m_{\tilde t_i},m_b,m_{\tilde
\chi^+_k},m_{\tilde t_j},m_{\tilde g},m_t, B^R_{k j},B^L_{k j},G^{L
*}_i,G^{R *}_i,G^{R}_j,G^{L}_j) \, .
\end{equation}

\subsection{Graph 2}

The second leading process is a vertex correction with a
sbottom-gluino-top in the loop.
\begin{center}
  \includegraphics[width=0.4\paperwidth]{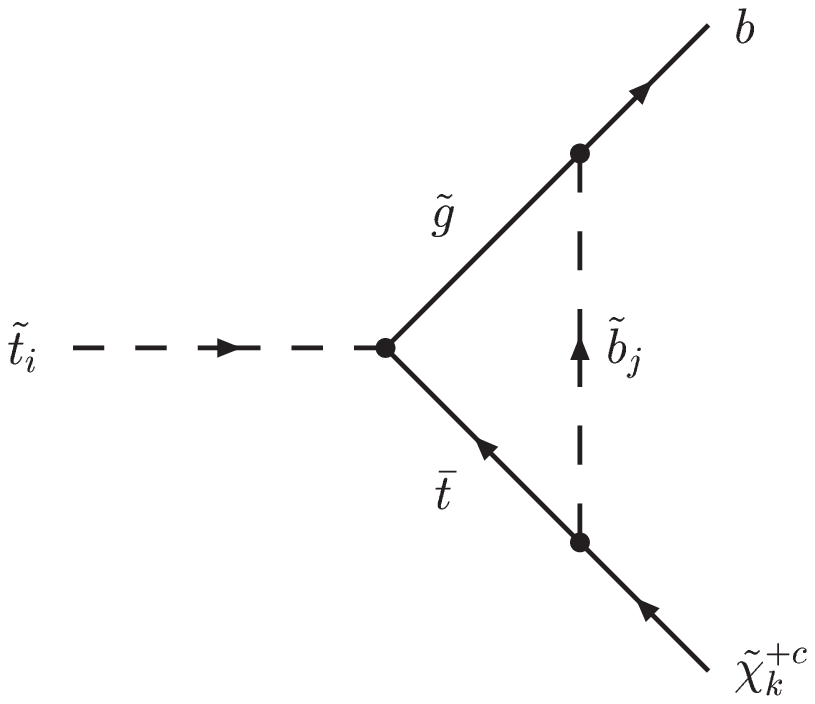}
\end{center}
The form factor takes the following set of arguments:
\begin{equation}
\Lambda^2_{R,L} = \sum_{j=1}^{2} \Lambda_{R,L}^{II}(m_{\tilde
t_i},m_b,m_{\tilde \chi^+_k},m_{\tilde b_j},m_{\tilde g},m_t,G^{L
*}_i,G^{R *}_i,G^{L *}_j,G^{R *}_j,A^{L *}_{k j},A^{R *}_{k j}) \, .
\end{equation}

\newpage
\chapter{Numerical Results and Conclusions}

In this chapter we finally study the decay rate asymmetry
$\delta^{CP}$ defined in Eq.~(\ref{eq:deltaCP}) and the combined
quantity $\delta^{CP} \times BR$ (branching ratio $BR$ defined in
Eq.~(\ref{eq:BR})) of the processes $\tilde t_1 \to b \, \tilde
\chi^+_1$ and $\tilde t_2 \to b \, \tilde \chi^+_1$ (and their CP
transformed counterparts) by varying the parameters $M_{\tilde Q}$,
$\tan \beta$, $\varphi_{A_t}$, $M_2$, $|A_t|$, $\mu$ and
$\varphi_\mu$.

\section{A Typical Scenario}

We fix the input parameters of the MSSM and the SM by the choice
given in Table~\ref{table:parameters}. The coupling of the strong
interaction force $\alpha_s$ is taken running in the dimensional
reduction regularization scheme
$\overline{DR}$~\cite{Siegel1979,Capper1980,Antoniadis1982,Jack1994},
renormalized at the scale of the decaying particle mass $m_{\tilde
t_i}$ ($i=\{1,2\}$) in the SPA
convention~\cite{Aguilar-Saavedra2006}. The gluino mass $m_{\tilde
g}$ calculated from $\alpha_s$ via GUT relations is therefore
running as well (calculated iteratively at the scale of $m_{\tilde
g}$ itself). Finally, for the values of the Yukawa couplings of
the third generation (s)quarks $(h_t, h_b)$, we again take the
running ones at the scale of the decaying particle mass.\\ The
SUSY breaking mass parameters $M_{\tilde Q}$, $M_{\bar u}$,
$M_{\bar d}$, $M_{\tilde L}$ and $M_{\bar e}$ are taken to be
equal in all generations. The trilinear breaking parameters of the
first and second generation are set to zero, i.e.
$A_{u,d,e}=A_{c,s,\mu}=0$. Furthermore, we simplify and set
$|M_1|=M_2/2$ (GUT relation), $M_{\tilde Q}=M_{\bar u}=M_{\bar
d}$, $M_{\tilde L}=M_{\bar e}$, $|A_t|=|A_b|=|A_{\tau}|$ and
$\varphi_{A_t}=\varphi_{A_b}=\varphi_{A_{\tau}}$.\\
To make sure that our chosen complex parameter set of the MSSM is
not already ruled out by the experimental limit on the electric
dipole moment of the electron (eEDM), we account for the constraint
by calculating and thus checking the eEDM automatically along the
way (see Appendix~\ref{chapter:edm}). Due to this stringent
constraint, we set $\varphi_\mu=0$ and focus just on the phase of
$A_f$ (the phase of $M_1$ is negligible in our case and thus set to
zero).\\
We study two different sets of processes, considering a) all
contributions at full one-loop level and b) the two contributions
with a gluino in the loop (see Graph~1 and 2 in
Chapter~\ref{chapter:contributions}).
\begin{table}[htbp]
\begin{center}
% MSSM parameters
\begin{tabular}[t]{|c|c|}
\hline
Parameter                &    Value               \\
\hline
$M_{A^0}$                &    $800$ GeV           \\
$\tan \beta$             &    $5$                 \\
$|M_1|$                  &    $M_2 / 2$           \\
$\varphi_{M_1}$          &    $0$                 \\
$M_2$                    &    $200$ GeV           \\
$|\mu|$                  &    $1000$ GeV          \\
$\varphi_{\mu}$          &    $0$                 \\
$M_{\tilde Q}$           &    $1000$ GeV          \\
$M_{\bar u}$             &    $M_{\tilde Q}$      \\
$M_{\bar d}$             &    $M_{\tilde Q}$      \\
$M_{\tilde L}$           &    $600$ GeV           \\
$M_{\bar e}$             &    $M_{\tilde L}$      \\
$|A_t|$                  &    $350$               \\
$\varphi_{A_t}$          &    $\pi / 4$           \\
$|A_b|$                  &    $|A_t|$             \\
$\varphi_{A_b}$          &    $\varphi_{A_t}$     \\
$|A_{\tau}|$             &    $|A_t|$             \\
$\varphi_{A_{\tau}}$     &    $\varphi_{A_t}$     \\
$A_{u,d,e}$              &    $0$                 \\
$A_{c,s,\mu}$            &    $0$                 \\
\hline
\end{tabular}
\hspace{10mm}
% SM parameters
\begin{tabular}[t]{|c|c|}
\hline
Parameter                &    Value                                 \\
\hline
$m_Z$                    &    $91.1875$ GeV                         \\ % ?!
$m_W$                    &    $80.45$ GeV                           \\ % ?!
$\cos \theta_W$          &    $m_W/m_Z$                             \\
$G_F$                    &    $1.16639 \times 10^{-5}$ GeV$^{-2}$   \\
$\alpha_{em}$            &    $1/127.9$                             \\ % ?!
$\alpha_{s}$             &    $9.1045846740771 \times 10^{-2}$      \\ % running?!
$m_u$                    &    $53.8 \times 10^{-3}$ GeV             \\ % ?!
$m_c$                    &    $1.5$ GeV                             \\ % ?!
$m_t$                    &    $171.4$ GeV                           \\ % ?!
$m_d$                    &    $53.8 \times 10^{-3}$ GeV             \\ % ?!
$m_s$                    &    $150 \times 10^{-3}$ GeV              \\ % ?!
$m_b$                    &    $4.2$ GeV                             \\ % ?!
$m_e$                    &    $0.51099907 \times 10^{-3}$ GeV       \\
$m_{\mu}$                &    $105.658389 \times 10^{-3}$ GeV       \\
$m_{\tau}$               &    $1777 \times 10^{-3}$ GeV             \\
\hline
\end{tabular}
\end{center}
\caption{The input parameters of the MSSM and the SM.}
\label{table:parameters}
\end{table}

\begin{figure}[htbp]
\begin{center}
\begin{tabular}{cc}
\includegraphics[width=0.45\textwidth]{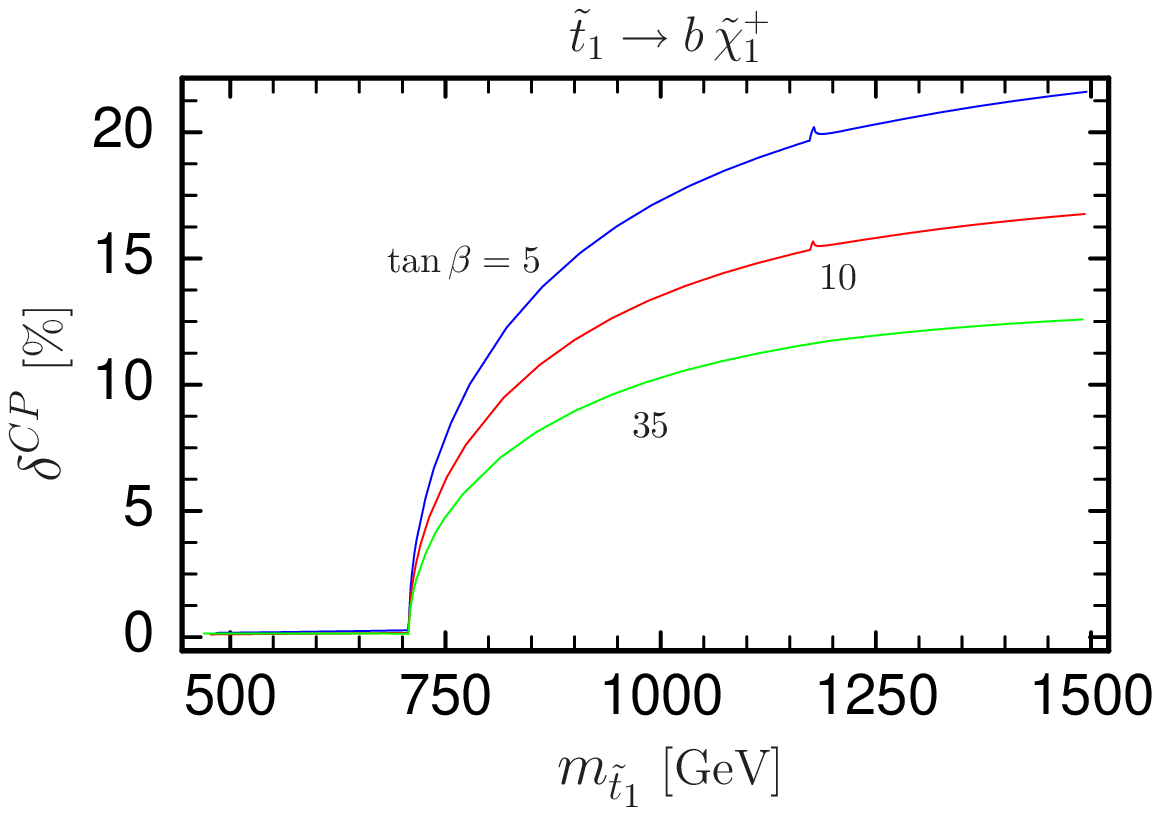} &
\includegraphics[width=0.45\textwidth]{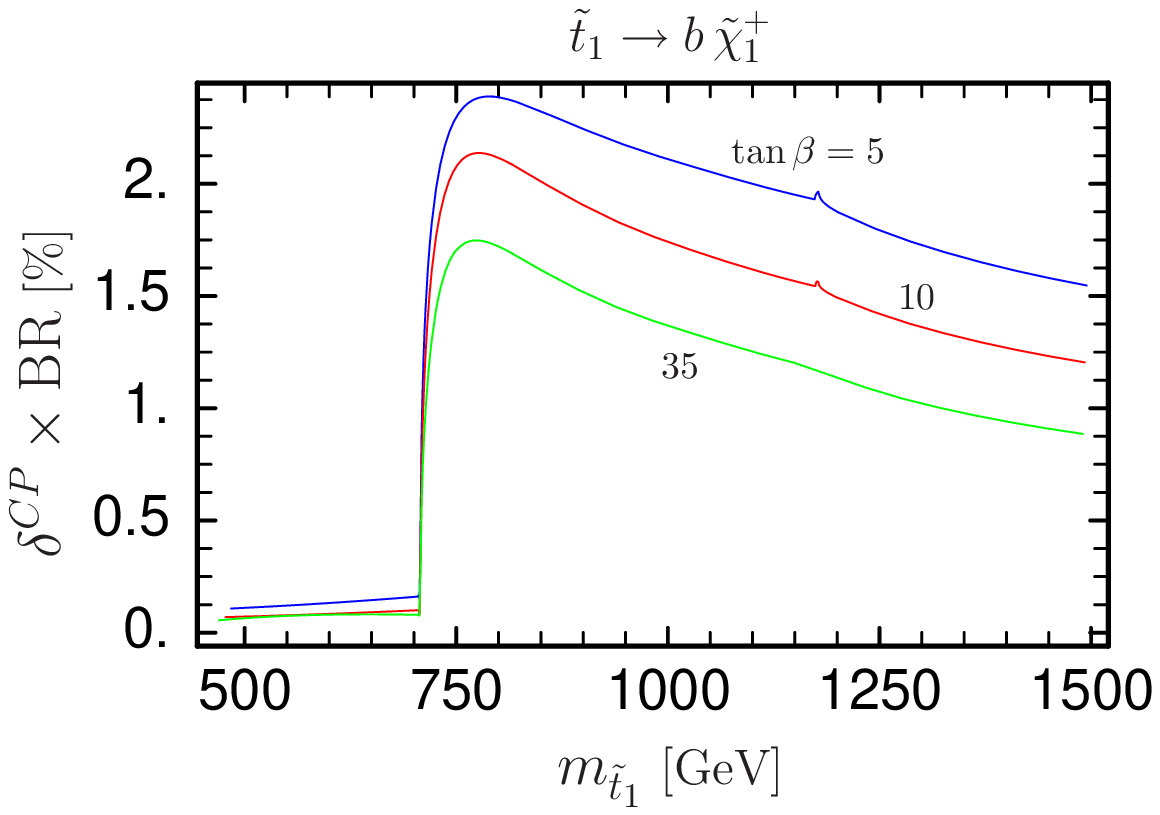} \\
\end{tabular}
\end{center}
\caption{$\delta^{CP}$ and $\delta^{CP} \times \mathrm{BR}$ as a
function of $m_{\tilde t_1}$ for various $\tan \beta$, considering
all contributions (full one-loop). The parameter $m_{\tilde t_1}$ is
shown for convenience, but the parameter actually varied is
$M_{\tilde Q}$ from $500$ to $1500$~GeV.}
\label{fig:delCP_Mstop1_TB_all}
\end{figure}

\noindent In Fig.~\ref{fig:delCP_Mstop1_TB_all} we show the
asymmetry $\delta^{CP}$ and $\delta^{CP} \times BR$, taking all
contributions. We vary the input parameter $M_{\tilde Q}$ from $500$
to $1500$~GeV (but show the output parameter $m_{\tilde t_1}$ for
better usability) for various $\tan \beta$. One can clearly see the
threshold of the decay $\tilde t_1 \to t \, \tilde g$ at $m_{\tilde
t_1} \sim 708$~GeV, after which the two gluino contributions
dominate over all other negligible processes. The asymmetry
$\delta^{CP}$ goes up to $\sim 22 \, \%$, the quantity $\delta^{CP}
\times BR$ has its maximum of $\sim 2.4 \, \%$ at $m_{\tilde t_1}
\sim 775$~GeV for $\tan \beta = 5$.
% delCP_Mstop1_TB_all & delCPxBR_Mstop1_TB_all

\begin{figure}[htbp]
\begin{center}
\includegraphics[width=0.6\textwidth]{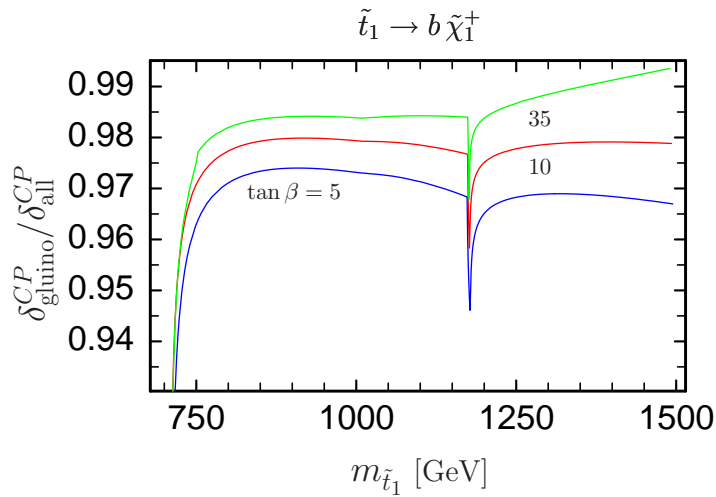}
\end{center}
\caption{Ratio of $\delta^{CP}$ between the two gluino contributions
and all contributions (full one-loop) as a function of $m_{\tilde
t_1}$ for various $\tan \beta$. The parameter actually varied is
$M_{\tilde Q}$ from $719$ to $1500$~GeV.}
\label{fig:ratio_gluino_all_Mstop1_TB}
\end{figure}

\noindent In Fig.~\ref{fig:ratio_gluino_all_Mstop1_TB} we show in
detail the dominance of the two gluino contributions over the
remaining ones by plotting the gluino-to-all ratio. After the
threshold at $m_{\tilde t_1} \sim 708$~GeV ($M_{\tilde Q} \sim
719$~GeV), the gluino processes account for $\sim 98 \, \%$ of all
processes, depending on $\tan \beta$. The kink at $m_{\tilde t_1}
\sim 1175$~GeV (which can be already seen in
Fig.~\ref{fig:delCP_Mstop1_TB_all}) comes from the threshold of
$\tilde t_1 \to t \, \tilde \chi^0_3$, where the graph with $\tilde
\chi^0_3 - t$ in the stop selfenergy-loop and the two graphs with
$H^\pm (G^\pm) - t - \tilde \chi^0_3$ in the vertex correction begin
to contribute.
%ratio_gluino_all_Mstop1_TB

\begin{figure}[htbp]
\begin{center}
\includegraphics[width=0.6\textwidth]{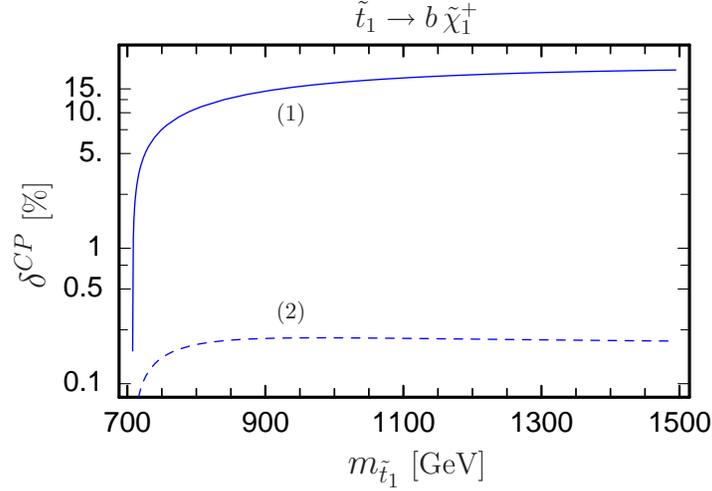}
\end{center}
\caption{Comparison of the two gluino contributions using
$\delta^{CP}$ as a function of $m_{\tilde t_1}$. Label (1)
represents the gluino in the selfenergy-loop and (2) the gluino in
the vertex-correction. The parameter actually varied is $M_{\tilde
Q}$ from $719$ to $1500$~GeV.}
\label{fig:delCP_Mstop1_gluino_compare}
\end{figure}

\noindent Fig.~\ref{fig:delCP_Mstop1_gluino_compare} shows the
comparison of the two gluino contributions by plotting their
respective $\delta^{CP}$ as a function of $M_{\tilde Q}$ from $719$
to $1500$~GeV ($m_{\tilde t_1}$ is again shown for convenience).
Contrary to the expectation that {\it both} gluino contributions
should dominate due to their strong coupling nature, only the
contribution with the gluino in the stop selfenergy-loop accounts
for $\delta^{CP}$ in a noteworthy manner. The reason why the vertex
correction is so suppressed lies in the $\tilde b_j - t - \tilde
\chi^+_1$ coupling of the Graph~2 in
Chapter~\ref{chapter:contributions}. However, as this coupling is
embedded in the one-loop vertex correction in a nontrivial way (see
Eq.~(\ref{eq:A_R-generic-vertex-correction})), there exists no
simple explanation for this feature.
% delCP_Mstop1_gluino_compare
% delCP_Mstop1_TB_except_gluino & delCPxBR_Mstop1_TB_except_gluino
\begin{figure}[htbp]
\begin{center}
\begin{tabular}{cc}
\includegraphics[width=0.45\textwidth]{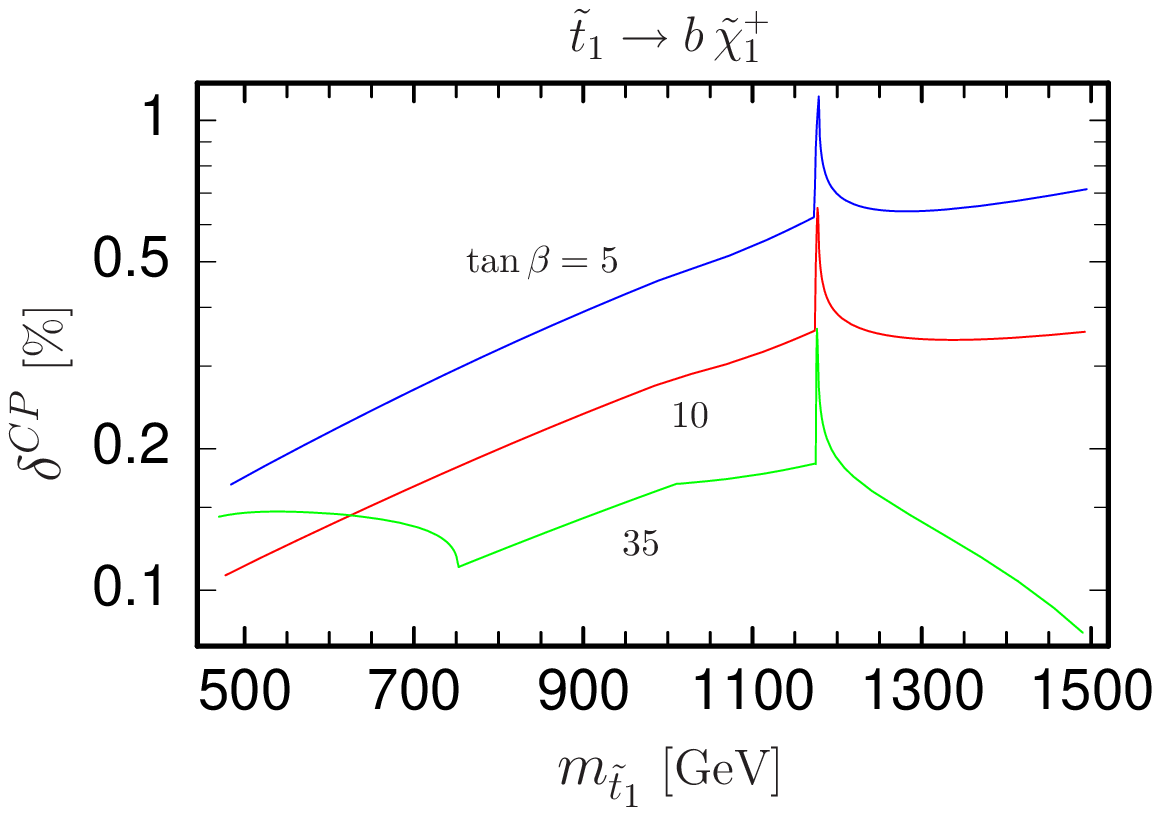} &
\includegraphics[width=0.45\textwidth]{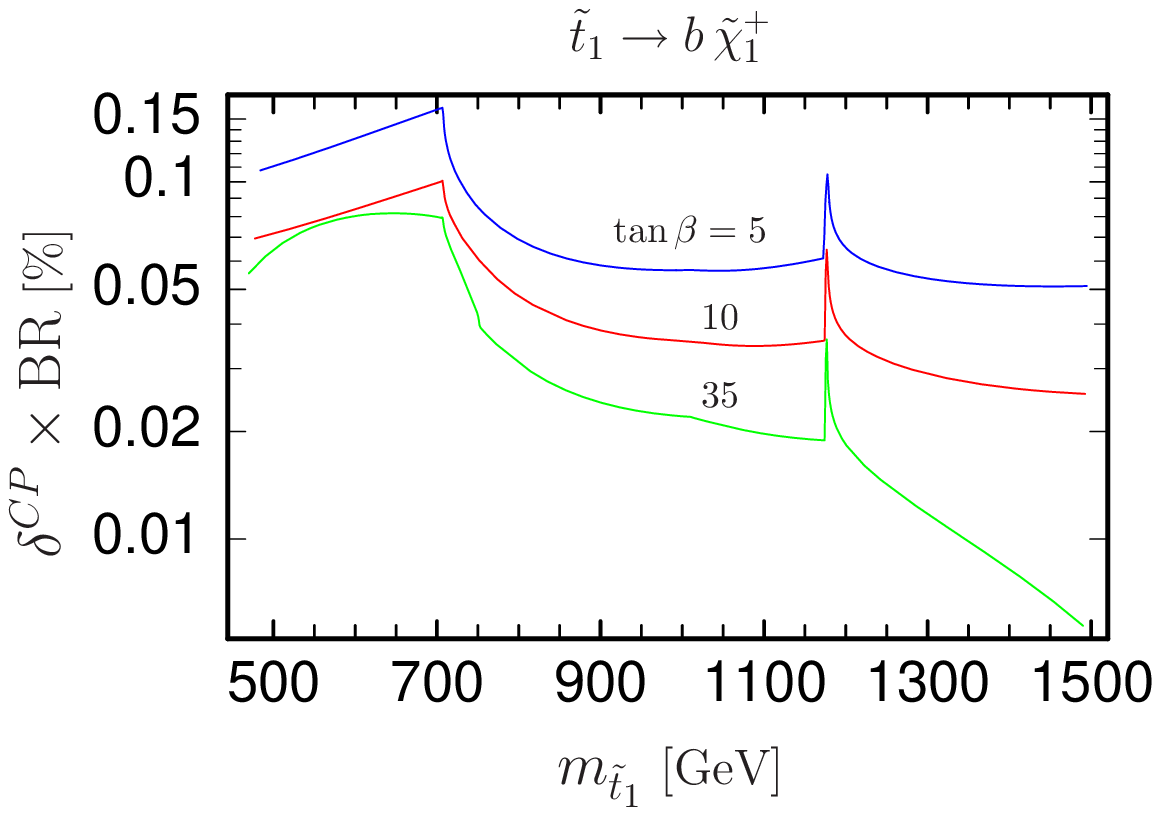} \\
\end{tabular}
\end{center}
\caption{$\delta^{CP}$ and $\delta^{CP} \times \mathrm{BR}$ as a
function of $m_{\tilde t_1}$ for various $\tan \beta$, considering
all contributions at full one-loop except the two dominant
contributions with a gluino. Again, the parameter actually varied is
$M_{\tilde Q}$ from $500$ to $1500$~GeV.}
\label{fig:delCP_Mstop1_TB_except_gluino}
\end{figure}
% delCP_Mstop1_phiAt_all & delCPxBR_Mstop1_phiAt_all
\begin{figure}[htbp]
\begin{center}
\begin{tabular}{cc}
\includegraphics[width=0.45\textwidth]{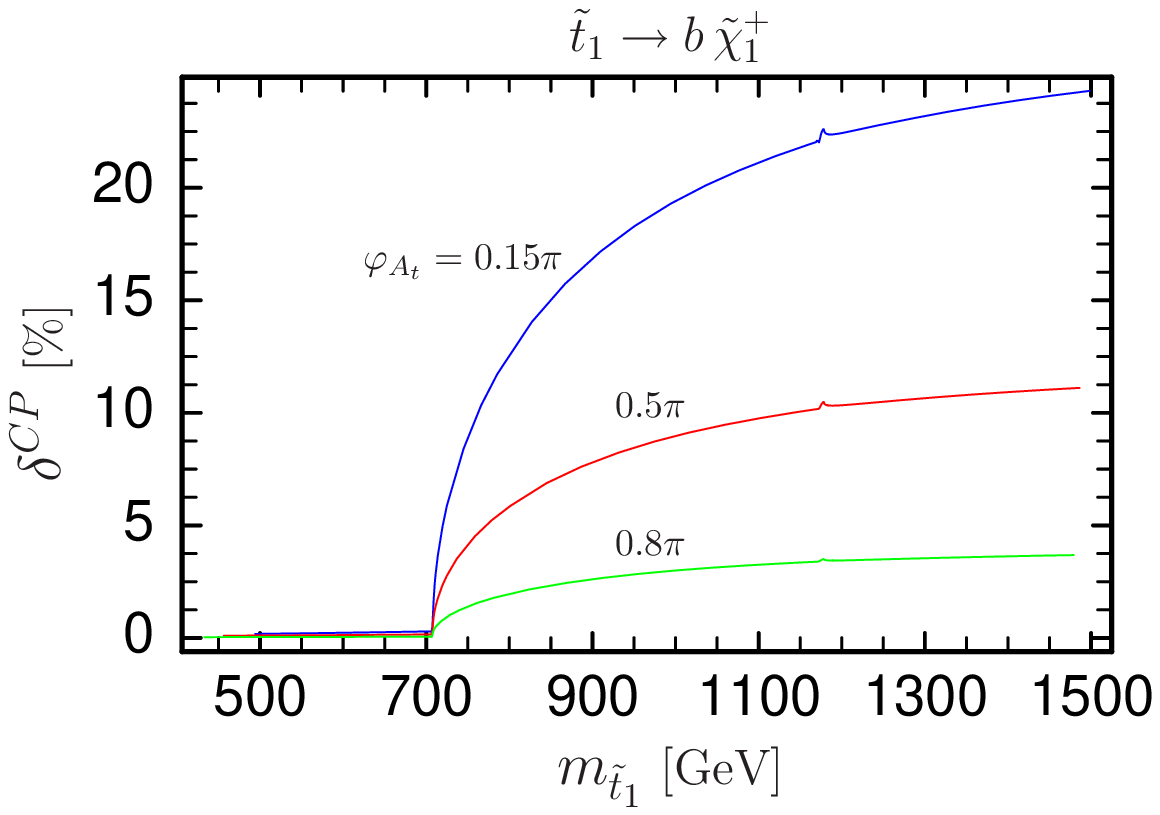} &
\includegraphics[width=0.45\textwidth]{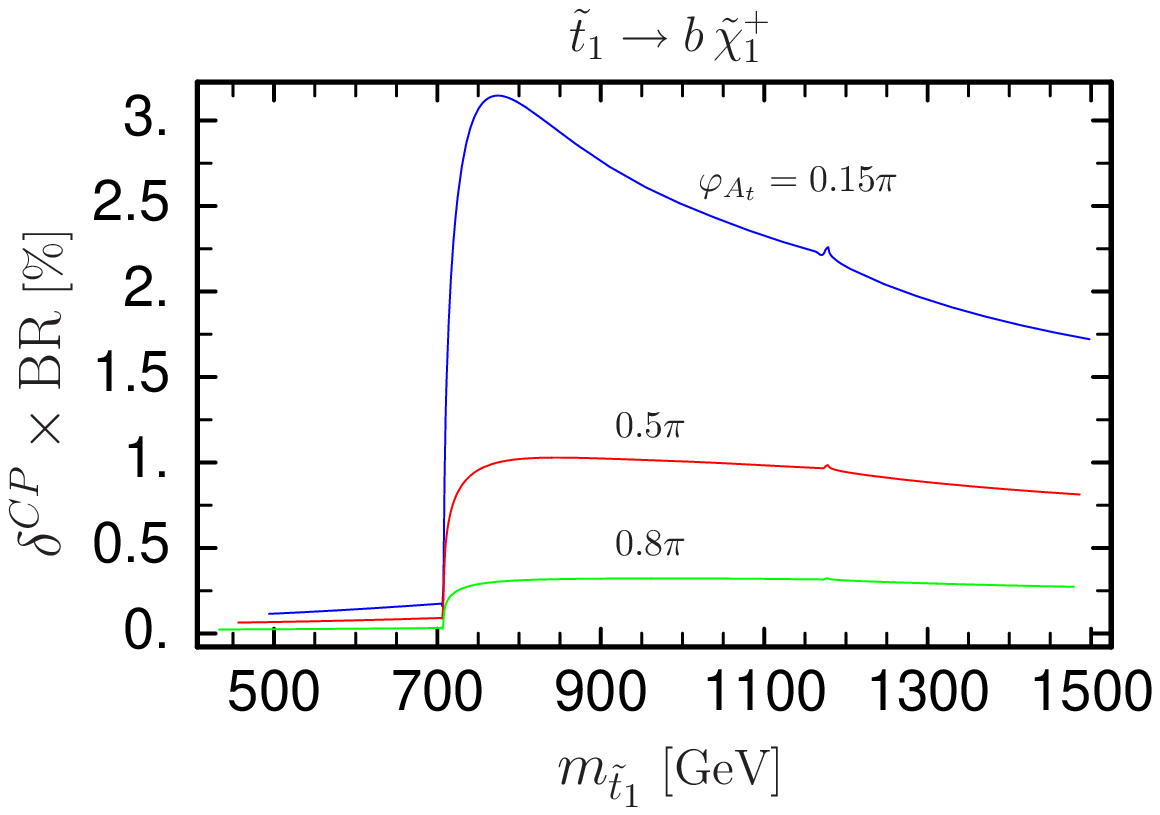} \\
\end{tabular}
\end{center}
\caption{$\delta^{CP}$ and $\delta^{CP} \times \mathrm{BR}$ as a
function of $m_{\tilde t_1}$ for various $\varphi_{A_t}$,
considering all contributions. As always, the parameter actually
varied is $M_{\tilde Q}$ from $500$ to $1500$~GeV.}
\label{fig:delCP_Mstop1_phiAt_all}
\end{figure}

\noindent In Fig.~\ref{fig:delCP_Mstop1_TB_except_gluino} we plot
$\delta^{CP}$ as well as $\delta^{CP} \times BR$ as a function of
$M_{\tilde Q}$ ($m_{\tilde t_1}$), this time taking all
contributions except the two contributions with a gluino. The kink
at $m_{\tilde t_1} \sim 1175$~GeV is the same feature already seen
in Fig.~\ref{fig:delCP_Mstop1_TB_all}
and~\ref{fig:ratio_gluino_all_Mstop1_TB}. The kink at $m_{\tilde
t_1} \sim 750$~GeV and $\tan \beta = 35$ comes from the closure of
the decay channel $\tilde t_1 \to W \, \tilde b_1$ in the graph with
$\tilde \chi^0_j - \tilde b_1 - W$ in the vertex correction.
Finally, the kink at $m_{\tilde t_1} \sim 708$~GeV seen in the
$\delta^{CP} \times BR$ plot comes from the branching ratio $BR$ of
the decay $\tilde t_1 \to b \, \tilde \chi^+_1$. Since the decay
channel into $t \, \tilde g$ is now open, it subtracts a lot from
$BR$. One again, we can see that the contributions without a gluino
can be neglected, if the decay into a gluino and a top is
kinematically possible.

\noindent In Fig.~\ref{fig:delCP_Mstop1_phiAt_all} we present
$\delta^{CP}$ and $\delta^{CP} \times BR$ as a function of
$M_{\tilde Q}$ ($m_{\tilde t_1}$), taking all contributions for
various $\varphi_{A_t}$. Because the complex phase of $A_t$
($=A_{b,\tau}$) is the only source of CP violation in our chosen
scenario, it highly influences $\delta^{CP}$. Taking $\varphi_{A_t}
= 0.15 \, \pi$, we obtain the highest value of $\delta^{CP}$ with
our parameter set, $\delta^{CP} \sim 24 \, \%$ at $m_{\tilde
t_1}=1500$~GeV.

\begin{figure}[htbp]
\begin{center}
\begin{tabular}{cc}
\includegraphics[width=0.45\textwidth]{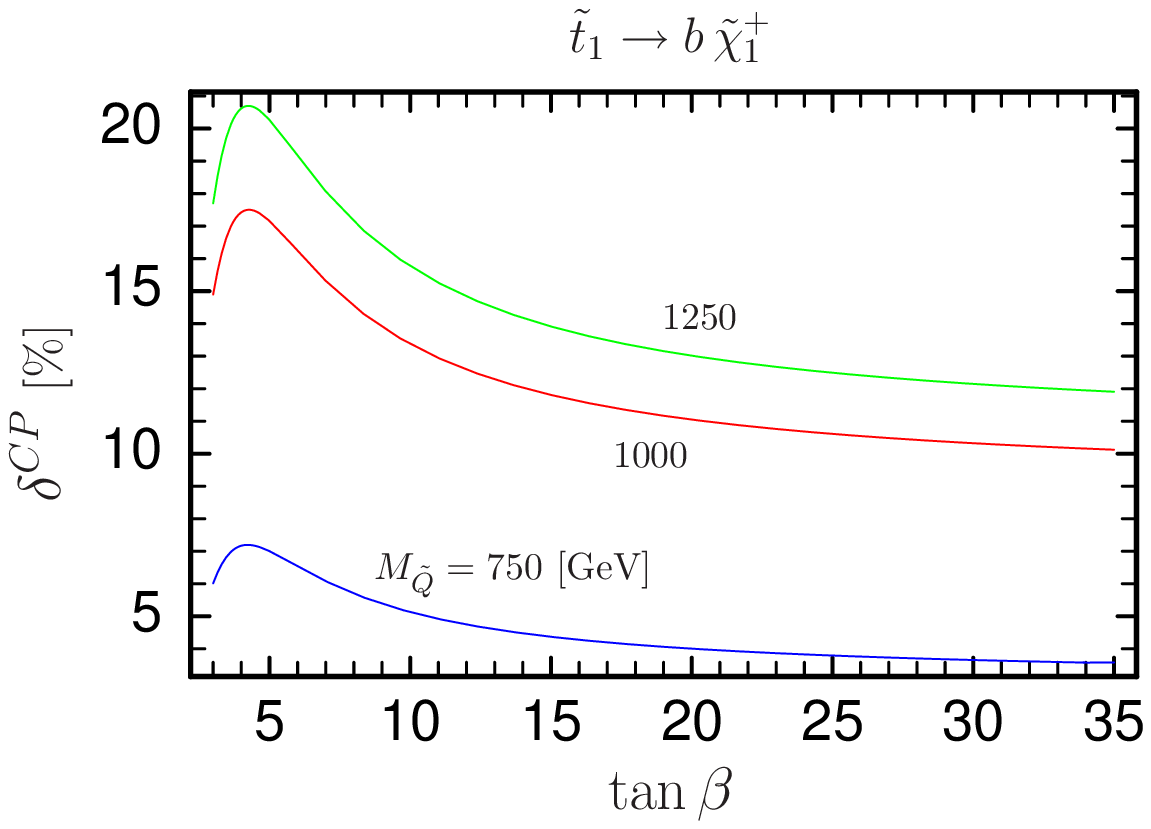} &
\includegraphics[width=0.45\textwidth]{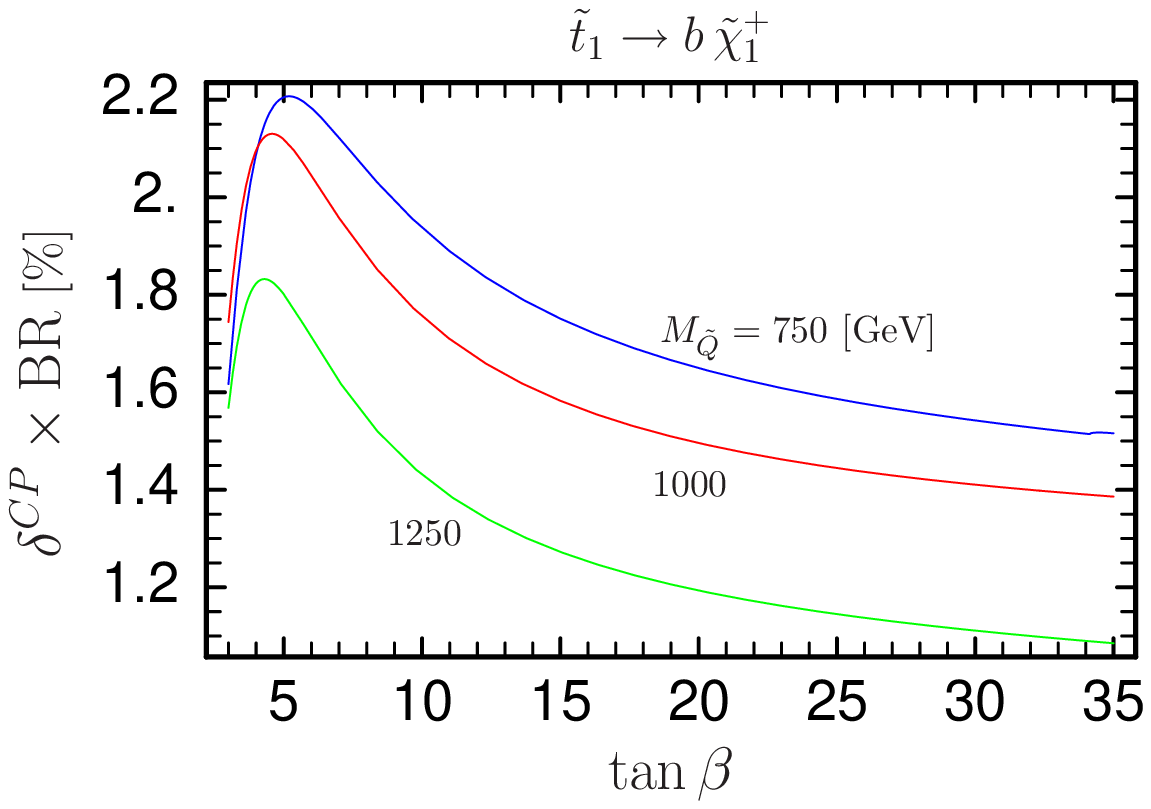} \\
\end{tabular}
\end{center}
\caption{$\delta^{CP}$ and $\delta^{CP} \times \mathrm{BR}$ as a
function of $\tan \beta$ for various $M_{\tilde Q}$, taking all
contributions.} \label{fig:delCP_TB_MSQ_all}
\end{figure}

\noindent Fig.~\ref{fig:delCP_TB_MSQ_all} shows the dependence of
$\delta^{CP}$ and $\delta^{CP} \times BR$ on $\tan \beta$ for
various $M_{\tilde Q}$, taking all contributions. The maximal
asymmetry lies around $\tan \beta \sim 4 \, \%$. The higher the
breaking mass parameter $M_{\tilde Q}$ (and thus the mass of the
decaying particle), the higher the asymmetry $\delta^{CP}$, because
more and more decay channels open up and hence more and more
processes start to contribute. On the other hand, the more decay
channels open up, the less is left for the branching ratio $BR$ of
the decay $\tilde t_1 \to b \, \tilde \chi^+_1$.
% delCP_TB_MSQ_all & delCPxBR_TB_MSQ_all

\begin{figure}[htbp]
\begin{center}
\begin{tabular}{cc}
\includegraphics[width=0.45\textwidth]{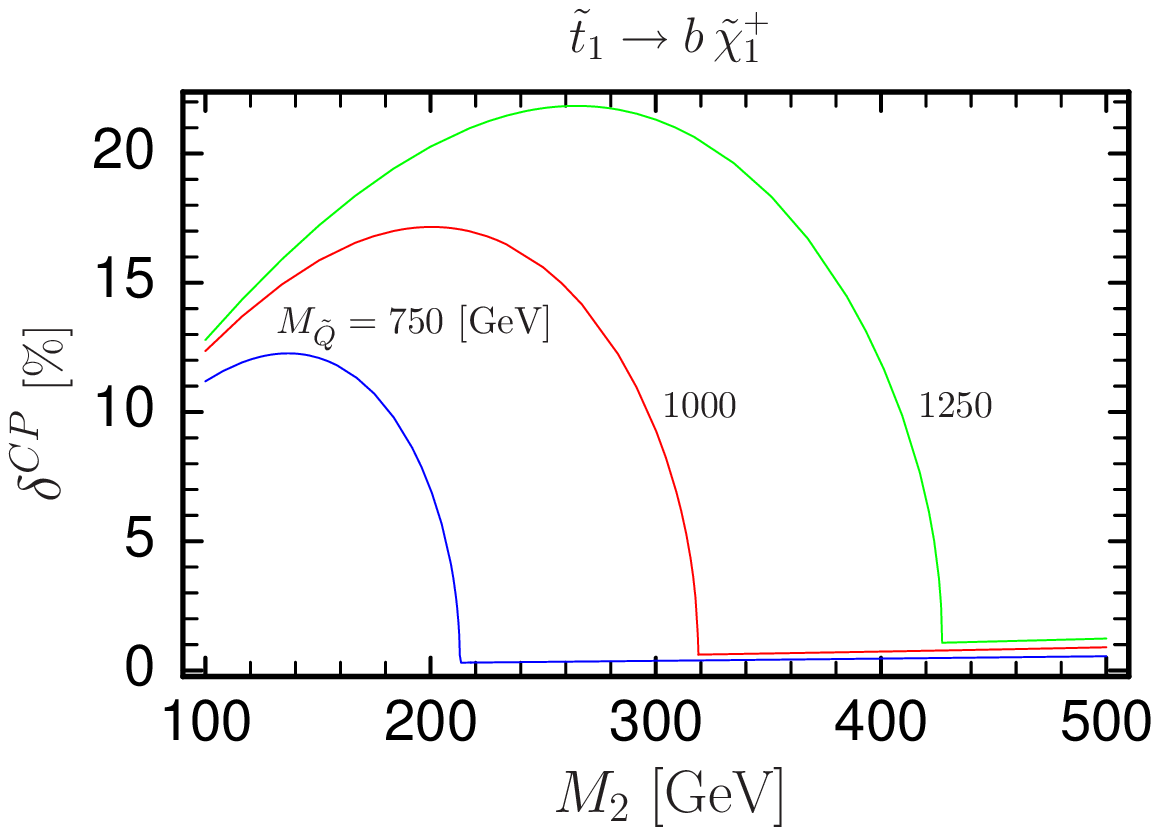} &
\includegraphics[width=0.45\textwidth]{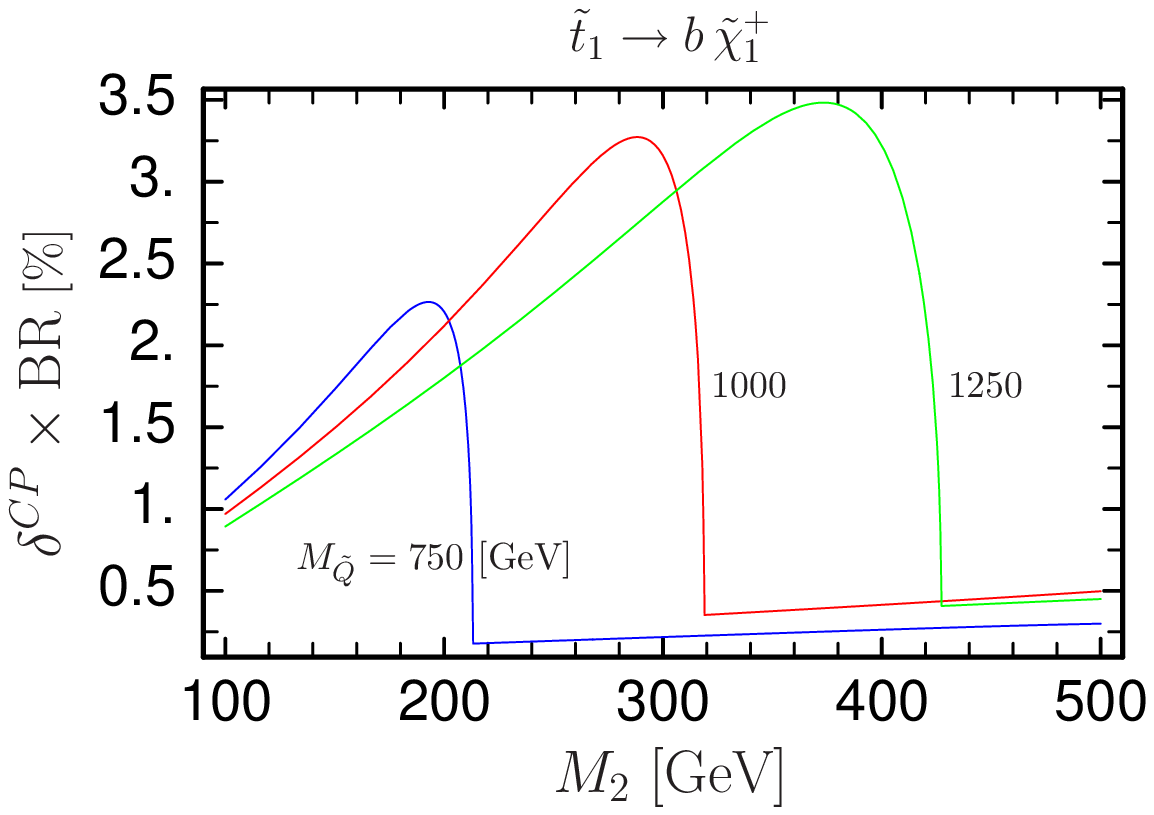} \\
\end{tabular}
\end{center}
\caption{$\delta^{CP}$ and $\delta^{CP} \times \mathrm{BR}$ as a
function of $M_2$ for various $M_{\tilde Q}$, taking all
contributions.} \label{fig:delCP_M2_MSQ_all}
\end{figure}

\noindent In Fig.~\ref{fig:delCP_M2_MSQ_all} we study the dependence
of $\delta^{CP}$ and $\delta^{CP} \times BR$ on $M_2$ for various
$M_{\tilde Q}$, taking all contributions. One can nicely see the
closure of the dominating decay channel $\tilde t_1 \to t \, \tilde
g$, because $M_2$ is related to the gluino mass $m_{\tilde g}$ via
GUT relations. The higher the mass $M_{\tilde Q}$ ($m_{\tilde
t_1}$), the later this closure happens. The maximal value of
$\delta^{CP} \times BR$ with our parameter set is $\delta^{CP}
\times BR \sim 3.5 \, \%$ at $M_2 \sim 370$~GeV and $M_{\tilde
Q}=1250$~GeV.
% delCP_M2_MSQ_all & delCPxBR_M2_MSQ_all

\begin{figure}[htbp]
\begin{center}
\includegraphics[width=0.6\textwidth]{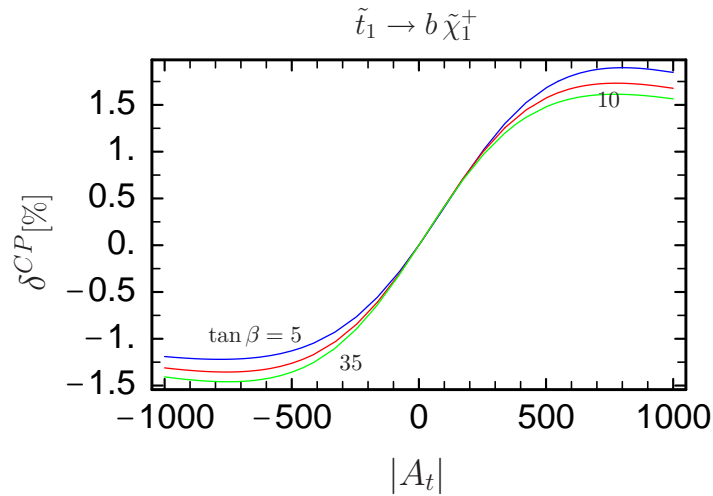}
\end{center}
\caption{$\delta^{CP}$ as a function of $|A_t|$ for various $\tan
\beta$, taking all contributions and setting $M_{\bar u}=1.1 \,
M_{\tilde Q}$, $M_{\bar d}=0.8 \, M_{\tilde Q}$.}
\label{fig:delCP_absAt_TB_all}
\end{figure}

\noindent Fig.~\ref{fig:delCP_absAt_TB_all} shows the dependence
on the absolute value of the trilinear breaking parameter $|A_t|$
for several $\tan \beta$. Here we alter our relation of the
breaking mass parameters to $M_{\bar u}=1.1 \, M_{\tilde Q}$,
$M_{\bar d}=0.8 \, M_{\tilde Q}$. Otherwise, we would obtain a
non-physical result, because of the degeneration of the stop
masses (see the main- and off-diagonal elements of the sfermion
mass matrix in Eq.~(\ref{eq:sfermion_mass})) which leads to a
singularity in the $\tilde t_j$ propagator of the Graph 1 in
Chapter~\ref{chapter:contributions}. On the other hand, this
degeneration would not bother us, if we have not used the
approximation done in Eq.~(\ref{eq:deltaCP_approx}). If we would
calculate $\Gamma^\pm$ of Eq.~(\ref{eq:deltaCP}) (and therefore
the sum $\Gamma^+ + \Gamma^-$ in $\delta^{CP}$) correctly at full
one-loop order using renormalization, this singularity becomes
harmless. Due to our relation of the breaking mass parameters, the
dependence of the asymmetry on $\tan \beta$ is small.
% delCP_absAt_TB_all

\begin{figure}[htbp]
\setlength{\unitlength}{1mm}
\begin{center}
\begin{picture}(73,50)
\put(0,0){\includegraphics[width=0.45\textwidth]{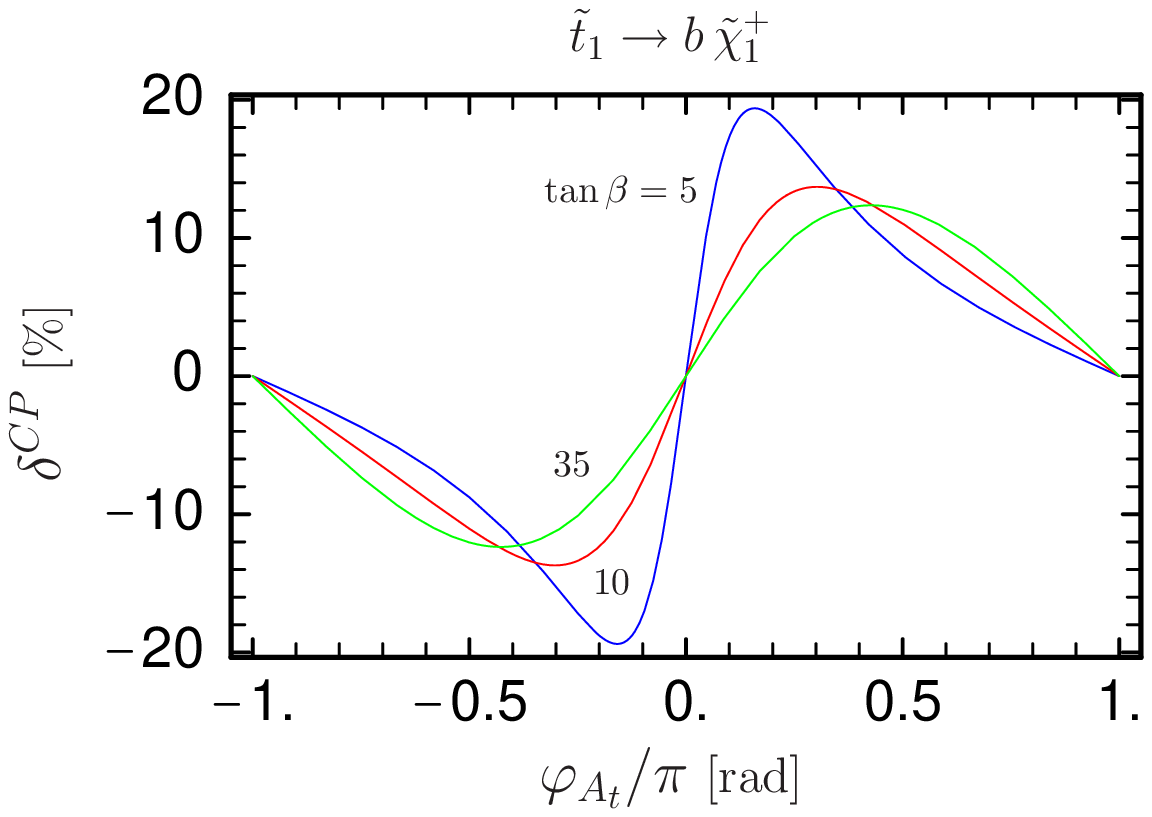}}
\put(0,45){(a)}
\end{picture}
\hspace{5mm}
\begin{picture}(73,50)
\put(0,0){\includegraphics[width=0.45\textwidth]{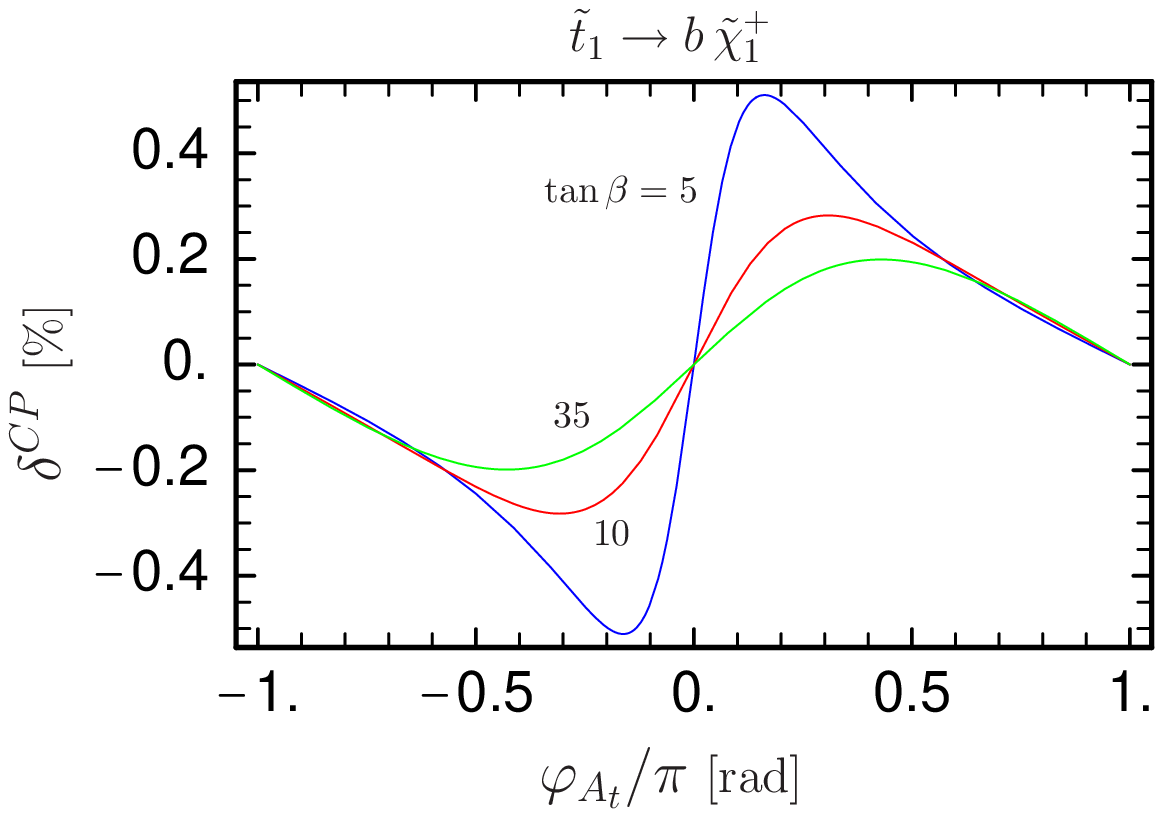}}
\put(0,45){(b)}
\end{picture}
\end{center}
\caption{$\delta^{CP}$ as a function of $\varphi_{A_t}$ for various
$\tan \beta$, taking (a) all contributions and (b) all contributions
except the gluino contributions.} \label{fig:delCP_phiAt_TB}
\end{figure}

\noindent In Fig.~\ref{fig:delCP_phiAt_TB} we plot the dependence of
$\delta^{CP}$ on $\varphi_{A_t}$, taking (a) all contributions and
(b) all contributions except the gluino contributions. One can
clearly see the periodic dependance on $\varphi_{A_t}$. The overall
maximum lies at $\varphi_{A_t} \sim 0.15 \, \pi$ with $\tan \beta =
5$.
% delCP_phiAt_TB_all & except_gluino

\begin{figure}[htbp]
\begin{center}
\includegraphics[width=0.6\textwidth]{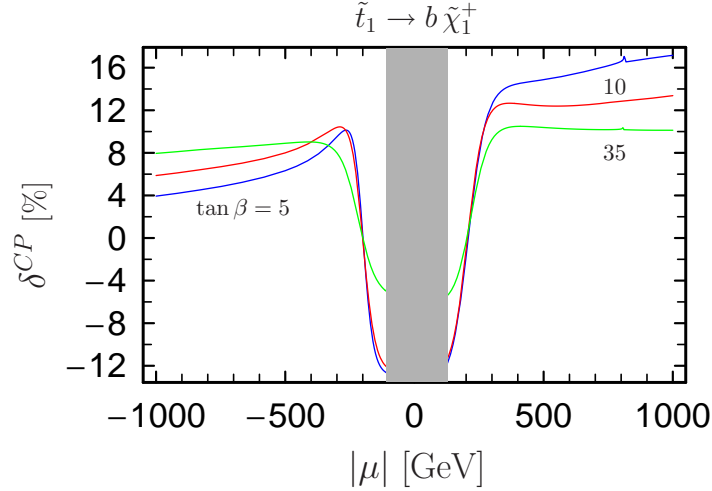}
\end{center}
\caption{$\delta^{CP}$ as a function of $|\mu|$ for various $\tan
\beta$, considering all contributions. The inner area is masked out
because the chargino mass falls below its lower bound.}
\label{fig:delCP_absMUE_TB_all}
\end{figure}

\noindent Fig.~\ref{fig:delCP_absMUE_TB_all} shows $\delta^{CP}$
as a function of $|\mu|$ for various $\tan \beta$, considering
again all contributions. Because the chargino mass falls below its
lower bound, the inner area is ruled out by experiment and thus
masked grey. For high $\tan \beta$ the dependence on $|\mu|$
becomes rather symmetric, because the low $\cot \beta$ in the
off-diagonal elements in Eq.~(\ref{eq:sfermion_mass_offdiag})
diminishes the dependence on $\mu$. At $|\mu| \sim \pm 300$~GeV we
can see the transition of the decaying particle $\tilde \chi^+_1$
between being wino-like (at higher $|\mu|$) and higgsino-like (at
lower $|\mu|$), resulting in a different behaviour of the coupling
and thus $\delta^{CP}$.
% delCP_absMUE_TB_all

\begin{figure}[htbp]
\begin{center}
\includegraphics[width=0.6\textwidth]{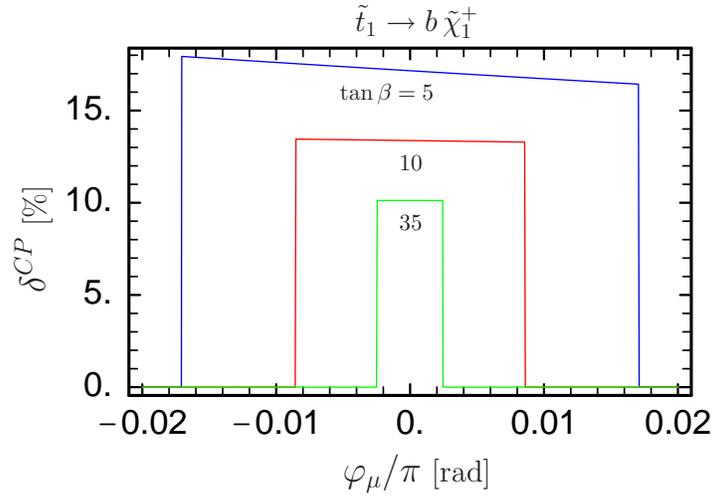}
\end{center}
\caption{$\delta^{CP}$ as a function of $\varphi_\mu$ for various
$\tan \beta$, considering all contributions. $\delta^{CP}$ is set to
zero in the regions not allowed by the experimental eEDM-limit.}
\label{fig:delCP_phiMUE_TB_all}
\end{figure}

\noindent Fig.~\ref{fig:delCP_phiMUE_TB_all} with $\delta^{CP}$ as a
function of $\varphi_\mu$ demonstrates the stringent constraint on
the phase of $\mu$, coming from the experimental eEDM-limit. Only a
very narrow area is allowed, outside this area we have set
$\delta^{CP}$ to zero. As $\tan \beta$ gets bigger, the allowed area
reduces even more and therefore we have chosen $\varphi_\mu=0$ in
our input parameter set in the first place. The negligible
dependence of $\delta^{CP}$ on $\varphi_\mu$ can be explained with
the very narrow parameter range of $\varphi_\mu$. Furthermore, the
higher $\tan \beta$ gets, the lower the dependence on $\mu$ (due to
$\cot \beta$ in Eq.~(\ref{eq:sfermion_mass_offdiag})) and thus
$\varphi_\mu$ becomes.
% delCP_phiMUE_TB_all

\begin{figure}[htbp]
\begin{center}
\begin{tabular}{cc}
\includegraphics[width=0.45\textwidth]{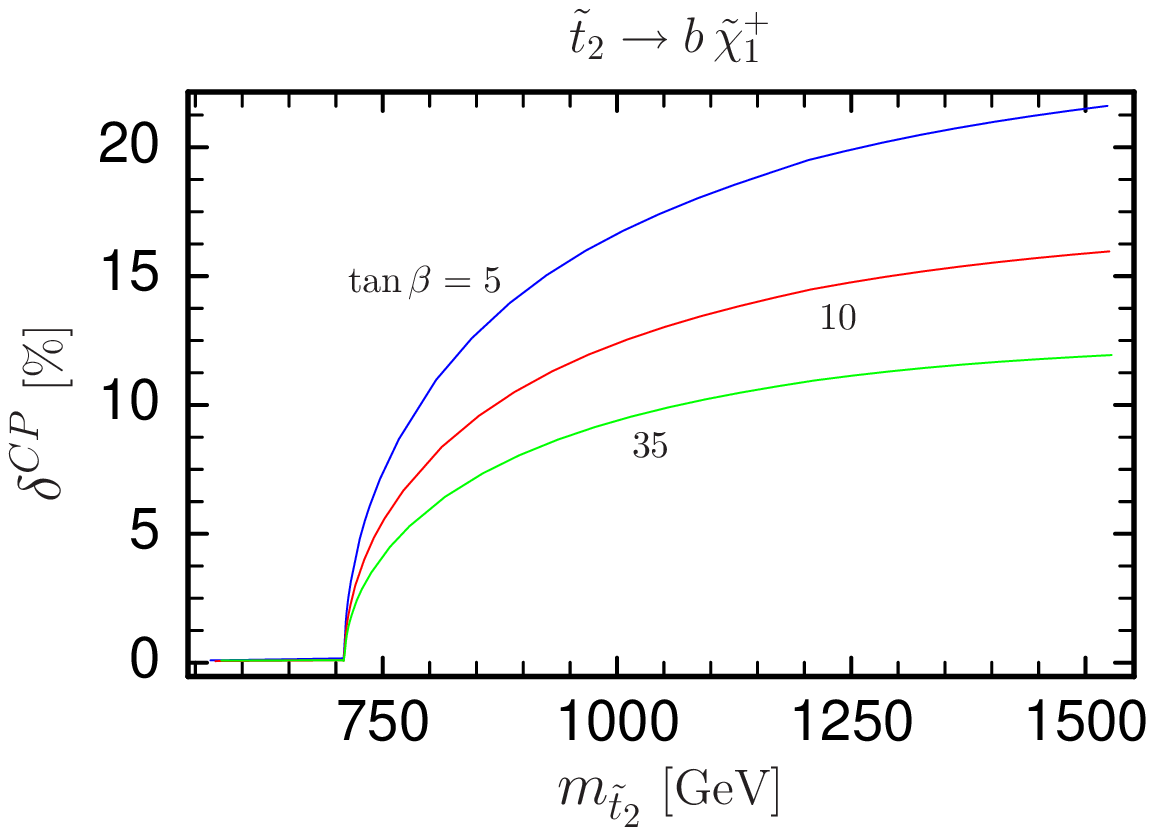} &
\includegraphics[width=0.45\textwidth]{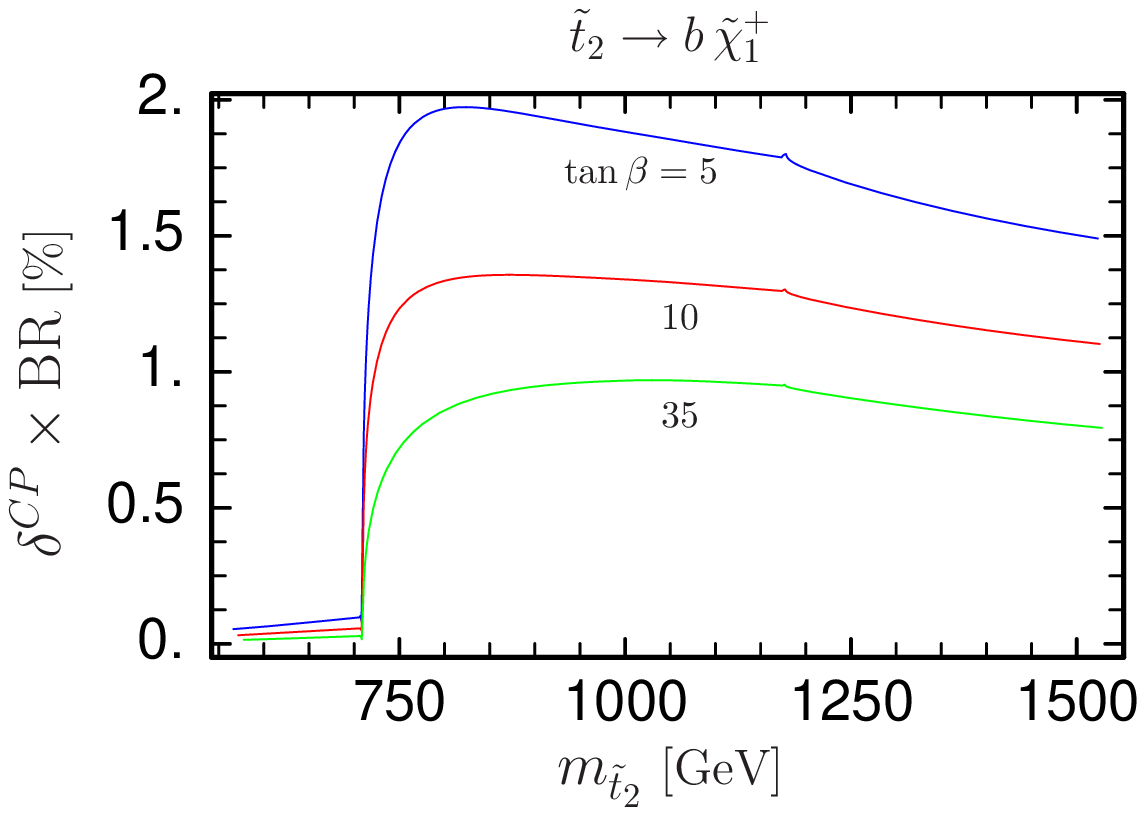} \\
\end{tabular}
\end{center}
\caption{$\delta^{CP}$ and $\delta^{CP} \times \mathrm{BR}$ of the
decaying particle $\tilde t_2$ as a function of $m_{\tilde t_2}$ for
various $\tan \beta$, considering all contributions (full one-loop).
The parameter $m_{\tilde t_2}$ is shown for convenience, but the
parameter actually varied is $M_{\tilde Q}$ from $500$ to
$1500$~GeV.} \label{fig:delCP_Mstop2_TB_all}
\end{figure}

\begin{figure}[htbp]
\begin{center}
\begin{tabular}{cc}
\includegraphics[width=0.45\textwidth]{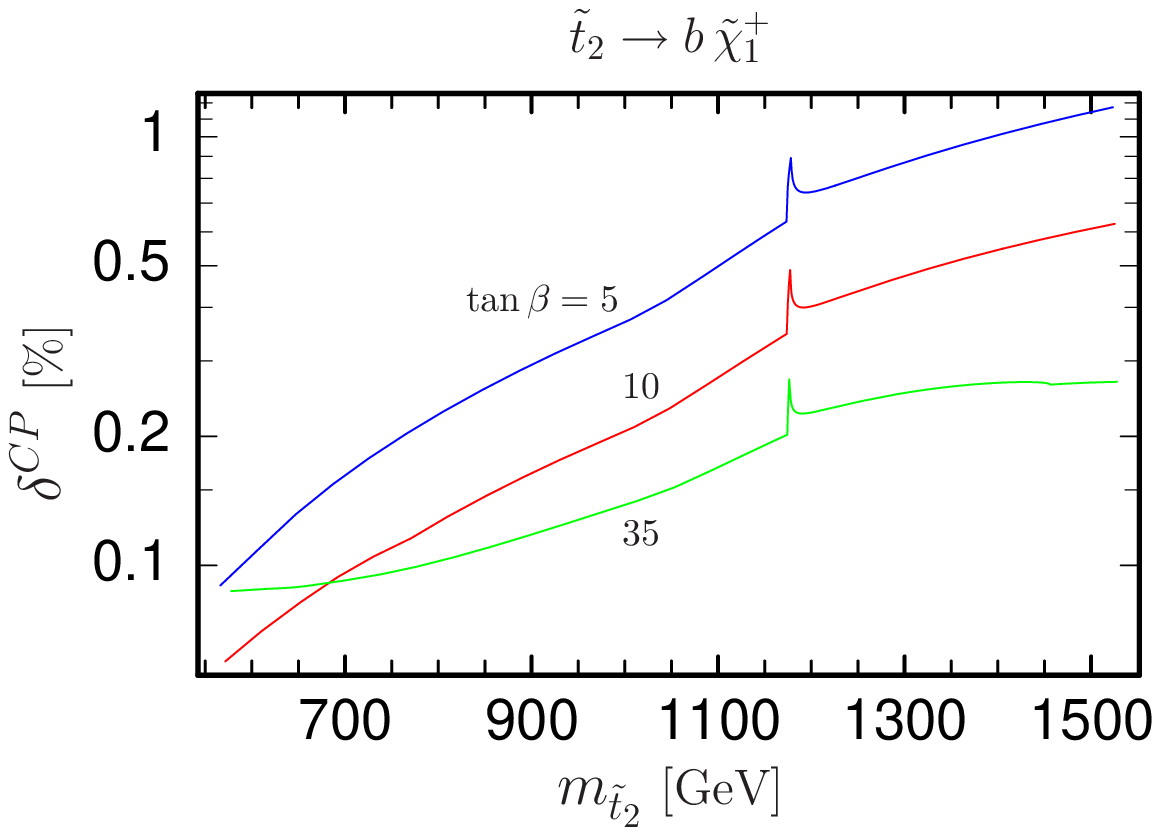} &
\includegraphics[width=0.45\textwidth]{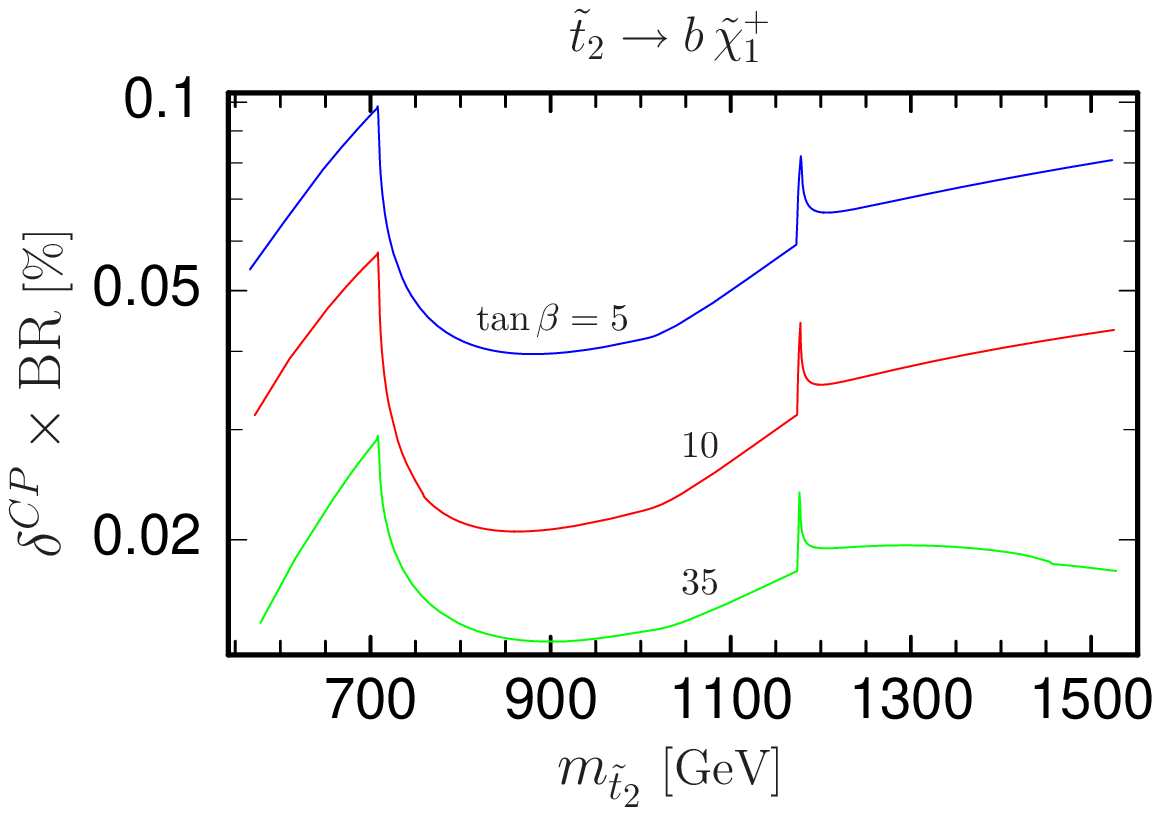} \\
\end{tabular}
\end{center}
\caption{$\delta^{CP}$ and $\delta^{CP} \times \mathrm{BR}$ of the
decaying particle $\tilde t_2$ as a function of $m_{\tilde t_2}$ for
various $\tan \beta$, considering all contributions at full one-loop
except the two dominant contributions with a gluino. Again, the
parameter actually varied is $M_{\tilde Q}$ from $500$ to
$1500$~GeV.} \label{fig:delCP_Mstop2_TB_except_gluino}
\end{figure}

\noindent For completeness, we also examined the decay of the
heavier stop $\tilde t_2$ into $b \, \tilde \chi^+_1$. Because the
two stop particles barely mix in our scenario, their masses are very
similar ($m_{\tilde t_1}=992.10$~GeV and $m_{\tilde
t_2}=1034.73$~GeV). Therefore, the resulting plots are alike, as one
can see in Fig.~\ref{fig:delCP_Mstop2_TB_all} in comparison with
Fig.~\ref{fig:delCP_Mstop1_TB_all} and also in
Fig.~\ref{fig:delCP_Mstop2_TB_except_gluino} compared with
Fig.~\ref{fig:delCP_Mstop1_TB_except_gluino}.
% delCP_Mstop2_TB_all & delCPxBR_Mstop2_TB_all
% delCP_Mstop2_TB_except_gluino & delCPxBR_Mstop2_TB_except_gluino

\noindent Finally, we also investigated the influence on the Yukawa
couplings $h_t$ and $h_b$ taken to be running. In our scenario, the
difference of the asymmetry $\delta^{CP}$ taken with running Yukawa
couplings (tested at two different scales, the mass of the decaying
particle and $1000$~GeV) and taken with not running ones is
negligible. Only at high values of $M_{\tilde Q}=1500$~GeV one
obtains a small deviation; not running Yukawa couplings yield a
slightly higher asymmetry of $\sim 5 \, \%$.

\clearpage

\section{Conclusions}

In this thesis we performed a detailed numerical analysis of the
CP violating decay rate asymmetry $\delta^{CP}$ and the quantity
$\delta^{CP} \times BR$ of the processes $\tilde t_1 \to b \,
\tilde \chi^+_1$ and $\tilde t_2 \to b \, \tilde \chi^+_1$ (and
their CP transformed counterparts), analyzing the dependence on
the parameters and phases involved.\\
The asymmetry $\delta^{CP}$ rises up to $\sim 24 \, \%$,
depending on the point in parameter space. The combined quantity
$\delta^{CP} \times \mathrm{BR}$ reaches up to $\sim 3.5 \, \%$.\\
Finally, we want to comment on the feasibility of measuring this
asymmetry at the Large Hadron Collider (LHC) at CERN, which will go
in operation in summer 2008. As the decaying stop is a strong
interacting particle, its production cross section will be large.
Therefore, measurement of our decay rate asymmetry $\delta^{CP}$
will be possible at LHC. However, the precise calculation of the
measurability is beyond the scope of this thesis, as it involves
monte-carlo simulations accounting for all the peculiarities of the
detector, among other things.\\
On the basis of our promising results of the CP violating decay rate
asymmetry, we suggest that experimenters should search for evidence
of these new CP violating asymmetries in the MSSM, which can be far
beyond the small CP violating effects in the SM. These new CP
violating sources are not only important in terms of baryogenesis
but also very interesting for the further understanding of the
subatomic world.

\appendix

\newpage
\chapter{Listing of All One-Loop Contributions}
\label{chapter:all-contributions}

Here we specify the complete list of all processes at full one-loop level
who can contribute to CP violating asymmetries.

% all vertex contributions (part 1)
\begin{figure}[htbp]
\begin{center}
\includegraphics[width=0.55\textwidth]{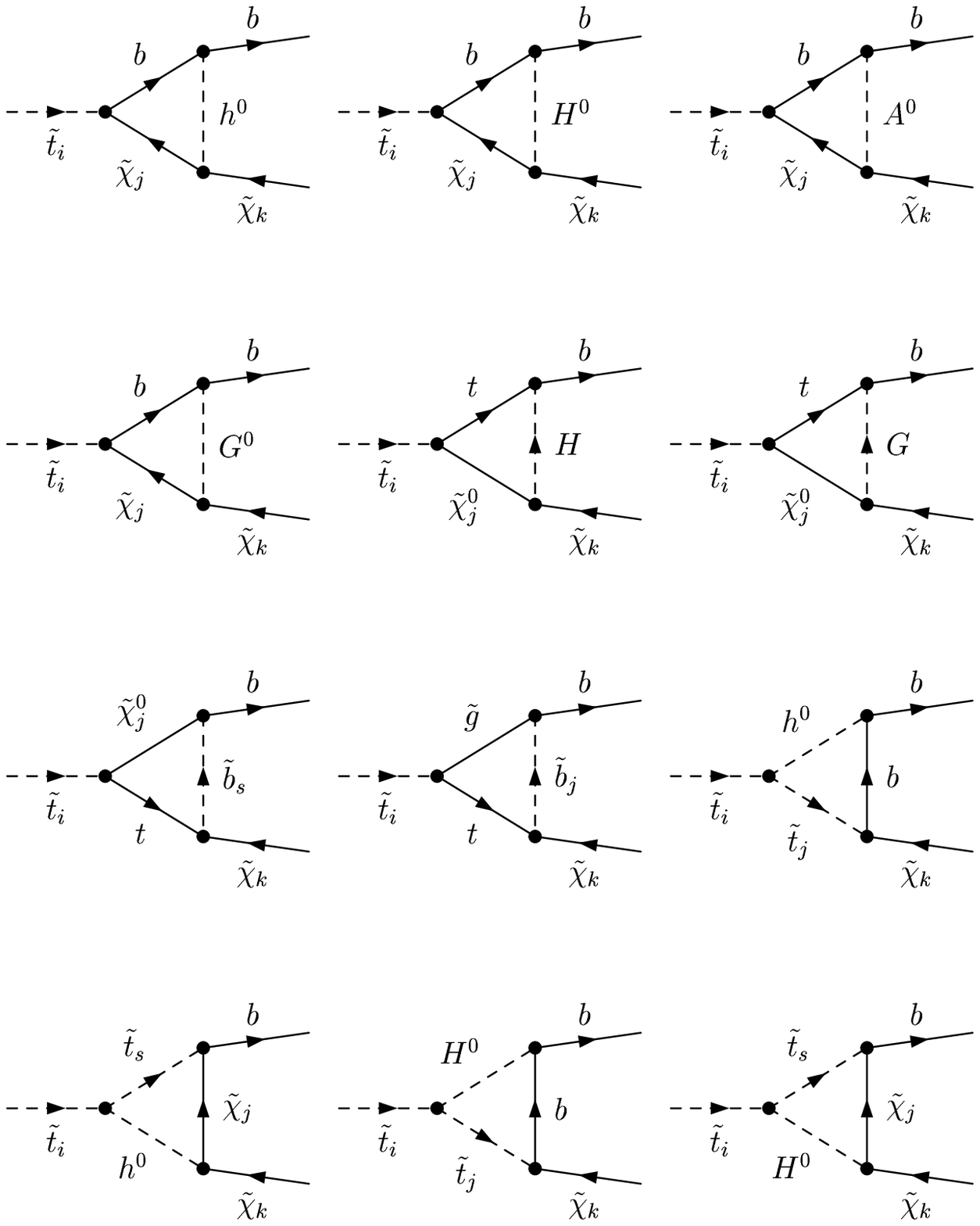}
\end{center}
\caption{All vertex contributions who add to the CP violation (Part 1).}
\label{fig:CP-violation-vertex1}
\end{figure}

% all vertex contributions (part 2)
\begin{figure}[htbp]
\begin{center}
\includegraphics[width=0.7\textwidth]{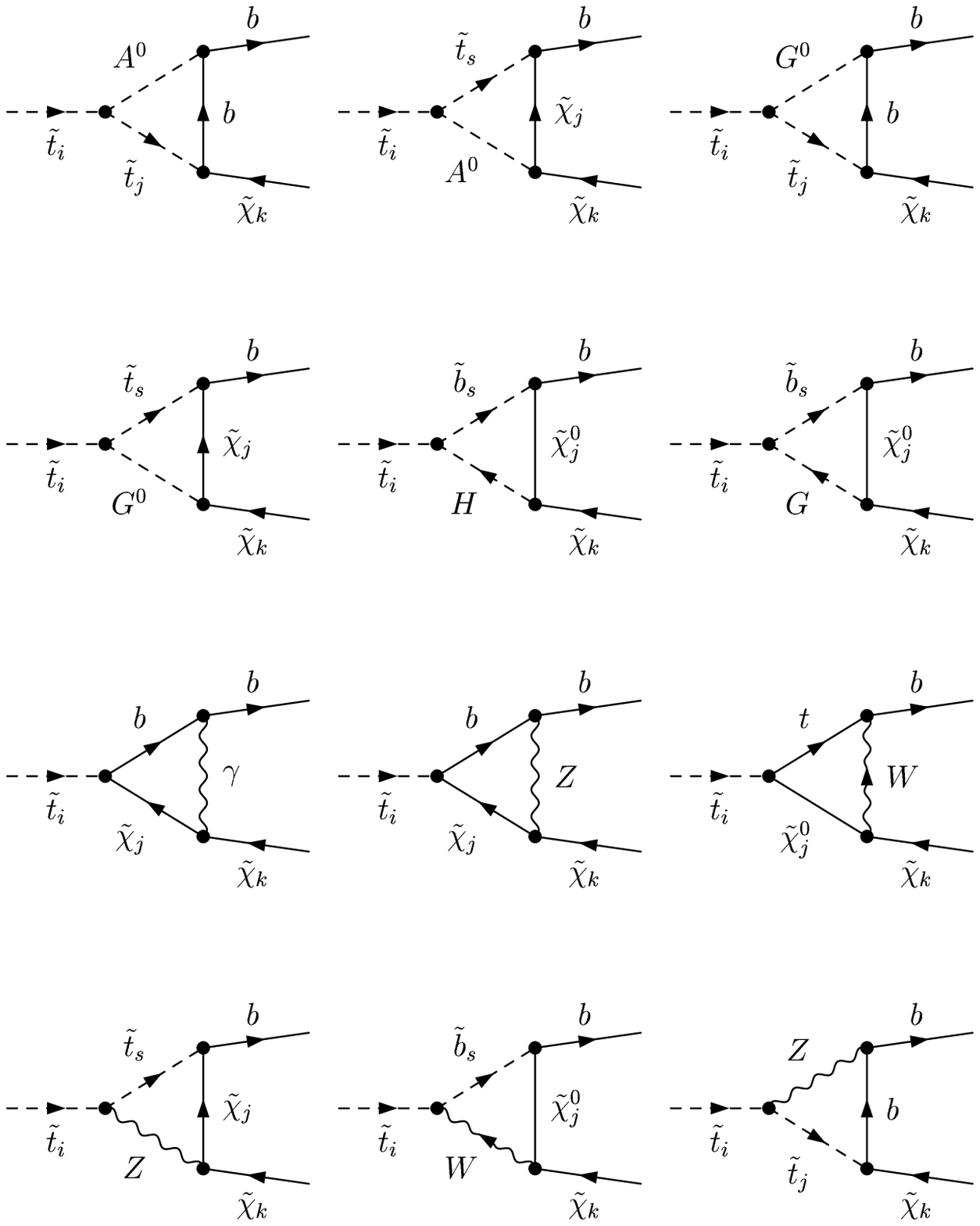}
\end{center}
\caption{All vertex contributions who add to the CP violation (Part 2).}
\label{fig:CP-violation-vertex2}
\end{figure}

% all stop-selfenergy contributions
\begin{figure}[htbp]
\begin{center}
\includegraphics[width=0.7\textwidth]{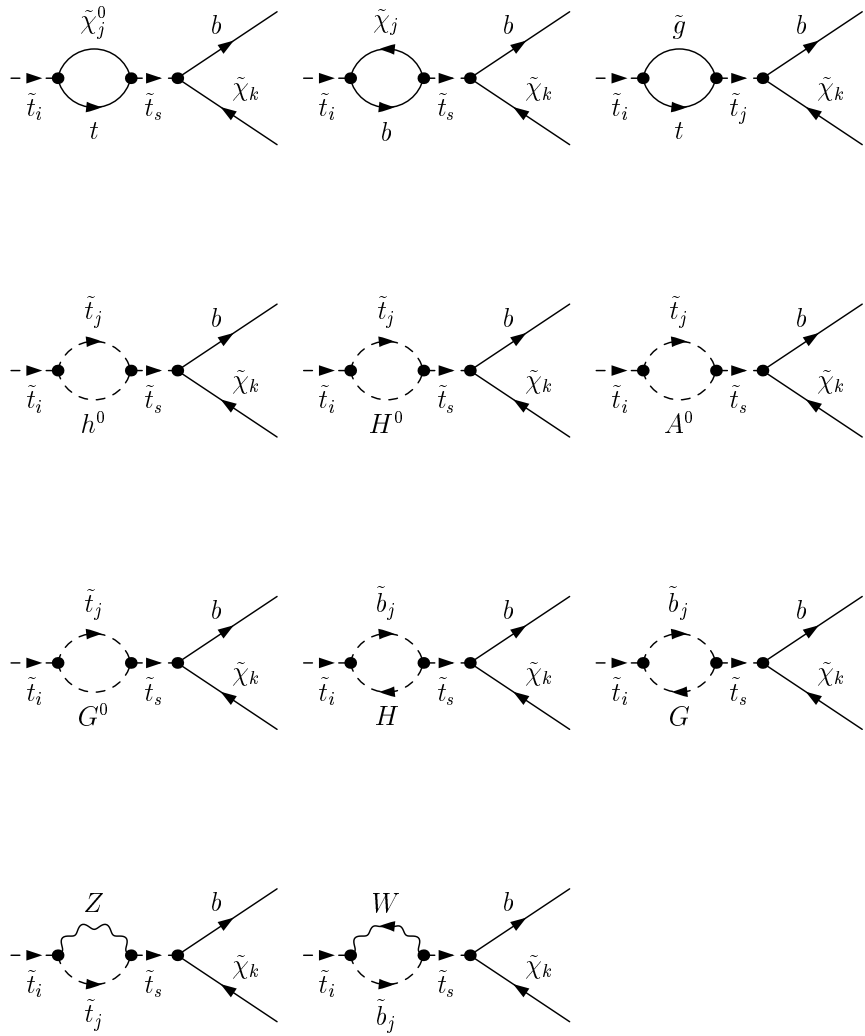}
\end{center}
\caption{All stop-selfenergy contributions who add to the CP violation.}
\label{fig:CP-violation-stopself}
\end{figure}

% all chargino-selfenergy contributions
\begin{figure}[htbp]
\begin{center}
\includegraphics[width=0.7\textwidth]{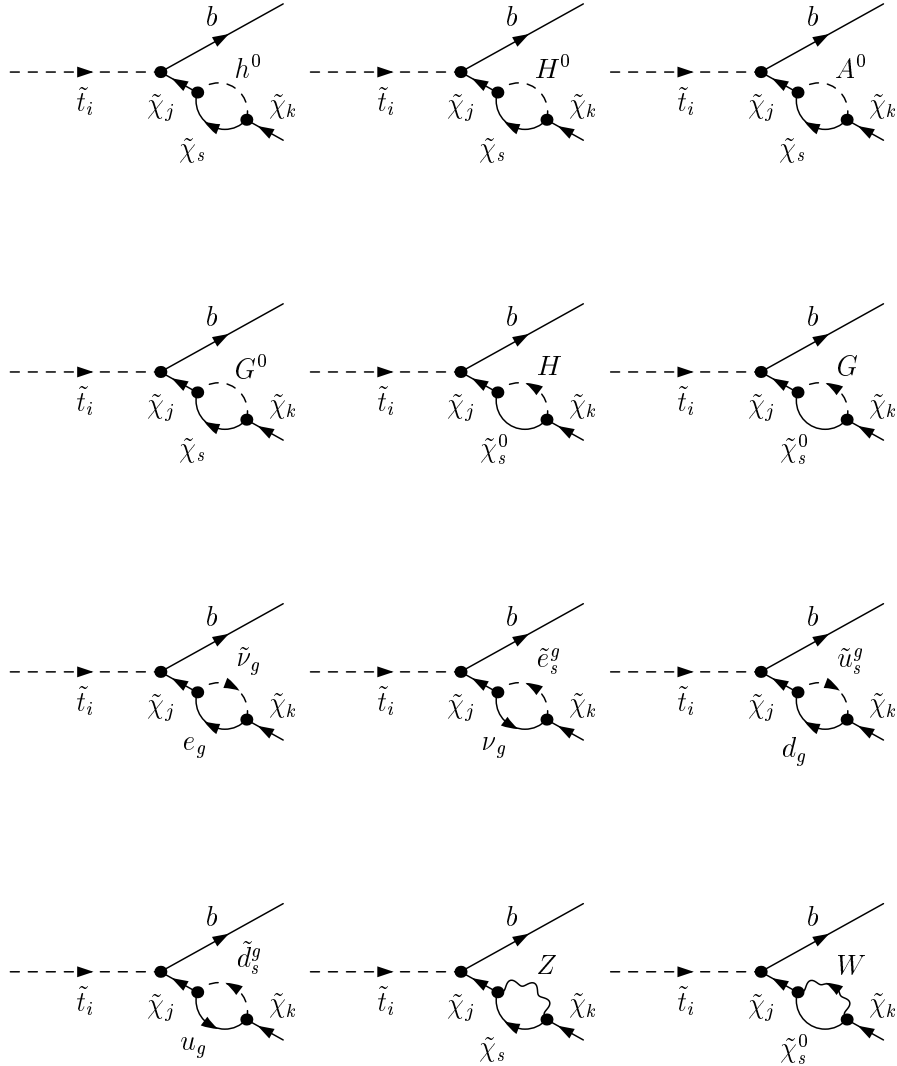}
\end{center}
\caption{All chargino-selfenergy contributions who add to the CP violation.}
\label{fig:CP-violation-chaself}
\end{figure}

\newpage
%\chapter{Passarino--Veltman integrals}
\chapter{Passarino--Veltman Integrals}
\label{chapter:pave}

In this chapter we give the definition of the Passarino--Veltman
one-, two-, and three-point functions~\cite{pave} and list some
functions with a special argument set.

\section{Definitions}

We define the Passarino--Veltman one-, two-, and three-point
functions in the convention of~\cite{Denner}. For the general
denominators we use the notation
\begin{equation}
  {\mathcal D}^{0} = q^{2} - m_{0}^{2}
  \quad \mbox{and} \quad
  {\mathcal D}^{j} = ( q + p_{j} )^{2} - m_{j}^{2}\,.
\end{equation}
Then the loop integrals in $D=4-\epsilon$ dimensions are as follows:
\begin{eqnarray}
  A_0(m_0^2) &=& \frac{1}{i\pi^2}\int d^{D}\!q\:\frac{1}{{\mathcal D}^0}\,,\\
  B_0(p_1^2,m_0^2,m_1^2)   &=& \frac{1}{i\pi^2} \int d^{D}\! q \:
          \frac{1}{{\mathcal D}^0 {\mathcal D}^1} \,, \label{eq:B_0-pave}\\
  B_\mu(p_1^2,m_0^2,m_1^2) &=& \frac{1}{i\pi^2} \int d^{D}\! q \:
      \frac{q_\mu}{{\mathcal D}^0 {\mathcal D}^1}
      = p_{1\mu}\, B_1 \,, \label{eq:B_mu-pave}\\
  B_{\mu \nu}(p_1^2,m_0^2,m_1^2) &=& \frac{1}{i\pi^2} \int d^{D}\! q \:
      \frac{q_\mu q_\nu}{{\mathcal D}^0 {\mathcal D}^1} = g_{\mu \nu} \, B_{00} + p_{1\mu} p_{1\nu} \, B_{11}
      \,, \label{eq:B_mu_nu-pave}
\end{eqnarray}
and
\begin{eqnarray}
  C_0 &=& \frac{1}{i\pi^2} \int d^{D}\! q \:
          \frac{1}{{\mathcal D}^0 {\mathcal D}^1 {\mathcal D}^2} \,,\\
  C_\mu &=& \frac{1}{i\pi^2} \int d^{D}\! q \:
      \frac{q_\mu}{{\mathcal D}^0 {\mathcal D}^1 {\mathcal D}^2}
      = p_{1\mu} C_1 + p_{2\mu} C_2  \,,\\
  C_{\mu\nu} &=& \frac{1}{i\pi^2} \int d^{D}\! q \:
      \frac{q_\mu q_\nu}{{\mathcal D}^0 {\mathcal D}^1 {\mathcal D}^2} \nn\\
      &=& g_{\mu\nu} C_{00} + p_{1\mu} p_{1\nu} C_{11}
          + ( p_{1\mu} p_{2\nu} + p_{2\mu} p_{1\nu} ) C_{12}
          + p_{2\mu} p_{2\nu} C_{22} \,,
\end{eqnarray}
where the $C$'s have
$(p_{1}^{2},(p_{1}-p_{2})^{2},p_{2}^{2},m_{0}^{2},m_{1}^{2},m_{2}^{2})$
as their arguments. For further details about the coefficient
functions, some reductions and relations, some analytical
expressions and some special argument sets see~\cite{helmut-diss}.

%\section{Special argument set}
\section{Special Argument Set}

Here we list the Passarino--Veltman integrals with a special
argument set needed in Appendix~\ref{chapter:edm}. All masses of the
external particles are set to zero ($p_{1}^{2} = (p_{1}-p_{2})^{2} =
p_{2}^{2} = 0$) and two of the internal particles are the same.\\
For the scalar B-function we obtain (see~\cite{helmut-diss})
\begin{equation}
B_0 (0, M_1^2, M_1^2) = \Delta + \log \left( \frac{Q^2}{M_1^2}
\right)
\end{equation}
% nicht ln statt log?!
with the real UV-divergence parameter $\Delta = \frac{2}{\epsilon} -
\gamma + \log (4 \pi)$, the Euler-Mascheroni Constant $\gamma \sim
0.577216$ and the scale parameter $Q$.\\
% evtl. für C_0 eine genauere Herleitung angeben (siehe Helmut-Zettel)
Using Feynman parametrization we derive for the scalar C-function
\begin{equation}
C_0 (f) = \frac{1 - x + \log(x)}{M_0^2 (1 - x)^2}
\end{equation}
with $x = M_1^2 / M_0^2$ and $f=(0,0,0,M_0^2,M_1^2,M_1^2)$.\\
The coefficient functions take the form
\begin{eqnarray}
C_1(f) & = & - \frac{3 - 4 x + x^2 + 2 \log (x)}{4 M_0^2 (1 - x)^3} \\
C_{00}(f) & = & \frac{1}{4} \bigg( \Delta + \log \left(
\frac{Q^2}{M_1^2} \right) + 1 + \frac{1 - x^2 + 2 \log (x)}{2 (1 -
x)^2} \bigg) \\
C_{11}(f) + C_{12}(f) & = & \frac{11 - 18 x + 9 x^2 - 2 x^3 + 6 \log
(x)}{12 M_0^2 (1 - x)^4} \, .
\end{eqnarray}
% wie lautet die Relation genau?!
In our special case, the relations $C_1 \sim C_2$ and $C_{11} +
C_{12} \sim C_{22} + C_{12}$ hold.

\newpage
%\chapter{Calculation of a generic structure}
\chapter{Calculation of a Generic Structure}

We show the calculation of a generic structure taking the generic
structure~I in
Section~\ref{section:most-important-generic-contributions} as an
example. First, we only take the matrix elements of the three
vertices alone:
\begin{eqnarray}
{\cal M}_0 & = & i \, \bar u(k_1) ( g_0^R P_R + g_0^L P_L ) v(-k_2) \label{eq:M_0-generic} \\
{\cal M}_1 & = & i \, \bar u(q+p) ( g_1^R P_R + g_1^L P_L ) v(-q) \\
{\cal M}_2 & = & i \, \bar u(q)   ( g_2^R P_R + g_2^L P_L )
v(-(q+p)).
\end{eqnarray}
Then we can write down the matrix element of the selfenergy loop
from ${\cal M}_1$ and ${\cal M}_2$ using Feynman rules. Because of
the closed fermion loop we introduce an additional overall factor
$(-1)$ (for a detailed derivation see \cite{helmut-diss}). We obtain
\begin{equation}
{\cal{M}}_\mathrm{self} = - \int \frac{\mathrm{d}^4 q}{(2 \pi)^4}
\frac{1}{D_0 D_1} X
\end{equation}
where we used the abbreviations
\begin{eqnarray}
D_0 & = & q^2 - M_1^2 \\
D_1 & = & ( q + p )^2 - M_2^2 \\
X & = & \mathrm{Tr} \bigl[ ( g_1^R P_R + g_1^L P_L ) ( \slashed{q} +
M_1 ) ( g_2^R P_R + g_2^L P_L ) ( \slashed{q} + \slashed{p} + M_2 )
\bigr]. \label{eq:M-self-generic}
\end{eqnarray}
We can modify $X$ to
\begin{eqnarray}
X & = & \mathrm{Tr} \bigl[ \slashed{q} ( \slashed{q} + \slashed{p} )
( g_1^R g_2^L P_R + g_1^L g_2^R P_L ) + \slashed{q} M_2 ( g_1^L
g_2^R P_R + g_1^R g_2^L P_L ) \nn \\
& & + ( \slashed{q} + \slashed{p} ) M_1 ( g_1^L g_2^L P_R + g_1^R
g_2^R P_L ) + M_1 M_2 ( g_1^R g_2^R P_R + g_1^L g_2^L P_L ) \bigr] \nn \\
& = & 2 ( q^2 + ( q.p ) ) ( g_1^R g_2^L + g_1^L g_2^R ) + 2 M_1 M_2
( g_1^R g_2^R + g_1^L g_2^L ) \label{eq:X-trace}
\end{eqnarray}
using the relations $P_{R,L}^2 = P_{R,L}$, $P_{R,L} P_{L,R} = 0$ and
$\mathrm{Tr} ( P_{R,L} ) = 2$, $\mathrm{Tr} ( \slashed{a}
\slashed{b} P_{R,L} ) = 2 ( a.b )$, $\mathrm{Tr} ( \slashed{a}
P_{R,L} ) = 0$.\\
Now we switch from $4$ to $D=4-\epsilon$ dimensions using the SUSY
invariant Dimensional Reduction regularization scheme $\overline
\mathrm{DR}$ resulting in $\mathrm{d}^4 q / (2 \pi)^4 \to
\mathrm{d}^D q / (2 \pi)^D$. Substituting this and
Eq.~(\ref{eq:X-trace}) into Eq.~(\ref{eq:M-self-generic}) results in
\begin{equation}
{\cal{M}}_\mathrm{self} = - \int \frac{\mathrm{d}^D q}{(2 \pi)^D}
\frac{1}{D_0 D_1} \Bigl[ 2 ( q^2 + ( q.p ) ) ( g_1^R g_2^L + g_1^L
g_2^R ) + 2 M_1 M_2 ( g_1^R g_2^R + g_1^L g_2^L ) \Bigr].
\end{equation}
For the calculation of the loop integrals we use the formalism of
Passarino--Veltman Integrals introduced in
Appendix~\ref{chapter:pave}. Using Eq.~(\ref{eq:B_0-pave}),
(\ref{eq:B_mu-pave}) and Eq.~(\ref{eq:B_mu_nu-pave}) reduced with
the metric tensor $g^{\mu \nu}$ we get
\begin{eqnarray}
\int \frac{\mathrm{d}^D q}{(2 \pi)^D} \frac{1}{D_0 D_1} & = &
\frac{i}{(4 \pi)^2} B_0 \\
\int \frac{\mathrm{d}^D q}{(2 \pi)^D} \frac{(q.p)}{D_0 D_1} & = & \frac{i}{(4 \pi)^2} p^2 B_1 \\
\int \frac{\mathrm{d}^D q}{(2 \pi)^D} \frac{q^2}{D_0 D_1} & = &
\frac{i}{(4 \pi)^2} ( 4 B_{00} + p^2 B_{11})
\end{eqnarray}
where the B-functions have $(p^2,M_1^2,M_2^2)$ as their arguments.
The matrix element then becomes
\begin{eqnarray}
{\cal{M}}_\mathrm{self} & = & - \frac{i}{(4 \pi)^2} \bigg[ 2 \Big( 4
B_{00} + m_0^2 (B_1 + B_{11}) \Big) ( g_1^R g_2^L + g_1^L g_2^R ) \nn \\
& & + 2 M_1 M_2 \Big( g_1^R g_2^R + g_1^L g_2^L \Big) B_0 \bigg] \nn \\
& = & - \frac{i}{(4 \pi)^2} \bigg[ \Big( A_0(M_1^2) + A_0(M_2^2) + ( M_1^2 + M_2^2 - m_0^2 ) B_0 \Big) ( g_1^R g_2^L + g_1^L g_2^R ) \nn \\
& & + 2 M_1 M_2 \Big( g_1^R g_2^R + g_1^L g_2^L \Big) B_0 \bigg]
\label{eq:M-self-generic-final}
\end{eqnarray}
where we used the on-shell relation $p^2=m_0^2$ and reduced the term
with $B_1$, $B_{00}$ and $B_{11}$ down to $A_0$ and $B_0$ using
equations found in \cite{helmut-diss}. The matrix element of the
final graph is then derived from Eq.~(\ref{eq:M_0-generic}) and
(\ref{eq:M-self-generic-final})
\begin{eqnarray}
\mathcal{M} & = & {\cal{M}}_\mathrm{self} \, \frac{i}{p^2 - M_0^2}
\, \mathcal{M}_0  \nn \\
& = & \frac{i}{(4 \pi)^2} \bar u(k_1) ( \Lambda_R P_R + \Lambda_L
P_L ) v(-k_2)
\end{eqnarray}
with the form factors
\begin{eqnarray}
\Lambda_R & = & \frac{1}{m_0^2 - M_0^2} g_0^R \bigg[ \Big(
A_0(M_1^2) +
A_0(M_2^2) + ( M_1^2 + M_2^2 - m_0^2 ) B_0 \Big) ( g_1^R g_2^L + g_1^L g_2^R ) \nn \\
& & + 2 M_1 M_2 \Big( g_1^R g_2^R + g_1^L g_2^L \Big) B_0 \bigg] \\
\Lambda_L & = & \Lambda_R \, (R \leftrightarrow L)
\end{eqnarray}
just like in Eq.~(\ref{eq:A_R-generic-scalar-selfenergy}).

\newpage
%\chapter{Calculation of a coupling}
\chapter{Calculation of a Coupling}

As an example for the calculation of couplings we derive the tree
level interaction of a chargino with a sfermion-fermion pair $\tilde
\chi^+ f \tilde f'$ as specified in
Chapter~\ref{chapter:couplings}.\\
Since charginos as well as sfermions are mass eigenstates that are
formed out of a mixing of interaction eigenstates, one has to
consider the Lagrangian of interaction eigenstates first. The
relevant terms are derived from the superpotential and the SUSY
gauge coupling terms (for a detailed derivation see
\cite{theoretikum}). We obtain
\begin{eqnarray}
\lagr_{\tilde \chi^+ f \tilde f'} & = & \lint_{\mathrm{Yuk}} + \lint_{\mathrm{weak}} \, , \\
\lint_{\mathrm{Yuk}} & = & h_t ( \tilde t^*_R b_L + \tilde b_L t^{\dagger}_R ) \psi^1_{H_2} + h_b ( \tilde b^*_R t_L + \tilde t_L b^{\dagger}_R ) \psi^2_{H_1} \nn \\
& & + h_{\tau} ( \tilde \tau^*_R \nu_{\tau} + \tilde \nu_{\tau} \tau^{\dagger}_R ) \psi^2_{H_1} + h.c. \, , \\
\lint_{\mathrm{weak}} & = &  i g \Bigl[ ( \tilde t^*_L b_L + \tilde \nu^*_{\tau} \tau_L ) \lambda^+ + ( \tilde b^*_L t_L + \tilde \tau^*_L \nu_{\tau} ) \lambda^- + h.c. \Bigr] \, .
\end{eqnarray}
Then we transform the Lagrangians to mass eigenstates using the
relations~(\ref{eq:sfermion-gauge2mass},
\ref{eq:chargino-gauge2mass}) resulting in
\begin{eqnarray}
\lint_{\mathrm{Yuk}} & = & h_t ( R^{\tilde t}_{i 2} \tilde t^*_i b_L + R^{\tilde b *}_{i 1} \tilde b_i t^{\dagger}_R ) V^*_{k 2} \chi^+_k + h_b ( R^{\tilde b}_{i 2} \tilde b^*_i t_L + R^{\tilde t *}_{i 1} \tilde t_i b^{\dagger}_R ) U^*_{k 2} \chi^-_k \nn \\
& & + h_{\tau} ( R^{\tilde \tau}_{i 2} \tilde \tau^*_i \nu_{\tau} + \tilde \nu_{\tau} \tau^{\dagger}_R
) U^*_{k 2} \chi^-_k + h.c. \, , \\
\lint_{\mathrm{weak}} & = & - g \bigl[ ( R^{\tilde t}_{i 1} \tilde t^*_i b_L + \tilde \nu^*_{\tau} \tau_L ) V^*_{k 1} \chi^+_k \nn \\
& & + ( R^{\tilde b}_{i 1} \tilde b^*_i t_L + R^{\tilde \tau}_{i 1} \tilde \tau^*_i \nu_{\tau} ) U^*_{k 1} \chi^-_k + h.c. \bigr] \, .
\end{eqnarray}
Summing up the results we gain
\begin{eqnarray}
\lagr_{\tilde \chi^+ f \tilde f'} & = & ( h_t R^{\tilde t}_{i 2} V^*_{k 2} - g R^{\tilde t}_{i 1} V^*_{k 1} ) \tilde t^*_i b_L \chi^+_k + h_t R^{\tilde b *}_{i 1} V^*_{k 2} \tilde b_i \tilde t^{\dagger}_R \chi^+_k - g V^*_{k 1} \tilde \nu^*_{\tau} \tau_L \chi^+_k \nonumber \\
& & + ( h_b R^{\tilde b}_{i 2} U^*_{k 2} - g R^{\tilde b}_{i 1} U^*_{k 1} ) \tilde b^*_i t_L \chi^-_k + h_b R^{\tilde t *}_{i 1} U^*_{k 2} \tilde t_i \tilde b^{\dagger}_R \chi^-_k \nonumber \\
& & + ( h_{\tau} R^{\tilde \tau}_{i 2} U^*_{k 2} - g R^{\tilde \tau}_{i 1} U^*_{k 1} ) \tilde \tau^*_i \nu_{\tau} \chi^-_k + h_{\tau} U^*_{k 2} \tilde \nu_{\tau} \tau^{\dagger}_R \chi^-_k + h.c. \, .
\end{eqnarray}
Finally we transform the Weyl spinors to Dirac spinors using the
relations~(\ref{eq:weyl-dirac-quark}, \ref{eq:weyl-dirac-lepton})
and separate the Lagrangian into a chargino-squark-quark and a
chargino-slepton-lepton part $\lagr_{\tilde \chi^+ f \tilde f'} =
\lagr_{\tilde \chi^+ q \tilde q'} + \lagr_{\tilde \chi^+ l \tilde
l'}$. We get
\begin{eqnarray}
{\cal L}_{\tilde \chi^+ q \tilde q'} & = & \bar t ( A^R_{k i} P_R + A^L_{k i} P_L ) \tilde \chi^+_k \tilde b_i + \bar b (B^R_{k i} P_R + B^L_{k i} P_L ) \tilde \chi^-_k \tilde t_i \nonumber \\
& & + \overline{\tilde \chi^+_k} ( A^{L *}_{k i} P_R + A^{R *}_{k i} P_L ) t \tilde b^*_i + \overline{\tilde \chi^-_k} ( B^{L *}_{k i}
P_R + B^{R *}_{k i} P_L ) b \tilde t^*_i \, , \\
{\cal L}_{\tilde \chi^+ l \tilde l'} & = & \bar \nu_{\tau} ( A'^R_{k i} P_R + A'^L_{k i} P_L ) \tilde \chi^+_k \tilde \tau_i + \bar \tau (B'^R_{k} P_R + B'^L_{k} P_L ) \tilde \chi^-_k \tilde \nu_{\tau} \nonumber \\
& & + \overline{\tilde \chi^+_k} ( A'^{L *}_{k i} P_R + A'^{R *}_{k i} P_L ) \nu_{\tau} \tilde \tau^*_i + \overline{\tilde \chi^-_k} ( B'^{L *}_{k} P_R + B'^{R *}_{k} P_L ) \tau \tilde \nu^*_{\tau} \, ,
\end{eqnarray}
with the projection operators $P_{R,L} = (1 \pm \gamma^5)/2$.\\
The abbreviated coupling matrices, which can be further transformed using $m_W = g \nu /2$, $m_{t} = ( h_{t} \nu \sin \beta ) / \sqrt 2$ and
$m_{b} = ( h_{b} \nu \cos \beta ) / \sqrt 2$, are
\begin{eqnarray}
A^R_{k i} & = & h_b R^{\tilde b *}_{i 2} U_{k 2} - g R^{\tilde b *}_{i 1} U_{k 1} \nonumber \\
& = & \frac{g}{\sqrt 2} \left( \frac{m_b}{m_W \cos \beta} U_{k 2} R^{\tilde b *}_{i 2} - \sqrt 2 U_{k 1} R^{\tilde b *}_{i 1} \right) \, , \nonumber \\
A^L_{k i} & = & h_t R^{\tilde b *}_{i 1} V^*_{k 2} = \frac{g m_t}{\sqrt 2 m_W \sin \beta} V^*_{k 2} R^{\tilde b *}_{i 1} \, , \nonumber \\
B^R_{k i} & = & h_t R^{\tilde t *}_{i 2} V_{k 2} - g R^{\tilde t *}_{i 1} V_{k 1} \nonumber \\
& = & \frac{g}{\sqrt 2} \left( \frac{m_t}{m_W \sin \beta} V_{k 2} R^{\tilde t *}_{i 2} - \sqrt 2 V_{k 1} R^{\tilde t *}_{i 1} \right) \, , \nonumber \\
B^L_{k i} & = & h_b R^{\tilde t *}_{i 1} U^*_{k 2} = \frac{g m_b}{\sqrt 2 m_W \cos \beta} U^*_{k 2} R^{\tilde t *}_{i 1} \, , \\
A'^R_{k i} & = & h_{\tau} R^{\tilde \tau *}_{i 2} U_{k 2} - g R^{\tilde \tau *}_{i 1} U_{k 1} \nn \\
& = & \frac{g}{\sqrt 2} \left( \frac{m_{\tau}}{m_W \cos \beta} U_{k 2} R^{\tilde \tau *}_{i 2} - \sqrt 2 U_{k 1} R^{\tilde \tau *}_{i 1} \right) \, , \nn \\
{A'}^L_{k i} & = & 0 \, , \nn \\
B'^R_{k} & = & - g V_{k 1} \, , \nn \\
B'^L_{k} & = & h_{\tau} U^*_{k 2} = \frac{g m_{\tau}}{\sqrt 2 m_W \cos \beta} U^*_{k 2} \, ,
\end{eqnarray}
just like in (\ref{eq:cha-sq-q-couplings}) and
(\ref{eq:cha-sl-l-couplings}). These coupling matrices as well as
the Lagrangians have the same convention as in
\cite{bernhard-dipl}.

\newpage
%\chapter{Transformation of Weyl to Dirac spinors}
\chapter{Transformation of Weyl to Dirac Spinors}

We define a four component Dirac spinor as follows
\begin{equation}
\Psi = { \chi \choose \bar \psi } ,
\end{equation}
where the two quantities $\chi$ and $\psi$ are two component Weyl spinors, standing for particle and anti-particle, respectively.\\
We choose the chiral representation so the gamma matrices  $\gamma^0$ and $\gamma^5$ and the projection operators $P_{R,L}$ as well as the adjoint Dirac spinor $\bar \Psi$ are:
\begin{equation}
\gamma^0 = \left( \begin{array}{cc}
0 & 1 \\
1 & 0
\end{array} \right)
\qquad
\gamma^5 = \left( \begin{array}{cc}
-1 & 0 \\
0 & 1
\end{array} \right) ,
\end{equation}
\begin{eqnarray}
P_R & = & \frac{1}{2} ( 1 + \gamma^5 ) = \left( \begin{array}{cc}
0 & 0 \\
0 & 1
\end{array} \right) , \nn \\
P_L & = & \frac{1}{2} ( 1 - \gamma^5 ) = \left( \begin{array}{cc}
1 & 0 \\
0 & 0
\end{array} \right) ,
\end{eqnarray}
\begin{equation}
\bar \Psi = \Psi^{\dagger} \gamma^0 = ( \psi \; \bar \chi ) \, .
\end{equation}
With these relations one can now carry out all sorts of transformations including products of Weyl spinors
\begin{equation}
\psi_1 \chi_2 = \bar \Psi_1 P_L \Psi_2   \qquad   \bar \chi_1 \bar \psi_2 = \bar \Psi_1 P_R \Psi_2 \, .
\label{eq:weyl-dirac-general}
\end{equation}
We define the fermions of the SM, the charginos as well as the neutralinos
\begin{eqnarray}
t = { t_L \choose t_R }  \quad  \bar t = ( t^{\dagger}_R \; t^{\dagger}_L )  & &  b = { b_L \choose b_R }  \quad  \bar b = ( b^{\dagger}_R \; b^{\dagger}_L ) \, , \nn \\
\nutau = { \nutau \choose 0 }  \quad  \bar \nutau = ( 0 \; \nutau^{\dagger} ) & & \tau = { \tau_L \choose \tau_R }  \quad  \bar \tau = ( \tau^{\dagger}_R \; \tau^{\dagger}_L ) \, ,
\end{eqnarray}
\begin{eqnarray}
\tilde \chi^+_k = { \chi^+_k \choose \bar \chi^-_k } \quad \overline{\tilde \chi^+_k} = ( \chi^-_k \; \bar \chi^+_k ) & & \tilde \chi^-_k = { \chi^-_k \choose \bar \chi^+_k } \quad \overline{\tilde \chi^-_k} = ( \chi^+_k \; \bar \chi^-_k ) \, ,
\end{eqnarray}
\begin{eqnarray}
\tilde \chi^0_k = { \chi^0_k \choose \bar \chi^0_k } & & \overline{\tilde \chi^0_k} = ( \chi^0_k \; \bar \chi^0_k ) \, .
\end{eqnarray}
Therefore we obtain for example the following transformations using Eq.~(\ref{eq:weyl-dirac-general})
\begin{eqnarray}
t_L \chi^-_k = \overline{\tilde \chi^+_k} P_L t  & &  b_L \chi^+_k = \overline{\tilde \chi^-_k} P_L b \, , \nn \\
t^{\dagger}_R \chi^+_k = \bar t P_L \tilde \chi^+_k  & &  b^{\dagger}_R \chi^-_k = \bar b P_L \tilde \chi^-_k \, , \nn \\
t_L \chi^0_k = \overline{\tilde \chi^0_k} P_L t & & b_L \chi^0_k = \overline{\tilde \chi^0_k} P_L b \, , \nn \\
t^{\dagger}_R \chi^0_k = \bar t P_L \tilde \chi^0_k & & b^{\dagger}_R \chi^0_k = \bar b P_L \tilde \chi^0_k \, ,
\label{eq:weyl-dirac-quark}
\end{eqnarray}
\begin{eqnarray}
\tau_L \chi^+_k = \overline{\tilde \chi^-_k} P_L \tau & \tau^{\dagger}_R \chi^-_k = \bar \tau P_L \tilde \chi^-_k & \nutau \chi^-_k = \overline{\tilde \chi^+_k} P_L \nutau \, , \nn \\
\tau_L \chi^0_k = \overline{\tilde \chi^0_k} P_L \tau & \tau^{\dagger}_R \chi^0_k = \bar \tau P_L \tilde \chi^0_k & \nutau \chi^0_k = \overline{\tilde \chi^0_k} P_L \nutau \, .
\label{eq:weyl-dirac-lepton}
\end{eqnarray}

\newpage
% evtl. kleine Fehler nicht ausgeschlossen! (aufgrund unterschiedlicher Definitionen)
%\chapter{The electric dipole moment (EDM)}
\chapter{The Electric Dipole Moment (EDM)}
\label{chapter:edm}

The additional CP-violating phases in the MSSM are new sources of CP
violation beyond the SM. From the point of view of baryogenesis, one
hopes that these phases are
large~\cite{Carena2003,Cohen1993,Riotto1999,Trodden1999}. But the
experimental limits on electron and neutron electric dipole moments
(EDMs), $|d_e| \leq 1.6 \times 10^{-27} \, [e \times
\mathrm{cm}]$~\cite{Regan2002} and $|d_n| \leq 2.9 \times 10^{-26}
\, [e \times \mathrm{cm}]$~\cite{Baker2006}, place constraints on
the CP violating phases of the MSSM. Especially the complex phase of
$\mu$ is severely constrained with about $\phi_\mu < \mathcal{O}
(10^{-2})$~\cite{Nath1991,Kizukuri1992,Garisto1997,Grossman1998} for
a typical SUSY mass scale of the order of a few hundred GeV. A
larger $\phi_\mu$ imposes fine-tuned relationships between this
phase and other SUSY
parameters~\cite{Ibrahim1998,Brhlik1999,Bartl1999}.\\
% Begründung fehlt, warum wir nur eEDM überprüfen!
In this chapter we calculate the EDM of an electron. We start by
giving the relevant generic structures and then derive the EDM of a
fermion. We apply this result to compute the EDM of the electron and
point out some numerical issues. Finally we use this calculation as
an automatic checkup-routine in our calculation of CP violating
asymmetries so as to make sure our chosen complex parameter set of
the MSSM is not already ruled out by experiment.

%\section{Generic structures}
\section{Generic Structures}

Here we list the two generic structures which are needed for the
calculation of the EDM of a fermion. Note that we rotated these two
generic vertex corrections to fit the generic structures found
in~\cite{helmut-diss}. The convention for the momenta and the masses
from Chapter~\ref{chapter:contributions} still applies but is of
course rotated as well.\\
The generic matrix element is
% stimmts?
\begin{equation}
\mathcal{M} = \frac{i}{(4 \pi)^2} \epsilon_\mu(p) \bar u(k_1) (
\gamma^\mu A_i + k^\mu B_i + p^\mu C_i ) P_i v(-k_2)
\label{eq:M-EDM-generic}
\end{equation}
with $i \in \{ R,L \}$. The form factors $A_L$, $B_L$ and $C_L$ can
be again easily obtained by exchanging right- and left-handed
couplings $(g^R \leftrightarrow g^L)$.\\
% compare Helmut-Diss (C.56-60)
The first generic structure is a scalar-fermion-fermion vertex
correction.
\begin{center}
\begin{tabular}{rl}
  \begin{tabular}{c}
  \includegraphics[width=0.4\paperwidth]{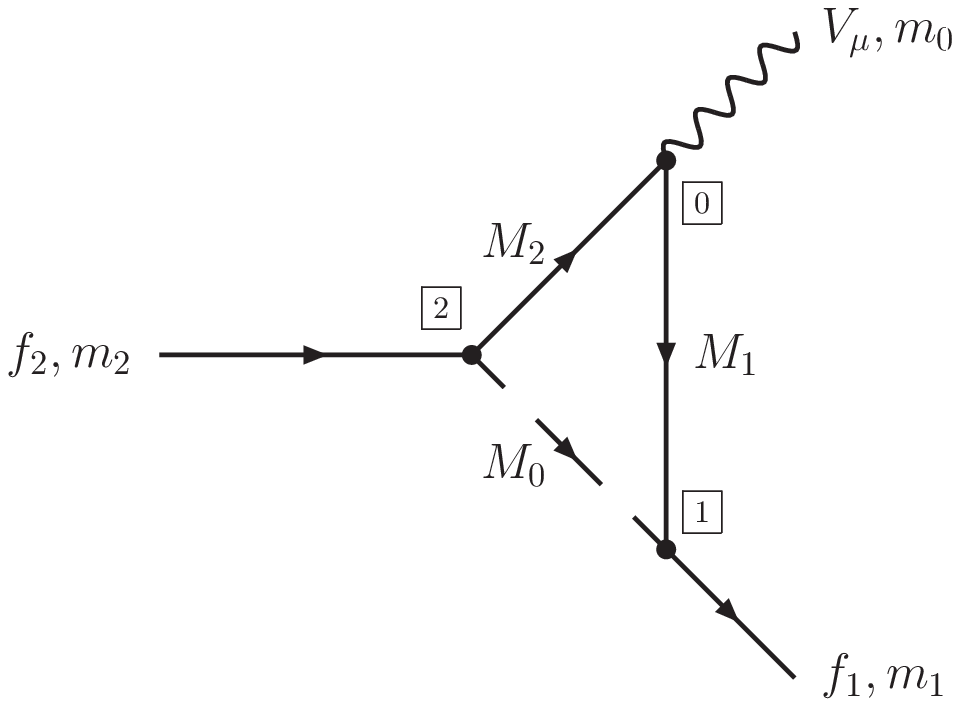}
  \end{tabular}
&
  \begin{tabular}{l}
  \framebox{0} : $i \gamma^\mu (g_0^R P_R + g_0^L P_L)$\\
  \framebox{1} : $i (g_1^R P_R + g_1^L P_L)$\\
  \framebox{2} : $i (g_2^R P_R + g_2^L P_L)$
  \end{tabular}
\end{tabular}
\end{center}
The appropriate form factors are
\begin{eqnarray}
A^I_R(f) & = & - \bigg( g_0^L g_1^L g_2^R \Big( 2 C_{00} -
B_0(m_0^2,M_1^2,M_2^2) \Big) + (g_0^R h_1^{RL} h_2^{LR} - g_0^L
g_1^L g_2^R M_0^2) C_0 \nn \\
& & + ( g_0^R g_1^R h_2^{LR} - g_0^L g_2^R h_1^{LR} ) m_1 C_1 - (
g_0^L g_1^L h_2^{RL} - g_0^R g_2^L h_1^{RL} ) m_2 C_2 \bigg) \, , \\
% evtl. hier ein Fehler! (evtl. g_0^L g_1^R h_2^{LR} C_2 statt g_0^R g_1^R h_2^{LR} C_2 oder so ...)
B^I_R(f) & = & - \Big( g_0^L g_2^R h_1^{LR} C_1 + g_0^R g_1^R
h_2^{LR}
C_2 \nn \\
& & + g_0^L g_1^L g_2^R m_1 (C_{11} + C_{12}) + g_0^R g_1^R g_2^L
m_2 (C_{22} + C_{12}) \Big) \, , \\
% evtl. hier ein Fehler! (siehe oben)
C^I_R(f) & = & - \Big( g_0^L g_2^R h_1^{LR} C_1 - g_0^R g_1^R
h_2^{LR}
C_2 \nn \\
& & + g_0^L g_1^L g_2^R m_1 (C_{11} - C_{12}) - g_0^R g_1^R g_2^L
m_2 (C_{22} - C_{12}) \Big)
\end{eqnarray}
with the auxiliary functions
\begin{equation}
h_i^{jk} = g_i^j m_i + g_i^k M_i \qquad i \in \{ 1,2 \} \, ; \, j,k
\in \{ L,R \}
\end{equation}
where we do not sum over the index $i$. The argument set is
\begin{equation}
f = (m_0,m_1,m_2,M_0,M_1,M_2,g_0^R,g_0^L,g_1^R,g_1^L,g_2^R,g_2^L) \,
.
\end{equation}
% compare Helmut-Diss (C.61-64)
The second generic structure is a fermion-scalar-scalar vertex
correction.
\begin{center}
\begin{tabular}{rl}
  \begin{tabular}{c}
  \includegraphics[width=0.4\paperwidth]{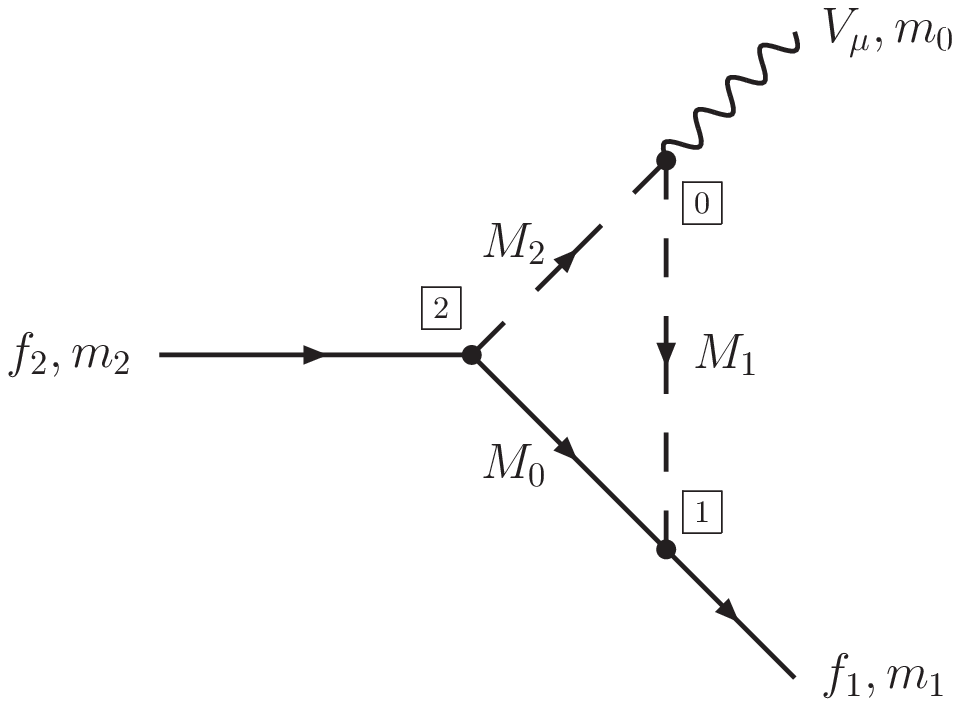}
  \end{tabular}
&
  \begin{tabular}{l}
  \framebox{0} : $i g_0 (2 q + k_1 + k_2)^\mu$\\
  \framebox{1} : $i (g_1^R P_R + g_1^L P_L)$\\
  \framebox{2} : $i (g_2^R P_R + g_2^L P_L)$
  \end{tabular}
\end{tabular}
\end{center}
The appropriate form factors are
\begin{eqnarray}
A^{II}_R(f) & = & 2 C_{00} g_0 g_1^L g_2^R \, , \\
B^{II}_R(f) & = & g_0 \Big( m_1 g_1^L g_2^R ( C_1 + C_{11} + C_{12} ) + m_2 g_1^R g_2^L ( C_2 + C_{22} + C_{12} ) \nn \\
& & - M_0 g_1^R g_2^R ( C_0 + C_1 + C_2 ) \Big) \, , \\
C^{II}_R(f) & = & g_0 \Big( m_1 g_1^L g_2^R ( C_{11} - C_{12} ) -
m_2 g_1^R g_2^L ( C_{22} - C_{12} ) \nn \\
& & - M_0 g_1^R g_2^R ( C_1 - C_2 ) \Big)
\end{eqnarray}
with the argument set $f =
(m_0,m_1,m_2,M_0,M_1,M_2,g_0,g_1^R,g_1^L,g_2^R,g_2^L)$.

%\section{Electric dipole moment of a fermion}
\section{Electric Dipole Moment of a Fermion}

Now we are ready for the calculation of the EDM of a fermion. We
start with the matrix element in Eq.~(\ref{eq:M-EDM-generic}) and
apply the Gordon decomposition in order to obtain the $\sigma^{\mu
\nu}$-terms which describe the spin exchange of the two fermions.
Then we use these terms to derive the EDM using the relations
found in~\cite{Bartl2003}.\\
The Gordon decomposition takes the general form
% stimmts?
\begin{equation}
\bar u(p') \gamma^\mu P_i u(p) = \frac{1}{2 m} \bar u(p') ( p' + p
)^\mu P_i u(p) + \bar u(p') i \frac{\sigma^{\mu \nu}}{2 m} ( p' - p
)_\nu P_i u(p)
\end{equation}
with $i \in \{ R,L \}$. Using this relation one can transform the
first two terms\footnote{The third term with $C_i$ vanishes in the
case of a photon due to the Lorentz condition $\epsilon_\mu(p) p^\mu
= 0$.} in Eq.~(\ref{eq:M-EDM-generic}) and we obtain
\begin{eqnarray}
\mathcal{M} & = & \frac{i}{(4 \pi)^2} \epsilon_\mu(p) \bigg( A_i
\Big( \frac{1}{2 m_f} \bar u(k_1) k^\mu P_i u(k_2) + \bar u(k_1) i
\frac{\sigma^{\mu \nu}}{2 m_f} p_\nu P_i u(k_2) \Big) \nn \\
& & + B_i \Big( 2 m_f \bar u(k_1) \gamma^\mu P_i u(k_2) - i \bar
u(k_1) \sigma^{\mu \nu} p_\nu P_i u(k_2) \Big) \bigg) \, .
\end{eqnarray}
Taking only the terms with $\sigma^{\mu \nu}$ and rearranging a bit
yields for the amplitude $T$ ($\mathcal{M}^{\sigma^{\mu \nu}} = i
T$) (see~\cite{Bartl2003})
% stimmts?
\begin{equation}
T = i e \epsilon_\mu(p) \frac{p_\nu}{2 m_f} \bar u(k_1) \sigma^{\mu
\nu} ( a^R P_R + a^L P_L ) u(k_2)
\end{equation}
with the coefficients
% stimmts?!
\begin{eqnarray}
a^R & = & \frac{1}{e (4 \pi)^2} (A_R - 2 m_f B_R) \label{eq:a^R-EDM} \\
a^L & = & \frac{1}{e (4 \pi)^2} (A_L - 2 m_f B_L) \label{eq:a^L-EDM}
\, .
\end{eqnarray}
The EDM of a fermion is then calculated by the simple
formula\footnote{On the right-hand-side of Eq.~(12)
in~\cite{Bartl2003} there is a factor $-1/(2 m_f)$ missing.}
\begin{equation}
d_f = - \frac{e}{4 m_f} \mathrm{Im} ( a^R - a^L ) \label{eq:d_f-EDM}
\, .
\end{equation}

%\section{Electric dipole moment of the electron (eEDM)}
\section{Electric Dipole Moment of the Electron (eEDM)}

\begin{figure}[htbp]
\begin{center}
\begin{tabular}{rl}
  \includegraphics[width=0.35\paperwidth]{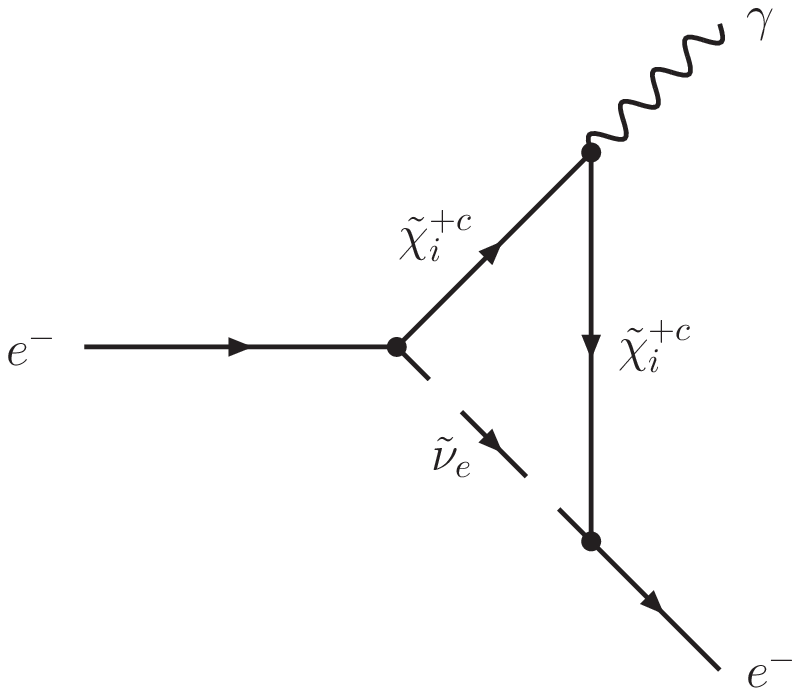}
&
  \includegraphics[width=0.35\paperwidth]{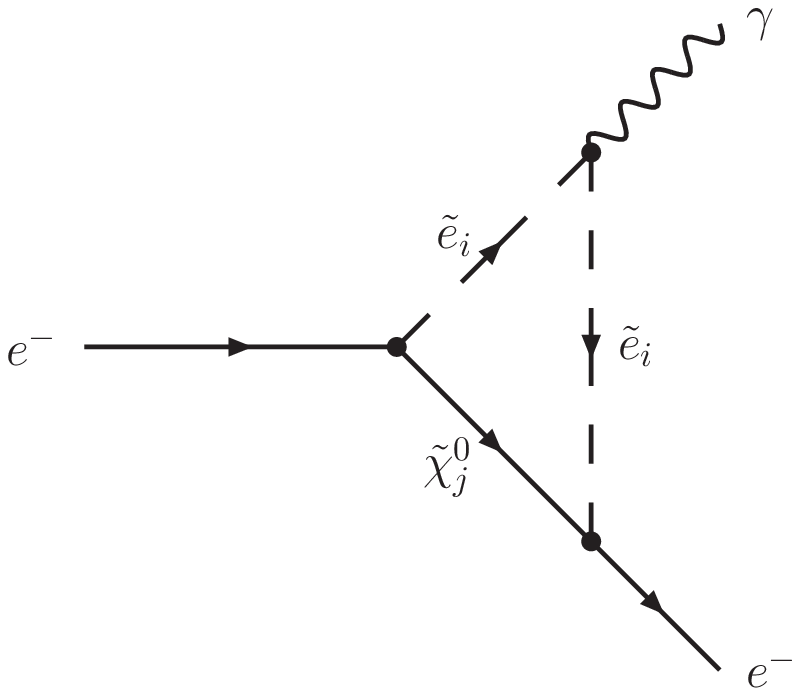}
\end{tabular}
\end{center}
\caption{The two processes who contribute to the eEDM. On the left a
sneutrino-chargino and on the right a selectron-neutralino in the
loop.} \label{fig:eEDM}
\end{figure}

There are two processes who contribute to the EDM of the electron,
one with a sneutrino-chargino and one with a selectron-neutralino in
the loop (see Fig.~\ref{fig:eEDM}).

\noindent The form factors for the sneutrino-chargino contribution
have the following arguments
% f = (m_0,m_1,m_2,M_0,M_1,M_2,g_0^R,g_0^L,g_1^R,g_1^L,g_2^R,g_2^L)
% stimmts?!
\begin{eqnarray}
A^1_{R,L} & = & \sum_{i=1}^{2} A^I_{R,L}( 0,m_e,m_e, m_{\tilde
\nu_e},m_{\tilde \chi^+_i},m_{\tilde \chi^+_i},
e,e,B'^R_{i},B'^L_{i},B'^{L*}_{i},B'^{R*}_{i} )
\end{eqnarray}
% es gibt (noch) einen anderen, triftigeren Grund für m_e=0, aber welcher?!
and analogously for $B^1_{R,L}$.\\
The form factors for the selectron-neutralino contribution have the
following arguments
% f = (m_0,m_1,m_2,M_0,M_1,M_2,g_0,g_1^R,g_1^L,g_2^R,g_2^L)
% stimmts?
\begin{eqnarray}
A^2_{R,L} & = & \sum_{i=1}^{2} \sum_{j=1}^{4} A^{II}_{R,L}(
0,m_e,m_e, m_{\tilde \chi^0_j},m_{\tilde e_i},m_{\tilde e_i},
e,D'^R_{ji},D'^L_{ji},D'^{L*}_{ji},D'^{R*}_{ji} )
\end{eqnarray}
and analogously for $B^2_{R,L}$.\\
Because of the huge masses of the particles in the loop compared to
the mass of the external electron, one has to carefully avoid
numerical instabilities. Especially the Passarino--Veltman
integrals, where we set the electron mass to zero, need special
treatment. As now all three external particles are massless, we have
to use the special Passarino--Veltman integrals derived in
Appendix~\ref{chapter:pave}
in order to get a numerically stable result.\\
After the careful calculation of the form factors we inserted them
in Eq.~(\ref{eq:a^R-EDM}) and (\ref{eq:a^L-EDM}) and calculated the
EDM of the electron with Eq.~(\ref{eq:d_f-EDM}). To compare the
result with the experimental boundaries given in units of $[ e
\times \mathrm{cm} ]$ we used the conversion factor $(m_e)^{-1}
[\mathrm{GeV}]^{-1} = 3.861592641934875 \times 10^{-11}
[\mathrm{cm}]$.\\
We checked our numerical result of $d_e$ (using our own coefficients
$a^R$ and $a^L$) with the result of $d_e$ obtained by using the
coefficients $a^R_{ii}$ and $a^L_{ii}$ found in~\cite{Bartl2003}.
Finally we incorporated this calculation of the eEDM as an automatic
checkup-routine in our calculation of CP violating asymmetries in
order to verify our complex parameter set of the MSSM.

\backmatter

\newpage
% ***************************************
% BibTex output file (.bbl) of document inserted here due to LaTeX processing issues with arXiv
% ***************************************
%\bibliographystyle{h-physrev}
%\bibliography{bibliography/thesis-bibliography}

% ***************************************
\newpage
% evtl. Berufserfahrung angeben (insbesonders Lehrtätigkeit!)
\chapter{Curriculum Vitae}

{
\setlength{\baselineskip}{1.4\baselineskip}

\textbf{Sebastian Frank}\\
Johann Wilhelm Kleinstra{\ss}e 72\\
A-4040 Linz, Austria\\
E-mail: sebastian.frank@gmail.com
\vspace{1.2cm}\\
\textbf{Personal Data}
\vspace{-0.4cm}\\
\hrule
\vspace{0.8cm}
\renewcommand{\arraystretch}{1.3}
\begin{tabular}{p{6cm}l}
Date/Place of Birth & September 27, 1978\\
                    & Innsbruck, Austria\\
Nationality         & Austrian\\
Marital Status      & Married
\end{tabular}
\vspace{1.2cm}\\
\textbf{Education}
\vspace{-0.4cm}\\
\hrule
\vspace{0.8cm}
\renewcommand{\arraystretch}{1.4}
\begin{tabular}{p{4cm}p{10cm}}
04/2007 --- 05/2008   & Diploma thesis at the Institute of High Energy Physics\\
                      & Austrian Academy of Sciences\\
09/2003 --- 02/2004   & Exchange semester (ERASMUS)\\
                      & Universidad de Granada, Spain\\
10/1999 --- 05/2008   & Study of Technical Physics (Dipl.Ing.)\\
                      & Johannes Kepler University Linz\\
                      & (Major subjects Theoretical Physics and\\
                      & Nanoscience \& -technology)\\
\end{tabular}
\newpage
\renewcommand{\arraystretch}{1.4}
\begin{tabular}{p{4cm}p{10cm}}
10/1997 --- 06/1998   & Study of Computer Science\\
                      & Johannes Kepler University Linz\\
                      & (after civilian service change to Technical Physics)\\
06/1997               & School leaving examination\\
                      & Bundesrealgymnasium Sillgasse, Innsbruck\\
%                      & (High school with an accent on natural sciences)
\end{tabular}
\vspace{1.2cm}\\
\textbf{Conference Attendance}
\vspace{-0.4cm}\\
\hrule
\vspace{0.8cm}
\renewcommand{\arraystretch}{1.4}
\begin{tabular}{p{4cm}p{10cm}}
01/2008               & Third Graduate School in Physics at Colliders\\
                      & Turin, Italy
\end{tabular}
\vspace{1.2cm}\\
\textbf{Commitments}
\vspace{-0.4cm}\\
\hrule
\vspace{0.8cm}
\renewcommand{\arraystretch}{1.4}
\begin{tabular}{p{4cm}p{10cm}}
10/2004 --- 06/2005   & Tutor for General Physics 1 and 2\\
07/2001 --- 06/2003   & Chairman of the physics students representation\\
                      & Austrian National Union of Students ({\"O}H)
\end{tabular}
\vspace{1.2cm}\\
\textbf{Skills}
\vspace{-0.4cm}\\
\hrule
\vspace{0.8cm}
\renewcommand{\arraystretch}{1.4}
\begin{tabular}{p{4cm}p{10cm}}
Languages             & German (native), English (fluent), Turkish (advanced),\\
                      & Spanish (advanced), French (basic)\\
Computer              & Windows, Mac OS X, Linux,\\
                      & Programming languages (Fortran, Mathematica, C,\\
                      & PHP, \dots), \LaTeX, XHTML/CSS
\end{tabular}
\vspace{1.2cm}\\
\textbf{Interests}
\vspace{-0.4cm}\\
\hrule
\vspace{0.8cm}
\renewcommand{\arraystretch}{1.4}
\begin{tabular}{p{14cm}}
Traveling, Trekking, Photography, Reading, Cinema
\end{tabular}

}

\end{document}